\documentclass[aps,prd,groupedaddress,nofootinbib,english,nobalancelastpage,floatfix,superscriptaddress,preprintnumbers]{revtex4-1}
\pdfoutput=1
\usepackage{flushend}
\DeclareMathAlphabet{\scr}{U}{rsfs}{m}{n}
\usepackage{graphicx}
\usepackage{amsmath,bm}
\usepackage{multirow}
\usepackage{rotating}
\usepackage{float}
\usepackage[normalem]{ulem}

\usepackage[usenames,dvipsnames]{xcolor}
\definecolor{naviBlue}{RGB}{0,0,128}
\usepackage[colorlinks  = true,
            linkcolor   = naviBlue,
            urlcolor    = naviBlue,
            citecolor   = naviBlue,
            anchorcolor = naviBlue]{hyperref}

\newcommand{\diff}{\mathrm{d}}

\newcommand{\KCl}{KC16}
\newcommand{\KCh}{KC21}
\newcommand{\DB}{DIFF.BRK}
\newcommand{\DR}{INJ.BRK$+v_{\rm A}$}

\begin{document}

\title{Testing the universality of cosmic-ray nuclei from protons to oxygen with AMS-02}

\author{Michael Korsmeier}
\email{michael.korsmeier@fysik.su.se}
\affiliation{The Oskar Klein Centre, Department of Physics, Stockholm University, AlbaNova, SE-10691 Stockholm, Sweden}
\author{Alessandro Cuoco}
\email{alessandro.cuoco@unito.it}
\affiliation{Dipartimento di Fisica, Universit\`a di Torino, Via P. Giuria 1, 10125 Torino, Italy}
\affiliation{Istituto Nazionale di Fisica Nucleare, Sezione di Torino, Via P. Giuria 1, 10125 Torino, Italy}

\begin{abstract}
The AMS-02 experiment has provided high-precision measurements of several cosmic-ray (CR) species. The achieved percent-level accuracy gives access to small spectral differences among the different species and, in turn, this allows scrutinizing the universality of CR acceleration, which is expected in the standard scenario of CR shock acceleration. While pre-AMS-02 data already indicated a violation of the universality between protons and helium, it is still an open question if at least helium and heavier nuclei can be reconciled. To address this issue, we performed a joint analysis using the AMS-02 CR measurements of antiprotons, protons, helium, helium 3, boron, carbon, nitrogen, and oxygen. We explore two competing propagation scenarios, one with a break in the diffusion coefficient at a few GVs and no reacceleration, and another one with reacceleration and with a break in the injection spectra of primaries. Furthermore, we explicitly consider the impact of the uncertainties in the nuclear production cross-sections of secondaries by including nuisance parameters in the fit. The resulting parameter space is explored with the help of Monte Carlo methods. We find that, contrary to the naive expectation, in the standard propagation scenarios CR universality is violated also for He, on the one hand, and C, N, and O, on the other hand, \emph{i.e.}, different injection slopes (at the level of $ \Delta \sim 0.05$) are required to explain the observed spectra. As an alternative, we explore further propagation scenarios, inspired by non-homogeneous diffusion, which might save universality. Finally, we also investigate the universality of CR propagation, \emph{i.e.}, we compare the propagation properties inferred using only light nuclei ($\bar{p}$, p, He, $^3$He) with the ones inferred using only heavier nuclei (B, C, N, O).
\end{abstract}

\maketitle

%===================================================================
\section{Introduction}
%===================================================================

In the last few years the AMS-02 experiment has provided precise measurements, at the level of a few percent, of several cosmic rays (CR) species in a large energy range between 1~GeV and a few TeV \cite{Aguilar:2021tos}. The consequences of these data for CR physics have been explored in several publications \cite{Korsmeier:2016kha,Tomassetti:2017hbe,Liu:2018ujp,Genolini:2019ewc,Weinrich:2020cmw,Weinrich:2020ftb,Evoli:2019wwu,Evoli:2019iih,Boschini:2018baj,Boschini:2019gow,Boschini:2020jty,Luque:2021nxb,DeLaTorreLuque:2021yfq,Schroer:2021ojh,Zhao:2021yzf}. 

An aspect that can be experimentally tested with this level of precision is the universality of CR acceleration. In the standard scenario of CR acceleration in supernova remnant (SNR) shocks, the spectrum of accelerated particles depends only on their rigidity not on the CR species itself, thus, a universal spectrum of CR is expected. Thus it was surprising when the first precise measurements of protons ($p$) and Helium (He) from PAMELA \cite{Adriani:2011cu} and CREAM \cite{Ahn:2010gv} showed that the two spectra have different slopes at the level of few percent but at a high significance. 
The origin of these ``discrepant hardenings" \cite{Ahn:2010gv} is still an open issue although several ideas have been explored.

One possibility is to explain the difference by propagation effects while keeping spectra universal \emph{at the source} as predicted by basic CR shock acceleration theory. The spectra observed at Earth are the ones after propagation in the Galaxy and the Heliosphere. Propagation is species-dependent due to the different energy losses, in particular, the inelastic energy losses due to fragmentation/spallation on the interstellar gas. These effects indeed alter the slope of the final propagated spectrum. 
In \cite{Thoudam:2014sta}, the authors show that it is in principle possible to explain the $p$-He difference by introducing reacceleration at weak (old) SNR shocks in the propagation framework. This explanation, however, could be problematic for the global CR energy budget \cite{Drury:2016ubm}. Other attempts to explain the difference with propagation and energy losses seem to require stretching the known uncertainties on either the inelastic cross sections or the gas density \cite{Vladimirov:2011rn} or on the propagation parameters \cite{Korsmeier:2016kha} (hereafter \KCl) by too much.

The alternative approach is to generate different $p$ and He spectra \emph{at the source level}. Although, as mentioned above, the basic CR shock acceleration theory predicts a species-independent spectrum, various ideas have been proposed to achieve different spectra at the source \cite{Tomassetti:2015xem,Tomassetti:2016bcf,Lagutin:2019nfn,Ohira:2010eq,Malkov:2011gb,Hanusch:2018bsk,Erlykin:2015mea,Fisk:2012zz,Ptuskin:2012qu}. For example, one could have different populations of source classes. A hydrogen-rich local source with a steep spectrum could explain the difference between $p$ and heavier nuclei \cite{Tomassetti:2015xem,Tomassetti:2016bcf,Lagutin:2019nfn}. Alternatively, Type I and Type II SNR have different hydrogen and helium compositions and can produce different accelerated CR spectra for $p$ and He when non-linear acceleration is considered \cite{Ptuskin:2012qu}.
On the other hand, more sophisticated studies of shock acceleration exploiting numerical simulations now predict an $A/Z$ dependence and shock Mach number dependence of the abundance of accelerated species~\cite{Hanusch:2018bsk,Caprioli:2017oun}. In particular, heavy particles are accelerated more efficiently in strong shocks~\cite{Hanusch:2018bsk,Caprioli:2017oun}. If this is combined with the fact that the Mach number of the shock decreases with time as the SNR expands, this can also quantitatively explain the deviation between $p$ ($A/Z=1$) and He ($A/Z=2$)~\cite{Ohira:2010eq,Malkov:2011gb,Hanusch:2018bsk}. For this class of explanations, we note that an $A/Z$-dependent spectrum still implies the same injection spectra for He, C, and O which all have $A/Z=2$.

With the recent precise measurements of AMS-02 it is, thus, natural to extend the test of CR universality to heavier CR species with the same $A/Z$ ratio, in particular, for Carbon (C) and Oxygen (O), i.e. the two most abundant species after Helium which are directly accelerated at the sources, the so-called primary CRs. Here, the measurements of AMS-02 find the same slope within uncertainties \cite{Aguilar:2017hno}, suggesting that CR universality is preserved. However, further inspection suggests the opposite conclusion as the  correct one. Indeed, if He, C, and O are injected at the sources with the same spectrum, the above-mentioned propagation effects should produce different slopes for the observed fluxes since these nuclei experience different energy losses. The situation is actually more complex because He and C are not completely primary but have a non-negligible (at the $\sim 10\%$ level) secondary component, i.e., He and C produced during propagation from the fragmentation of heavier nuclei. The presence of this secondary component also alters the slope of the observed spectrum, which is the sum over all isotopes. Separate isotopic measurements are available only in a few cases and for limited energy ranges. One of the main goals of the present study is thus to address this issue in detail and investigate whether CR universality for He, C, and O holds or not.

In parallel with the universality of CR acceleration, a similar universality can be discussed for CR propagation in the Galaxy. Historically, propagation of CR has mainly been investigated using Boron (B) and the B/C ratio, since B is almost completely absent in the sources and is instead produced from the fragmentation of heavier nuclei during propagation, i.e., it is a \emph{secondary} CR nucleus. Lithium (Li) and beryllium (Be) are also secondaries, but only recently they have been measured precisely by AMS-02 and have started to be used to constrain CR propagation (see e.g., \cite{Korsmeier:2021brc}, hereafter \KCh). Nonetheless, there are further lighter secondary nuclei which can be used to study propagation, namely $\bar{p}$ \cite{Korsmeier:2016kha}, and the light CR isotopes Deuterium ($^2\mathrm{H}$) and Helium 3 ($^3\mathrm{He}$). These three are all secondaries and their production involves mainly only $p$ and He interactions, thus, the system $p$, He, $\bar{p}$, $^2\mathrm{H}$, $^3\mathrm{He}$ constitutes an almost closed system that can be used to study CR propagation with light nuclei only. Some studies in this sense have indeed been performed pre-AMS-02 \cite{Coste:2011jc,Wu:2018lqu}. Again, the recent measurements from AMS-02 of $^3\mathrm{He}$ \cite{Aguilar:2019eiz} demands to revisit the issue. The new AMS-02 measurements are taken at higher energies than the previous PAMELA measurements \cite{PAMELA:2013tuo} and thus less affected by solar modulation. Consequently, they are expected to provide more robust constraints. AMS-02 measurements of $^2\mathrm{H}$ are not yet available but expected soon. Thus, the second main issue we will address in this work concerns the universality of CR propagation for light and heavy CR species. While this universality is expected in the standard scenario of CR propagation, it can be, however, violated in more complex scenarios, for example in presence of a non-homogenous propagation medium. Some indications in this direction have indeed been suggested in past analyses with older data \cite{Johannesson:2016rlh}. 

Following \KCh, to address the above issues we will consider CR propagation within two main complementary frameworks. In the first one, we do not consider reacceleration, and the main ingredient is CR diffusion described by a smoothly broken power-law in rigidity with a break at about $\sim$ 5--10~GeV. The spectrum of primaries at the sources (the injection spectrum) is described with a single power law. In the second one, instead, reacceleration is included and diffusion is described by a single power law. A break at $\sim$ 5--10~GeV is instead present in the injection spectrum of the primaries. The second key aspect which was introduced in \KCh, and which we reiterate here, is taking into account the systematic uncertainties in the nuclear cross-section for the production of secondaries, which are included in the analysis by introducing nuisance parameters. Most of these cross sections are poorly measured and thus the propagation of their uncertainties on the propagated CR spectra can be significant. As stressed in \KCh, including these uncertainties is crucial. In particular, propagation scenarios that at first seem excluded remain still viable once these uncertainties are considered. Finally, from a technical point of view as in \KCh, we will perform a simultaneous exploration of the propagation and cross-section uncertainties in a global fit. We will rely on Monte Carlo scanning techniques to explore this large joint parameter space.

As clear from the above description, we will rely heavily on \KCh\ for the methodologies, thus, in the following we will provide a short summary of the analysis steps, referring to \KCh\ for a more detailed description. The remainder of the article is structured as follows: In the next section, we give a general overview of the CR propagation framework used in this work. We then specify the employed CR data sets in Sec.~\ref{sec::data}, the treatment of nuisance parameters related to secondary production cross sections in Sec.~\ref{sec::cs}, and the Monte Carlo based fit methods in Sec~\ref{sec::methods}. A comparison of the results for light vs heavy nuclei is provided in Sec.~\ref{sec::motivation}, while the scenario investigated in the main fits are detailed in Sec.~\ref{sec::strategy} and the results presented in Sec.~\ref{sec::results}. Finally, we compare our results to other works in Sec.~\ref{sec::comparison} before concluding in Sec.~\ref{sec::conclusions}.

%===================================================================
\section{CR propagation} \label{sec::cr_prop}
%===================================================================

The AMS-02 experiment measures CRs between one GeV and a few TeV. These CRs move in the magnetic field of our Galaxy and are deflected by it. Effectively, this process can be modeled with a diffusion equation and propagation in the so-called diffusive halo which extends a few kpc above and below the Galactic plane. While diffusion is the dominant effect at high energies, CRs below $\sim$10 GV can experience a significant impact from convection, reacceleration, or energy losses. Furthermore, so-called primary CRs can fragment by collisions on the interstellar medium (ISM) and produce so-called secondary CRs. All these processes can be modeled by a chain of coupled diffusion equations for the CR density $\psi_i$ of species $i$. More specifically, indicating with $\psi_i(\bm{x}, p, t)$ the CR number density per volume and absolute momentum at the place $\bm{x}$, momentum $p$ and time $t$, the diffusion equation reads~\cite{StrongMoskalenko_CR_rewview_2007}:
\begin{eqnarray}
 \label{eqn::propagationEquation}
 \frac{\partial \psi_i (\bm{x}, p, t)}{\partial t} = 
 q_i(\bm{x}, p) &+&  
 \bm{\nabla} \cdot \left( D_{xx} \bm{\nabla} \psi_i - \bm{V} \psi_i \right) \nonumber \\ 
 &+& \frac{\partial}{\partial p} p^2 D_{pp} \frac{\partial}{\partial p} \frac{1}{p^2} \psi_i - 
 \frac{\partial}{\partial p} \left( \frac{\diff p}{\diff t} \psi_i  
 - \frac{p}{3} (\bm{\nabla \cdot V}) \psi_i \right) -
 \frac{1}{\tau_{f,i}} \psi_i - \frac{1}{\tau_{r,i}} \psi_i.
\end{eqnarray} 
We model the diffusion coefficient $D_{xx}$ as a double broken power law in rigidity $R$ with the two break positions at $R_{D,0}$ and $R_{D,1}$. The power-law index below, between, and above the breaks are labeled $\delta_l$, $\delta$, and $\delta_h$, respectively. Furthermore, we allow for a smoothing of the breaks mediated by the parameters $s_{D,0}$ and $s_{D,1}$. Explicitly, the diffusion coefficient is given by
\begin{equation}
 \label{eqn:diffusion_coefficient}
 D_{xx} = \beta R^{\delta_l}
 	 \cdot \left( 1 + \left(\frac{R}{R_{D,0}}\right)^{1/s_{D,0}} \right)^{s_{D,0}\,( \delta - \delta_l) }  
 	 \cdot \left( 1 + \left(\frac{R}{R_{D,1}}\right)^{1/s_{D,1}} \right)^{s_{D,1}\,( \delta_h - \delta) }
\end{equation}
where $\beta$ is the CR velocity in units of speed of light. We allow for convection of CRs away from the Galactic plane with a constant convection velocity $\bm{V}(\bm{x})= {\rm sign}(z)\, v_{0,{\rm c}}\,{\bm e}_z$. Reacceleration is modeled as diffusion in momentum space with $D_{pp} \sim v_\mathrm{A}/D_{xx}$. Here $v_\mathrm{A}$ is the speed of Alfv\'en magnetic waves. CRs can experience continuous energy losses included in the term ${\diff p}/{\diff t}$ or catastrophic losses by fragmentation or decay with the respective interaction and decay times $\tau_{f,i}$ and $\tau_{r,i}$. 

Finally, $q_i(\bm{x}, p)$ is the source term for each species $i$. For primary CRs, which are accelerated and injected in astrophysical sources, we assume that the spatial distribution follows the distribution of SNRs \cite{Green:2015isa}. The energy spectrum is assumed to be a smoothly broken power-law in rigidity where we denote the spectral indices above and below the break at position $R_\mathrm{inj,0}$ by $\gamma_1$ and $\gamma_2$. The amount of smoothing is regulated by the parameter $s$ defined similarly to Eq.~\ref{eqn:diffusion_coefficient}. In principle, the injection and thus parameters $\gamma_1$ and $\gamma_2$ can be different for different species, thus an additional subscript can be present. For secondary CRs, which are produced in the interaction and fragmentation of heavier primary CRs, the source term is given by the convolution of the primary fluxes, the ISM density, and the fragmentation cross-sections (see Sec.~\ref{sec::cs}). An additional ingredient is given by CR propagation in the Solar system. This so-called Solar modulation is treated in the force-field approximation \cite{Fisk:1976aw} which is fully determined by a single parameter, the solar modulation potential $\varphi$. We allow also allow for a charge-sign dependence, namely, the freedom to adjust the solar modulation parameter for antiparticles, which in our case are given only by antiprotons. For more details, we refer to \KCl\ and \KCh.

We employ the \textsc{Galprop} code\footnote{http://galprop.stanford.edu/} \cite{Strong:1998fr,Strong:2015zva} to solve the chain of coupled of diffusion equations, see Eq.~\eqref{eqn::propagationEquation}, numerically.%
\footnote{
 \textsc{Dragon}~\cite{Evoli:2008dv,Evoli:2017vim} and \textsc{Picard}~\cite{Kissmann:2014sia} are alternative numerical codes solving the diffusion equations. 
 Assuming additional simplifications a (semi-)analytically treatment is also possible \cite{Putze:2010zn,Maurin:2018rmm}.
}
The diffusion is approximated to be cylindrically symmetric with a radial extent of 20 kpc and the half-height of the diffusion halo $z_h$ is fixed to 4 kpc. There is a well-known degeneracy between $z_h$ and $D_0$, and the Be/B ratio measured by AMS-02 disfavors values of $z_h$ smaller 3 kpc (\KCh), thus, for simplicity, $z_h$ is fixed to 4 kpc. To solve the chain of equations, we include CR nuclei up to silicon in our analysis. As a starting point for analysis, we use \textsc{Galprop} version~56.0.2870 combined with \textsc{Galtoollibs}~855\footnote{https://galprop.stanford.edu/download.php} but we have implemented some custom modification as detailed in \KCh. As described above we have implemented smoothly broken power laws with up to two breaks both for the primary injection spectra and the diffusion coefficient. Furthermore, we implemented the possibility to adjust the injection spectrum individually for each species. Then, as discussed more in detail later, we will consider models where diffusion may vary from species to species and thus we have also implemented the possibility to adjust the diffusion coefficient individually for each species. Finally, as detailed in Sec.~\ref{sec::cs}, we have implemented nuisance parameters to allow freedom in the default fragmentation cross-sections for the production of secondary CRs and we have changed the default cross-section for secondary antiproton production to the one from Ref.~\cite{Korsmeier:2018gcy} (Parm. IIB). 

\medskip

In \KCh, we have studied five different setups of CR propagation and confronted them with the latest AMS-02 data from lithium to oxygen. These setups differ regarding the inclusion of a diffusion break at a few GV, the inclusion of a break in the injection spectrum, and the presence of reacceleration. Here we consider the two frameworks which are, firstly, most different from the physics point of view and, secondly, minimal in the sense that they have the smallest number of free parameters: 

\subsubsection*{\DB}
This setup was named BASE in the \KCh\ and is renamed \DB\ here. The setup employs a simple power law for the injection spectrum of primary CRs. It uses a double broken power law for the diffusion coefficient, where the first break is at low energies ($R\sim5$ GeV), and the second one models the observed hardening of CR spectra around 300 GV. This setup does not include reacceleration. In total it depends on 10 free CR parameters which we briefly repeat for completeness. They are the slopes of proton and heavier nuclei injection spectra ($\gamma_{2,p}$ and $\gamma_2$), the normalization, slopes, breaks and smoothing of the diffusion coefficient ($D_{0}$, $\delta_{l}$, $\delta$, $\delta_{h}$, $R_{0,D}$, $R_{1,D}$ and $s_{D}$) and the convection velocity ($v_{0,\mathrm{c}}$). Finally, there are the solar modulation potentials. Modulo the presence of convection this setup is often called in the literature as Plain Diffusion. We will see, nonetheless, that although convection is included, the fit does not provide evidence for it. 

\subsubsection*{\DR}
This setup corresponds to the one labeled BASE+inj+$v_A$$-$diff.brk in \KCh\ and it is renamed \DR\ here. Instead of a break in the diffusion coefficient at low energies, this setup employs a break in the injection spectrum of primary CRs. Furthermore, it allows for diffusive reacceleration. The free parameters are the slopes, break and smoothing of the primary injection spectra ($\gamma_{1,p}$, $\gamma_{1}$, $\gamma_{2,p}$, $\gamma_{2}$ $R_{\rm inj,0}$ and $s$), the normalization, slopes and break of the diffusion coefficient ($D_{0}$, $R_{1,D}$, $\delta$, and $\delta_{h}$) as well as the Alfv\'en and convection velocities ($v_A$ and $v_{0,\mathrm{c}}$) for a total of 12 parameters. Exactly like in the setup \DB\, there are further parameters to describe solar modulation.

%=====================================================================
%    \                                                                                        |
%      \                                                                                      |
%        \                                                                                    |
\begin{table*}
  \caption{
    CR data sets used in the fits. For each data set, we state the experiment, 
    the number of data points included in the fits, and the reference.
  }
  \centering
  \renewcommand{\arraystretch}{1.5}
    \begin{tabular}{c @{\hspace{25px}} c @{\hspace{25px}} c @{\hspace{25px}} c @{\hspace{25px}}c @{\hspace{25px}} c}
    \hline \hline
    {CR species} & {experiment} & {number of data points} &  {Ref.} \\ \hline
    $p$            & AMS-02  &      67       &    \cite{Aguilar:2015ooa}       \\
    $p$            & Voyager &       9       &    \cite{2013Sci...341..150S}   \\
    He             & AMS-02  &      68       &    \cite{Aguilar:2017hno}       \\
    He             & Voyager &       5       &    \cite{2013Sci...341..150S}   \\
    $\bar p/p$     & AMS-02  &      48       &    \cite{Aguilar:2016kjl}       \\
    $^3$He/$^4$He  & AMS-02  &      26       &    \cite{Aguilar:2019eiz}       \\
    C              & AMS-02  &      68       &    \cite{Aguilar:2017hno}       \\
    N              & AMS-02  &      67       &    \cite{Aguilar:2018keu}       \\
    O              & AMS-02  &      67       &    \cite{Aguilar:2017hno}       \\
    B/C            & AMS-02  &      67       &    \cite{Aguilar:2018njt}       \\ \hline \hline
    \end{tabular}
  \renewcommand{\arraystretch}{1.0}
  \label{tab::data_sets}
\end{table*}
%                                                                                  \         |
%                                                                                    \       |
%                                                                                      \     |
%=====================================================================

%===================================================================
\section{CR Data}\label{sec::data}
%===================================================================

The goal of the analysis is to provide a single propagation framework for light and heavy nuclei from proton to oxygen. Following the strategy of \KCh, we use the AMS-02 measurements of the carbon and oxygen fluxes from Ref.~\cite{Aguilar:2017hno}, and nitrogen from Ref.~\cite{Aguilar:2018keu}. Instead of using the absolute flux of boron, we employ the B/C ratio from Ref.~\cite{Aguilar:2018njt}, since in the ratio some systematic uncertainties cancel out. Furthermore, the secondary-to-primary flux ratio is only mildly sensitive to the injection parameters and primarily sensitive to the propagation parameters providing a noticeable simplification in the exploration of the parameter space during the fit. All the above measurements correspond to a data-taking period of 5 years from 2011 to 2016, and, therefore, the fluxes are equally affected by solar modulation. We do not fit here lithium and beryllium. In \KCh\ it has been already shown that they can be fit together with B/C, C, N, and O within the same propagation scenario although this might require some freedom in the production cross-sections, which is nonetheless within the allowed cross-section uncertainties. For simplicity here, thus, we do not include them in the fit. 

The heavy nuclei from above are complemented with the light primaries, protons and helium, from AMS-02 \cite{Aguilar:2015ooa, Aguilar:2017hno} and Voyager \cite{2013Sci...341..150S}. Additionally, we exploit the secondary-to-primary ratios of antiprotons-to-protons \cite{Aguilar:2016kjl} and $\mathrm{^3He/^4He}$ \cite{Aguilar:2019eiz}, both determined by the AMS-02 experiment. We note that the helium flux corresponds to the same data-taking period as the heavy nuclei and thus experiences the same solar modulation potential. We will thus use a unique modulation potential $\varphi_{\rm HeBCNO}$. The proton and antiproton-over-proton flux data corresponds to a shorter period of data taking of 3 and 4 years, respectively. We use the data-driven method described in \cite{Cuoco:2019kuu} to obtain an effective 4-year proton flux. In this way, we can use the same solar modulation potential $\varphi_p$ for the protons of the primary flux and those in the antiproton-to-proton flux ratio. We allow for a charge-sign dependence of the modulation potential, and thus use a different potential for the antiprotons in the ratio, $\varphi_{\bar{p}}$. In the fits we vary the difference $\Delta_{\varphi, \bar p} = \varphi_{\bar p}-\varphi_p$ for which we impose a weak Gaussian prior of 100 MV. The $\mathrm{^3He/^4He}$ data refer to a period of 6.5 years but we use for the $\mathrm{^3He}$ and $\mathrm{^4He}$ in the ratio the potential $\varphi_{\rm HeBCNO}$ neglecting in first approximation the difference between 6.5 yr and 5 yrs. This is justified also by the fact that the effect of modulation in a ratio is milder than on absolute fluxes. Since the Voyager data are taken outside the Heliosphere we do not apply any modulation to them.

We use Voyager data only above a kinetic energy-per-nucleon of 0.1 GeV. This is to avoid further complications which might arise at very low energies, like effects of stochasticity from local sources or the possible presence of a further low energy break in the spectra \cite{Phan:2021iht}. A list of all the data used is given in Tab.~\ref{tab::data_sets}. In a few cases, we exclude some data points at low rigidity, where the effect of Solar modulation is largest. In particular, we exclude proton data below 2 GV and $\bar{p}/p$ data below 3 GV.  

%=====================================================================
%    \                                                                                        |
%      \                                                                                      |
%        \                                                                                    |
\begin{table*}[t!]
  \caption{
    Cross section related nuisance parameters which are included in the CR fits. 
  }
  \centering
\renewcommand{\arraystretch}{1.5}
\begin{tabular}{c @{\hspace{15px}} c}
\hline\hline
{fit parameter} & {nuisance parameters} \\ \hline

$\delta_{\mathrm{XS}\rightarrow \mathrm{^3He}}$ &
$\delta_{\,^{4}_{2}\mathrm{He} \rightarrow \,^{3}_{2}\mathrm{He}  } $ \\

$\delta_{\mathrm{XS}\rightarrow \mathrm{B}}$ & 
$\delta_{\,^{16}_{\phantom{1}8}\mathrm{O } \rightarrow \,^{10}_{\phantom{1}5}\mathrm{B} } $ \hspace{0.3cm}
$\delta_{\,^{12}_{\phantom{1}6}\mathrm{C } \rightarrow \,^{10}_{\phantom{1}5}\mathrm{B} } $ \hspace{0.3cm}
$\delta_{\,^{16}_{\phantom{1}8}\mathrm{O } \rightarrow \,^{11}_{\phantom{1}5}\mathrm{B} } $ \hspace{0.3cm}
$\delta_{\,^{12}_{\phantom{1}6}\mathrm{C } \rightarrow \,^{11}_{\phantom{1}5}\mathrm{B} } $ \\

$\delta_{\mathrm{XS}\rightarrow \mathrm{C}}$ & 
$\delta_{\,^{16}_{\phantom{1}8}\mathrm{O } \rightarrow \,^{12}_{\phantom{1}6}\mathrm{C} } $ \hspace{0.3cm}
$\delta_{\,^{16}_{\phantom{1}8}\mathrm{O } \rightarrow \,^{13}_{\phantom{1}6}\mathrm{C} } $ \\

$\delta_{\mathrm{XS}\rightarrow \mathrm{N}}$ & 
$\delta_{\,^{16}_{\phantom{1}8}\mathrm{O } \rightarrow \,^{14}_{\phantom{1}7}\mathrm{N} } $ \hspace{0.3cm}
$\delta_{\,^{16}_{\phantom{1}8}\mathrm{O } \rightarrow \,^{15}_{\phantom{1}7}\mathrm{N} } $  \\

$A_{\mathrm{XS}\rightarrow {\bar p}}$ &
$A_{p \rightarrow \,\bar p  } $\hspace{0.3cm}
$A_{\,^{4}_{2}\mathrm{He} \rightarrow \,\bar p  } $ \\

$A_{\mathrm{XS}\rightarrow \mathrm{^3He}}$ &
$A_{\,^{4}_{2}\mathrm{He} \rightarrow \,^{3}_{2}\mathrm{He}  } $ \\

$A_{\mathrm{XS}\rightarrow \mathrm{B}}$ & 
$A_{\,^{16}_{\phantom{1}8}\mathrm{O } \rightarrow \,^{10}_{\phantom{1}5}\mathrm{B} } $ \hspace{0.3cm}
$A_{\,^{12}_{\phantom{1}6}\mathrm{C } \rightarrow \,^{10}_{\phantom{1}5}\mathrm{B} } $ \hspace{0.3cm}
$A_{\,^{16}_{\phantom{1}8}\mathrm{O } \rightarrow \,^{11}_{\phantom{1}5}\mathrm{B} } $ \hspace{0.3cm}
$A_{\,^{12}_{\phantom{1}6}\mathrm{C } \rightarrow \,^{11}_{\phantom{1}5}\mathrm{B} } $ \\

$A_{\mathrm{XS}\rightarrow \mathrm{C}}$ & 
$A_{\,^{16}_{\phantom{1}8}\mathrm{O } \rightarrow \,^{12}_{\phantom{1}6}\mathrm{C} } $ \hspace{0.3cm}
$A_{\,^{16}_{\phantom{1}8}\mathrm{O } \rightarrow \,^{13}_{\phantom{1}6}\mathrm{C} } $ \\  \vspace{0.3em}

$A_{\mathrm{XS}\rightarrow \mathrm{N}}$ & 
$A_{\,^{16}_{\phantom{1}8}\mathrm{O } \rightarrow \,^{14}_{\phantom{1}7}\mathrm{N} } $ \hspace{0.3cm}
$A_{\,^{16}_{\phantom{1}8}\mathrm{O } \rightarrow \,^{15}_{\phantom{1}7}\mathrm{N} } $ \\
\hline\hline
\end{tabular}
\renewcommand{\arraystretch}{1.0}
  \label{tab::nuisance_param}
\end{table*}
%                                                                                  \         |
%                                                                                    \       |
%                                                                                      \     |
%=====================================================================

%===================================================================
\section{Nuclear Cross Sections} \label{sec::cs}
%===================================================================

The sizable uncertainties in nuclear fragmentation cross sections~\cite{Genolini:2018ekk} at the level of 20\% to 30\% pose a serious systematic for the interpretation of the AMS-02 data. Many of these cross sections are poorly measured, thus the effect of their uncertainties on the prediction of the abundance of secondary elements can be significant. Several recent works  have investigated this aspect for CR nuclei \cite{Genolini:2018ekk,Weinrich:2020cmw,Weinrich:2020ftb,Evoli:2019wwu,Evoli:2019iih,Schroer:2021ojh,Boschini:2018baj,Boschini:2019gow} and for CR antiprotons \cite{Korsmeier:2018gcy,Donato:2017ywo,Winkler:2017xor,Kachelriess:2015wpa}. These have triggered various experimental efforts to perform new cross-section measurements both at accelerators \cite{Aaij:2018svt,Unger:2019nus}, and with the  AMS-02 detector itself \cite{Aguilar:2021tos}.

In general, there are many contributing channels. We denote the cross-section for the fragmentation of a CR species $i$ on the ISM component $j$ to the new species $i$ by $\sigma_{k+j\rightarrow i}$. We follow the approach introduced in \KCh\ to profile over the cross-section uncertainties. To this end we introduce nuisance parameters that allow us to change the default cross-section parametrization by adjusting the overall normalization and the slope at low energies as follows:
\begin{eqnarray}
 \label{eqn::nuisance_XS} 
 \sigma_{k+j\rightarrow i}(T_k/A) = \sigma^{\mathrm{default}}_{k+j\rightarrow i} (T_k/A) 
          \cdot A_{k+j\rightarrow i} \cdot
          \begin{cases} 
         	 (T_k/A)^{\delta_{k+j\rightarrow i}} & T_k/A < T_\mathrm{ref}/A\\
          	1 & \mathrm{otherwise}
          \end{cases}
 \quad . 
\end{eqnarray}
Here $A_{k+j\rightarrow i}$ is the renormalization constant and ${\delta_{k+j\rightarrow i}}$ adjusts the slope of the default cross section as a function of kinetic energy per nucleon $T_k/A$ below $T_\mathrm{ref}/A$. Our default cross section parametrization is taken from \textsc{Galprop}, option \texttt{kopt=12}, and we choose the reference energy to be $T_\mathrm{ref}/A = 5$\ GeV/n. 

In order to keep the number of additional nuisance parameters feasible we follow two strategies. First, we focus on the most dominant fragmentation and production cross sections that have the largest impact on the CR spectra and, secondly, we build groups of cross sections with similar effects and only vary one global parameter for the whole group. For example, boron is mostly produced by the four reactions
$^{16}_{\phantom{1}8}\mathrm{O } + {\rm H} \rightarrow \,^{10}_{\phantom{1}5}\mathrm{B} $,
$^{12}_{\phantom{1}6}\mathrm{C } + {\rm H} \rightarrow \,^{10}_{\phantom{1}5}\mathrm{B} $, 
$^{16}_{\phantom{1}8}\mathrm{O } + {\rm H} \rightarrow \,^{11}_{\phantom{1}5}\mathrm{B} $, and
$^{12}_{\phantom{1}6}\mathrm{C } + {\rm H} \rightarrow \,^{11}_{\phantom{1}5}\mathrm{B} $. 
Instead of introducing a renormalization and slope parameter for each of them we introduce two effective fit parameters $A_{\mathrm{XS}\rightarrow \mathrm{B}}$ and $\delta_{\mathrm{XS}\rightarrow \mathrm{B}}$. The parameters of every single reaction are then set to the effective parameter. Furthermore, we assume that the spallation cross-section on ISM helium is proportional to the one on hydrogen. Hence, those cross-sections are indirectly changed in the same way. The most important cross-sections and the effective parameters are summarized in Tab.\ \ref{tab::nuisance_param}.

Furthermore, we will perform some fits varying the {\it total inelastic fragmentation cross sections}. We follow the same approach as above introducing a parametrization similar to the one of Eq.~\ref{eqn::nuisance_XS} in order to vary a normalization and a slope. We will consider total inelastic cross sections for fragmentation of He, C, N, and O.

%===================================================================
\section{Methods}\label{sec::methods}
%===================================================================

To perform the fit we adopt the same procedure as the one detailed in \KCh. The CR log-likelihood is given by the sum of $\chi^2$s for each CR species: 
\begin{eqnarray}
 	\label{eqn::likelihood_CR}
 -2\,\log\left({{\cal L}_{{\rm CR}} }(\boldsymbol\theta_{\rm CR}, \boldsymbol\theta_{\rm XS})\right) 
    = \chi^2_{\rm CR} (\boldsymbol\theta_{\rm CR}, \boldsymbol\theta_{\rm XS}) =  
    \sum\limits_{s,i}
							\left(\frac{\phi^{}_{{ s},i} - \phi^{(\text{m})}_{s,i} 
							(\boldsymbol\theta_{\rm CR}, \boldsymbol\theta_{\rm XS})}
							{ \sigma_{s,i} }\right)^2 
							\, .
\end{eqnarray}
Here $\phi^{}_{{ s},i}$ is the measured CR flux of species $s$ at the rigidity $R_i$ which is then compared with the corresponding model prediction labeled $\phi^{(\text{m})}_{s}$. The uncertainty $\sigma_{s,i}$ as given by the experiments includes statistical and systematic uncertainties added in quadrature. We note that, although no covariance matrix is provided by the AMS-02 collaboration, the systematic uncertainties of the AMS-02 data are expected to exhibit correlation in rigidity. Such correlations can have an important impact for example on the interpretation of a putative DM signal \cite{Boudaud:2019efq,Heisig:2020nse,Cuoco:2019kuu}. We checked, however, in previous works (\KCh) that such correlations only have a marginal impact on the inferred propagation parameters. So, we do not consider correlations in this work. The model predictions depend on two types of parameter sets. The first set, $\boldsymbol \theta_{\rm CR}$, contains the CR propagation and solar modulation parameters as detailed in Sec.~\ref{sec::cr_prop}. We perform fits in different setups and with different combinations of free parameters (see below). In the maximal case, this first set comprises a total of 17 free parameters. The second set of free parameters relates to the cross-sections uncertainties and contains the cross-section nuisance parameters. These parameters are always included in our fits. 

We explore a huge parameter space of up to 30 free parameters comprising both CR propagation and cross-section uncertainties. This strategy allows us to profile over the uncertainties in the fragmentation and production cross-sections. However, such a large parameter space constitutes a computational challenge. We rely on a hybrid strategy to sample the parameter space efficiently: First, we use the \textsc{MultiNest}~\cite{Feroz:2008xx} algorithm to sample all parameters that depend on the evaluation of \textsc{Galprop}. \textsc{MultiNest} employs a nested sampling algorithm within ellipsoidal regions of the sample space such that the algorithm quickly focuses on the most relevant parameter space. Secondly, parameters that do not depend on the evaluation of \textsc{Galprop} (for example, parameters equivalent to a linear rescaling) are treated in a simplified way. For each evaluation of \textsc{Galprop}, we profile over those parameters on-the-fly with respect to the likelihood of Eq.~\eqref{eqn::likelihood_CR} and directly pass the maximum value to \textsc{MultiNest}~\cite{Rolke:2004mj}. The profiling is performed using \textsc{Minuit}~\cite{James:1975dr}. We provide a summary of all fit parameters in Tab.~\ref{tab:free_param}. The on-the-fly parameters are marked with $\ast$. For \textsc{MultiNest} settings we use 400 live points, an enlargement factor \textsc{efr=0.7}, and a stopping criterion of \textsc{tol=0.1}. A typical MultiNest run requires about 2 million likelihood evaluations and a single likelihood evaluation takes between 150 and 200 sec on a single core. While the nested sampling algorithm of \textsc{MultiNest} formally is a Bayesian inference tool, the results can also be used to construct profiled likelihood for different parameters or parameter combinations which can be interpreted in a frequentist framework. All plots and tables in this work follow the frequentist interpretation. Nonetheless, in \KCh\ we showed that calculations following a Bayesian interpretation of the fit provide very similar parameter constraints and contours. This is expected since we are in a regime where the parameter space is well-constrained given the good constraining power of the data, and in this regime Bayesian and frequentist interpretations tend to converge.

\medskip

In \KCh\ we used the input abundances of primary CRs as fit parameters, which has the advantage of giving a fully self-consistent analysis, but increases significantly the amount of computational resources required to perform the fit. In the present analysis thus, we revert to a simplified approach where instead, we use as parameters the normalizations of the propagated final CR spectra. This simplifies noticeably the fit since these parameters can be treated as fast on-the-fly parameters. On the other hand, care is needed since if the final preferred normalization is significantly different from 1 the fit is formally not self-consistent, and the fit should be in principle repeated and iterated with new input abundances until the final normalizations converge to 1. In the following fits, the fitted normalizations will be always close to 1, except in one case which we will discuss more in detail later.
The input abundances of the primaries are fixed to $1.06\times 10^{ 6}$ for protons,
 $9.65   \times 10^{ 4}$ for $^{ 4}_{      2}\mathrm{He} $, 
 $3.56   \times 10^{ 3}$ for $^{12}_{\phantom{1}6}\mathrm{C } $, 
 $0.35   \times 10^{ 3}$ for $^{14}_{\phantom{1}7}\mathrm{N } $, and
 $4.30   \times 10^{ 3}$ for $^{16}_{\phantom{1}8}\mathrm{O } $.

%===================================================================
\section{Light vs Heavy Nuclei -- some preliminary considerations} \label{sec::motivation}
%===================================================================

%=====================
%    \                                           |
%      \                                         |
%        \                                       |
\begin{figure*}[t!]
\centering
\includegraphics[width=0.99\textwidth]{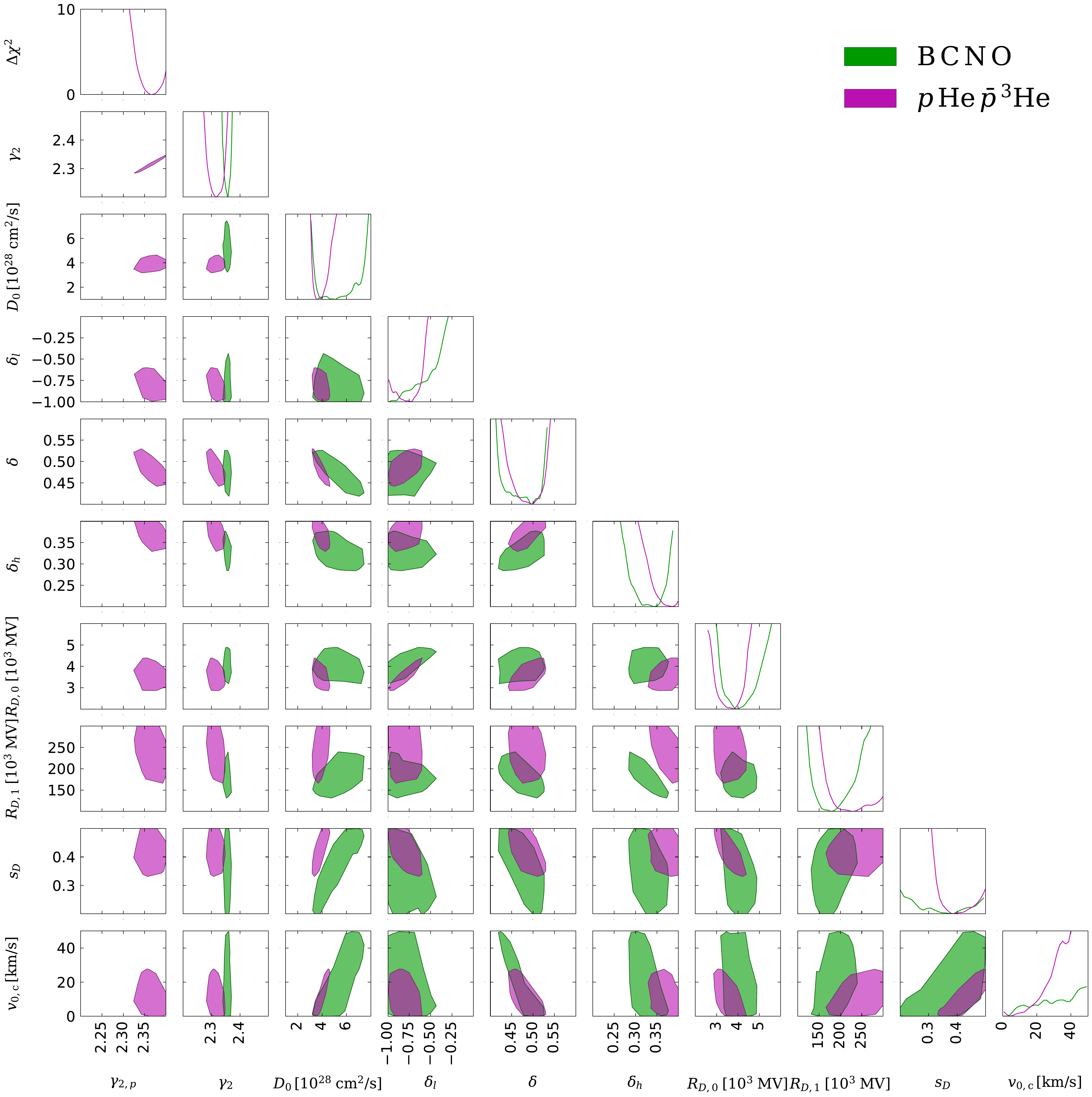}
\caption{ 
          Triangle plot comparing the results of the fit to the light nuclei ($p$, He, $\bar p/p$, and $^3$He/$^4$He) 
          with the fit to the heavier nuclei (B/C, C, N, O) in the \DB\ setup. Results are shown for a selection of the most relevant CR propagation parameters. 
          The contours represent the 2$\sigma$ statistical uncertainties derived from the two-dimensional $\chi^2$-profiles for each 
          combination of two parameters, while the diagonal shows the $\chi^2$-profiles for each individual parameter.
          \label{fig:Light_Heavy_triangel}
        }
\end{figure*}
%                                      \         |
%                                        \       |
%                                          \     |
%=====================

In previous works, we have separately studied the propagation of the light nuclei $p$, He, and $\bar p$ (\KCl) and the heavier nuclei from Li to O (\KCh). We have also further investigated the combination of $p$, He, and $\bar p$ in relation to a possible DM contribution to the $\bar p$ flux \cite{Cuoco:2016eej,Cuoco:2017rxb,Cuoco:2017iax,Cuoco:2019kuu,Heisig:2020nse}. While the individual studies and fits provide a good match to the observed CR fluxes measured by AMS-02, it is apparent that the source and propagation parameters of the two data sets point to slightly different values. A similar result was indeed already pointed out in Ref.~\cite{Johannesson:2016rlh} with older data. To illustrate this point more in detail we compare here the two fits. Since $\mathrm{^3He}$ has been recently made available by AMS-02 we perform an updated fit of all available light nuclei i.e., $p$, He, $\bar p/p$, and $\mathrm{^3He/^4He}$. The fit is performed using the methodology described in the previous sections and for the \DB\ setup. This is compared with a similar fit using B/C, C, N, and O nuclei data. This second fit is directly taken from \KCh. The results of the two fits are compared in Fig.~\ref{fig:Light_Heavy_triangel} in a triangle plot which shows 2$\sigma$ contours for a subset of the most relevant parameter. The complete result of the fit for the light nuclei is given in the tabular form in the appendix.

The most obvious tension concerns the slope of the injection spectrum, $\gamma_2$: The fit of heavy nuclei converges to a value of ${2.357}^{+0.003}_{-0.005}$, while the lighter nuclei prefer a value of ${2.31}^{+0.02}_{-0.01}$. To be more precise, in the case of light nuclei the slope refers to $^4$He and for the heavier nuclei it refers to the primary components of C, N, and O. We remind that we allow for an individually different injection slope for primary protons, $\gamma_{2,p}$, and helium in the fit of light nuclei, which is unavoidable given the steeper slope of the measured CR proton flux w.r.t the helium flux as discussed in the introduction. On the other hand, the spectra of the measured He, C, and O flux show very similar spectral behavior at high energies~\cite{Aguilar:2017hno} above $\sim 50$ GV. So, at first glance, it makes sense to assume a universal injection spectrum for all nuclei heavier than protons. However, the result of the fit tells otherwise, with the injection of helium and CNO being different at a significance of about 2$\sigma$, and with the incompatibility of the two datasets being actually larger, as one can see from the fact that in Fig.~\ref{fig:Light_Heavy_triangel} some contours do not overlap at the 2$\sigma$ level.
The difference in $\gamma_{2}$ is actually not dramatically large in absolute terms, about 2\%, but the high precision of AMS-02 data turns this into a quite significant difference. 

The key ingredients to understand this difference are three propagation effects that are able to change the spectra of primaries. The main effect is diffusion itself, which, if taken alone, creates a steepening of the injection slopes by an amount of $\delta$. The second ingredient is related to the secondary contributions. Although they are mainly primary, the helium and carbon fluxes contain a significant contribution from secondary production which are at the level of 15\% and 10\% respectively, while oxygen is almost a pure primary. Since the secondaries have a softer spectrum with respect to primaries we expect the total observed spectra, which are the sum of both contributions, to become harder the larger the secondary contribution. The third effect is given by the total inelastic cross sections. Inelastic energy losses tend to affect CRs mainly at low energies, fragmenting, i.e., removing low energy nuclei, while at higher energies diffusion dominates and nuclei can escape the Galaxy before being significantly affected by energy losses. The overall effect is again a hardening of the spectrum. The cross section of helium on ISM protons ($\sigma \sim 100$ mb \cite{Mazziotta:2015uba,Coste:2011jc}) and, for example, oxygen on ISM proton ($\sigma \sim 300$ mb \cite{Mazziotta:2015uba,Genolini:2018ekk}) are rather different by a factor of about 3. Thus, starting with the same injection and experiencing the same diffusion the final observed spectra will be unavoidably different due to the different energy losses experienced. Thus, contrary to the first intuition, observing the same helium and oxygen spectrum is actually an unexpected outcome, and in the standard scenario implies different injections as quantified in the above fit. A closer look at Fig.~\ref{fig:Light_Heavy_triangel} reveals that there is also some further discrepancy in the value of the normalization of the diffusion coefficient. The 2D best-fit contours of $D_0$ vs. $v_{c,0}$ and $D_0$ vs. $\delta$ show that the light nuclei prefer smaller values of $D_0$. On the other hand, these differences are not at a strong significance, thus are less worrisome and it is conceivable that in a combined light+heavy fit an intermediate value can be preferred still providing a reasonably good fit. The above outcomes for the \DB\ fit also similarly apply to the \DR\ setup although we do not discuss it here in detail. The results of the \DR\ fit are also reported in tabular form in the appendix.

The above considerations already indicate which path should be followed in order to achieve a combined light+heavy fit and, thus, which assumptions have to be relaxed to obtain a consistent picture of CR nuclei from $p$ to O. Clearly, the two obvious routes are to violate the universality between light and heavy nuclei either for the primary injection spectrum or for diffusion. These options will be explored more thoroughly in the following. A third option is to modify the total inelastic cross sections, but the required modification is large and we deem the possibility unlikely. Nonetheless, we will also briefly explore and quantify this possibility with a specific fit.

%=====================================================================
%    \                                                                                        |
%      \                                                                                      |
%        \                                                                                    |
\begin{table*}[t!]
    \caption{\label{tab:free_param}
    	Summary of free parameters and parameter dependencies in the various fits. The on-the-fly parameters (see text for more details) are marked with an asterisk ($\ast$).
    }
  \centering
\scalebox{0.78}{
\renewcommand{\arraystretch}{1.5}
\begin{tabular}{cccccccccccccccccccc}
\hline \hline
Parameter                                                    &        &$\;$&  \multicolumn{7}{c}{ $\mathrm{DIFF.BRK}$ }                                                                                                                                                                           &$\;$& \multicolumn{7}{c}{ $\mathrm{INJ.BRK}+v_A$ }                                                                                                                                                                 &    & Prior             \\  \cline{4-10} \cline{12-18}
$\,$                                                         &        &    &  default                                        &    &   free He inj                                   &    & free $D_{0,{\rm light}}$                        &    &  free sec. norms                                &    &  default                                        &    &  free He inj                                    &    & free $D_{0,{\rm light}}$                        &    &  free sec. norms                        &$\;$&                   \\ \cline{1-1} \cline{4-4} \cline{6-6} \cline{8-8} \cline{10-10} \cline{12-12} \cline{14-14} \cline{16-16} \cline{18-18} \cline{20-20}
%%&
%%&
$\gamma_{1,p}$                                               &        &    & $\gamma_{1,p}=\gamma_{2,p}$                     &    & $\gamma_{1,p}=\gamma_{2,p}$                     &    & $\gamma_{1,p}=\gamma_{2,p}$                     &    & $\gamma_{1,p}=\gamma_{2,p}$                     &    & free                                            &    & free                                            &    & free                                            &    & free                                    &    & [1.0, 2.4]        \\
$\gamma_{2,p}$                                               &        &    & free                                            &    & free                                            &    & free                                            &    & free                                            &    & free                                            &    & free                                            &    & free                                            &    & free                                    &    & [2.25, 2.45]      \\
$\gamma_{1,\mathrm{He}}$                                     &        &    & $\gamma_{1,\mathrm{He}}=\gamma_2$               &    & $\gamma_{1,\mathrm{He}}=\gamma_{2,\mathrm{He}}$ &    & $\gamma_{1,\mathrm{He}}=\gamma_2$               &    & $\gamma_{1,\mathrm{He}}=\gamma_2$               &    & $\gamma_{1,\mathrm{He}}=\gamma_1$               &    & free                                            &    & $\gamma_{1,\mathrm{He}}=\gamma_1$               &    & $\gamma_{1,\mathrm{He}}=\gamma_1$       &    & [1.0, 2.4]        \\
$\gamma_{2,\mathrm{He}}$                                     &        &    & $\gamma_{2,\mathrm{He}}=\gamma_2$               &    & free                                            &    & $\gamma_{2,\mathrm{He}}=\gamma_2$               &    & $\gamma_{2,\mathrm{He}}=\gamma_2$               &    & $\gamma_{2,\mathrm{He}}=\gamma_2$               &    & free                                            &    & $\gamma_{2,\mathrm{He}}=\gamma_2$               &    & $\gamma_{2,\mathrm{He}}=\gamma_2$       &    & [2.25, 2.45]      \\
$\gamma_1$                                                   &        &    & $\gamma_{1}=\gamma_{2}$                         &    & $\gamma_{1}=\gamma_{2}$                         &    & $\gamma_{1}=\gamma_{2}$                         &    & $\gamma_{1}=\gamma_{2}$                         &    & free                                            &    & free                                            &    & free                                            &    & free                                    &    & [1.0, 2.4]        \\
$\gamma_2$                                                   &        &    & free                                            &    & free                                            &    & free                                            &    & free                                            &    & free                                            &    & free                                            &    & free                                            &    & free                                    &    & [2.25, 2.45]      \\
$R_{\rm inj,0} \;\mathrm{[ 10^{3}\;MV]}$                     &        &    & -                                               &    & -                                               &    & -                                               &    & -                                               &    & free                                            &    & free                                            &    & free                                            &    & free                                    &    & [1.0, 10.0]       \\
$s$                                                          &        &    & -                                               &    & -                                               &    & -                                               &    & -                                               &    & free                                            &    & free                                            &    & free                                            &    & free                                    &    & [0.1, 0.5]        \\
$D_{0,{\rm light}}\;\mathrm{[ 10^{28}\;cm^2/s]}$             &        &    & $D_{0,{\rm light}}=D_{0}$                       &    & $D_{0,{\rm light}}=D_{0}$                       &    & free                                            &    & $D_{0,{\rm light}}=D_{0}$                       &    & $D_{0,{\rm light}}=D_{0}$                       &    & $D_{0,{\rm light}}=D_{0}$                       &    & free                                            &    & $D_{0,{\rm light}}=D_{0}$               &    & [1.0, 10.0]       \\
$D_{0}\,\mathrm{[ 10^{28}\;cm^2/s]}$                         &        &    & free                                            &    & free                                            &    & free                                            &    & free                                            &    & free                                            &    & free                                            &    & free                                            &    & free                                    &    & [1.0, 10.0]       \\
$\delta_{l}$                                                 &        &    & free                                            &    & free                                            &    & free                                            &    & free                                            &    & $\delta_{l}=\delta$                             &    & $\delta_{l}=\delta$                             &    & $\delta_{l}=\delta$                             &    & $\delta_{l}=\delta$                     &    & [-1.0, 0.0]       \\
$\delta_{}$                                                  &        &    & free                                            &    & free                                            &    & free                                            &    & free                                            &    & free                                            &    & free                                            &    & free                                            &    & free                                    &    & [0.3, 0.6]        \\
$\delta_{h}$                                                 &        &    & free                                            &    & free                                            &    & free                                            &    & free                                            &    & free                                            &    & free                                            &    & free                                            &    & free                                    &    & [0.2, 0.5]        \\
$R_{0,D}\,\mathrm{[ 10^{3}\;MV]}$                            &        &    & free                                            &    & free                                            &    & free                                            &    & free                                            &    & -                                               &    & -                                               &    & -                                               &    & -                                       &    & [1.0, 10.0]       \\
$R_{1,D}\,\mathrm{[ 10^{3}\;MV]}$                            &        &    & free                                            &    & free                                            &    & free                                            &    & free                                            &    & free                                            &    & free                                            &    & free                                            &    & free                                    &    & [100, 400]        \\
$s_{D}$                                                      &        &    & free                                            &    & free                                            &    & free                                            &    & free                                            &    & -                                               &    & -                                               &    & -                                               &    & -                                       &    & [0.1, 0.5]        \\
$v_{0,\mathrm{c}}\,\mathrm{[km/s]}$                          &        &    & free                                            &    & free                                            &    & free                                            &    & free                                            &    & free                                            &    & free                                            &    & free                                            &    & free                                    &    & [0, 50]           \\
$v_{\mathrm{A}}\,\mathrm{[km/s]}$                            &        &    & -                                               &    & -                                               &    & -                                               &    & -                                               &    & free                                            &    & free                                            &    & free                                            &    & free                                    &    & [0, 30]           \\
$\delta_{\mathrm{XS}\rightarrow \mathrm{^3He}}$              &        &    & free                                            &    & free                                            &    & free                                            &    & free                                            &    & free                                            &    & free                                            &    & free                                            &    & free                                    &    & [-0.3, 0.3]       \\
$\delta_{\mathrm{XS}\rightarrow \mathrm{B}}$                 &        &    & free                                            &    & free                                            &    & free                                            &    & free                                            &    & free                                            &    & free                                            &    & free                                            &    & free                                    &    & [-0.2, 0.2]       \\
$\delta_{\mathrm{XS}\rightarrow \mathrm{C}}$                 &        &    & free                                            &    & free                                            &    & free                                            &    & free                                            &    & free                                            &    & free                                            &    & free                                            &    & free                                    &    & [-0.2, 0.3]       \\
$\delta_{\mathrm{XS}\rightarrow \mathrm{N}}$                 &        &    & free                                            &    & free                                            &    & free                                            &    & free                                            &    & free                                            &    & free                                            &    & free                                            &    & free                                    &    & [-0.2, 0.2]       \\
$A_{\mathrm{XS}\rightarrow \bar p}$                          & $\ast$ &    & -                                               &    & -                                               &    & -                                               &    & free                                            &    & -                                               &    & -                                               &    & -                                               &    & free                                    &    & [0.1, 2.0]        \\
$A_{\mathrm{XS}\rightarrow \mathrm{^3He}}$                   &        &    & free                                            &    & free                                            &    & free                                            &    & free                                            &    & free                                            &    & free                                            &    & free                                            &    & free                                    &    & [0.1, 1.9]        \\
$A_{\mathrm{XS}\rightarrow \mathrm{B}}$                      & $\ast$ &    & free                                            &    & free                                            &    & free                                            &    & free                                            &    & free                                            &    & free                                            &    & free                                            &    & free                                    &    & [0.1, 2.0]        \\
$A_{\mathrm{XS}\rightarrow \mathrm{C}}$                      &        &    & free                                            &    & free                                            &    & free                                            &    & free                                            &    & free                                            &    & free                                            &    & free                                            &    & free                                    &    & [0.5, 1.5]        \\
$A_{\mathrm{XS}\rightarrow \mathrm{N}}$                      &        &    & free                                            &    & free                                            &    & free                                            &    & free                                            &    & free                                            &    & free                                            &    & free                                            &    & free                                    &    & [0.5, 2.0]        \\
Abd. $p$                                                     & $\ast$ &    & free                                            &    & free                                            &    & free                                            &    & free                                            &    & free                                            &    & free                                            &    & free                                            &    & free                                    &    & [0.1, 2.0]        \\
Abd. He                                                      & $\ast$ &    & free                                            &    & free                                            &    & free                                            &    & free                                            &    & free                                            &    & free                                            &    & free                                            &    & free                                    &    & [0.1, 2.0]        \\
Abd. C                                                       & $\ast$ &    & free                                            &    & free                                            &    & free                                            &    & free                                            &    & free                                            &    & free                                            &    & free                                            &    & free                                    &    & [0.1, 2.0]        \\
Abd. N                                                       & $\ast$ &    & free                                            &    & free                                            &    & free                                            &    & free                                            &    & free                                            &    & free                                            &    & free                                            &    & free                                    &    & [0.1, 2.0]        \\
Abd. O                                                       & $\ast$ &    & free                                            &    & free                                            &    & free                                            &    & free                                            &    & free                                            &    & free                                            &    & free                                            &    & free                                    &    & [0.1, 2.0]        \\
$\varphi_p\,\mathrm{[GV]}$                                   & $\ast$ &    & free                                            &    & free                                            &    & free                                            &    & free                                            &    & free                                            &    & free                                            &    & free                                            &    & free                                    &    & [0.3, 0.9]        \\
$\varphi_{\rm HeBCNO}\;\mathrm{[GV]}$                        & $\ast$ &    & free                                            &    & free                                            &    & free                                            &    & free                                            &    & free                                            &    & free                                            &    & free                                            &    & free                                    &    & [0.3, 0.9]        \\
$\varphi_{\bar p}-\varphi_p\,\mathrm{[GV]}$                  & $\ast$ &    & free                                            &    & free                                            &    & free                                            &    & free                                            &    & free                                            &    & free                                            &    & free                                            &    & free                                    &    & [-0.3, 0.3]       \\
\# free parameters                                           & $    $ &    &   26                                            &    &   27                                            &    &   27                                            &    &   27                                            &    &   28                                            &    &   30                                            &    &   29                                            &    &   29                                    &    &                   \\
\hline
\hline
\end{tabular}
\renewcommand{\arraystretch}{1.0}
}
  \label{tab::frameworks}
\end{table*}
%                                                                                  \         |
%                                                                                    \       |
%                                                                                      \     |
%=====================================================================

%===================================================================
\section{Analysis strategy}\label{sec::strategy}
%===================================================================

Following the discussion from the previous section, we will extend the two default propagation scenarios including extra degrees of freedom. We will perform a total of eight different fits, four for each CR propagation setup: \DB\ and \DR. In each fit, both the CR propagation parameters and cross-section-related nuisance parameters are varied together in a single combined parameter space. In the \emph{default} cases we strictly employ the setups outlined at the end of Sec.~\ref{sec::cr_prop}, with the only restriction that the antiproton cross-section normalization is fixed to the nominal one, i.e. the parameter $A_{{\rm XS} \rightarrow \bar{p}}$ is fixed to 1.0. This is motivated by the degeneracy present between the normalization of the secondary CRs and the normalization of the diffusion coefficient, $D_0$. Thus, if we would allow all normalizations of the secondary production cross-sections to be free, we would introduce an overall global degeneracy, which we instead avoid fixing $A_{{\rm XS} \rightarrow \bar{p}}$. Beyond the \emph{default} setup, the scenario which we will explore are the following:

\begin{itemize}
\item
In the first case, we allow for additional freedom in the injection spectra of primaries. As we have seen in Fig.~\ref{fig:Light_Heavy_triangel} the separate fit of light and heavier nuclei points to different slopes of helium on the one hand and carbon, nitrogen, and oxygen on the other hand. So we allow helium to have a different shape of the injection spectrum from carbon, nitrogen, and oxygen thus violating the universality of CR injection spectra. In the case of \DB\ this means one additional free parameter ($\gamma_{2,{\rm He}}$), while there are two new free parameters for \DR\ ($\gamma_{1,{\rm He}}$ and $\gamma_{2,{\rm He}}$). This extension is labeled \emph{free $\mathrm{He}\ inj$}.
\item
Another alternative is to allow more freedom in the propagation. We thus investigate the possibility that the average diffusion coefficient observed for light and heavier nuclei is different. Hence, we allow a different normalization of the diffusion coefficient for the light nuclei ($p$, He and $\bar p$), although enforcing the same spectral behavior. In both setups, \DB\ and \DR, this leads to one additional free parameter labeled $D_{0,{\rm light}}$, and the setup is called \emph{free} $D_{0,{\rm light}}$. The diffusion coefficient $D_{0,{\rm light}}$ is applied to all light species, i.e. $p$, $^3$He, $^4$He, and $\bar p$. A physical justification for this scenario can be provided by inhomogeneous diffusion. Light and heavy nuclei travel on average at different distances in the Galaxy since they have different energy losses due to the different inelastic cross-sections. Thus, if the propagation medium is non-homogeneous, light and heavy nuclei could experience a different average diffusion, and a different diffusion coefficient might be an effective way to catch this effect. Furthermore, gamma-ray observations indicate a softening of the CR spectra towards the Galactic center, which could be related to non-homogeneous diffusion \cite{Acero:2016qlg,Yang:2016jda,Pothast:2018bvh}.
\item 
Finally, in reality, the above degeneracy between secondary CRs normalization and $D_0$ is actually broken since we also fit CR primaries for which this degeneracy is not valid. 
Thus, we consider a third extension in which we leave the cross-section normalization of the antiproton production free in the fit to investigate whether this can bring an improvement in the fit. This setup is called \emph{free sec. norms} 
\end{itemize}

A summary of all the setups and all the free parameters in each setup is given in Tab.~\ref{tab:free_param}.

%=====================
%    \                                           |
%      \                                         |
%        \                                       |
\begin{figure*}[!b]
\centering
\setlength{\unitlength}{1\textwidth}
\begin{picture}(1,1.25)
 \put(-0.025, -0.0){\includegraphics[width=0.249\textwidth ]{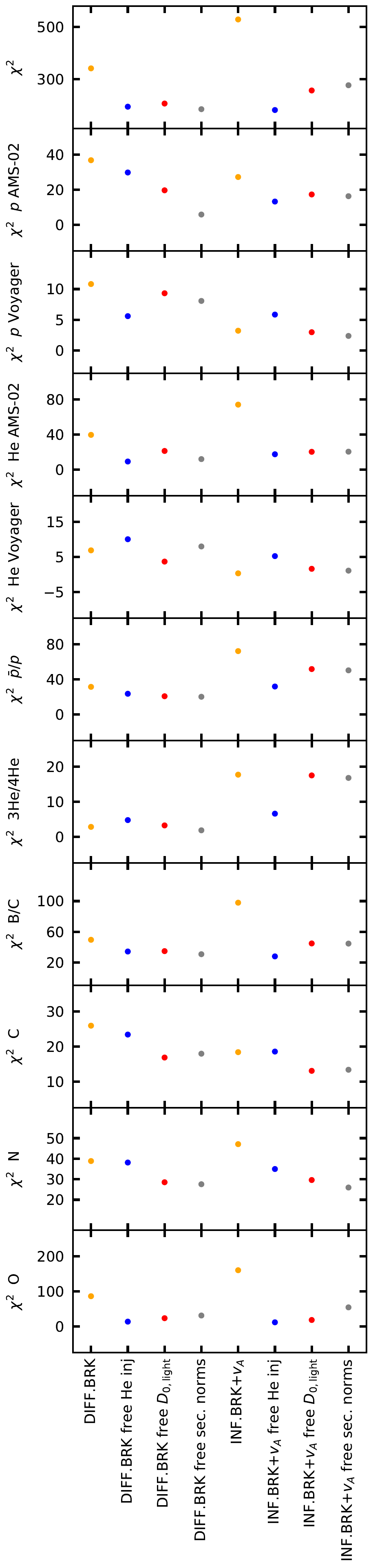}}
 \put( 0.24 , -0.0){\includegraphics[width=0.25\textwidth  ]{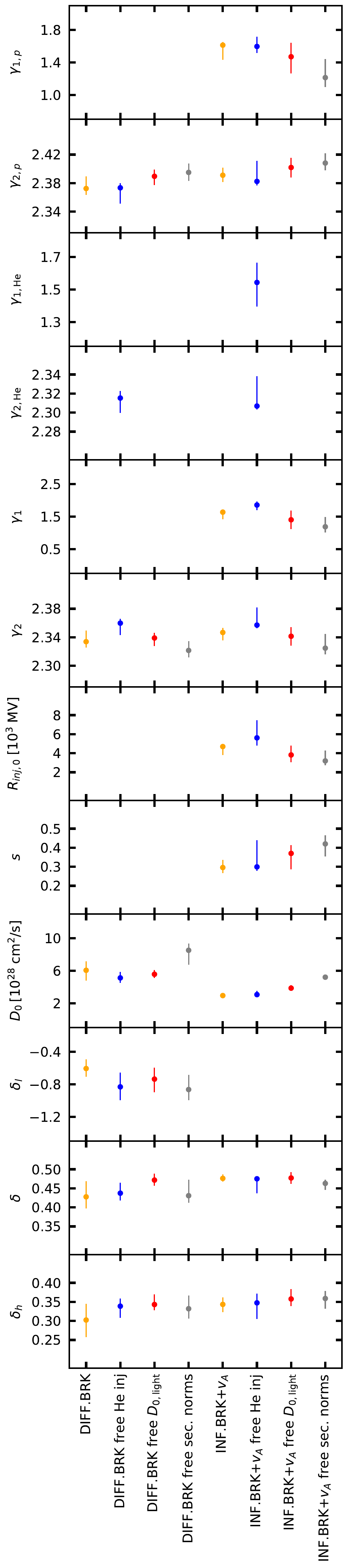}}
 \put( 0.505, -0.0){\includegraphics[width=0.25\textwidth  ]{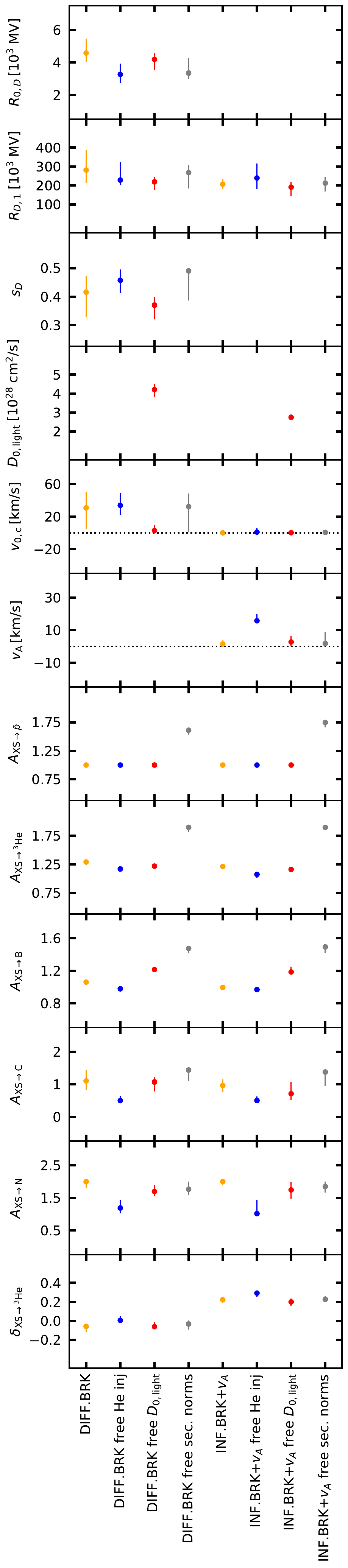}}
 \put( 0.77 , -0.0){\includegraphics[width=0.249\textwidth ]{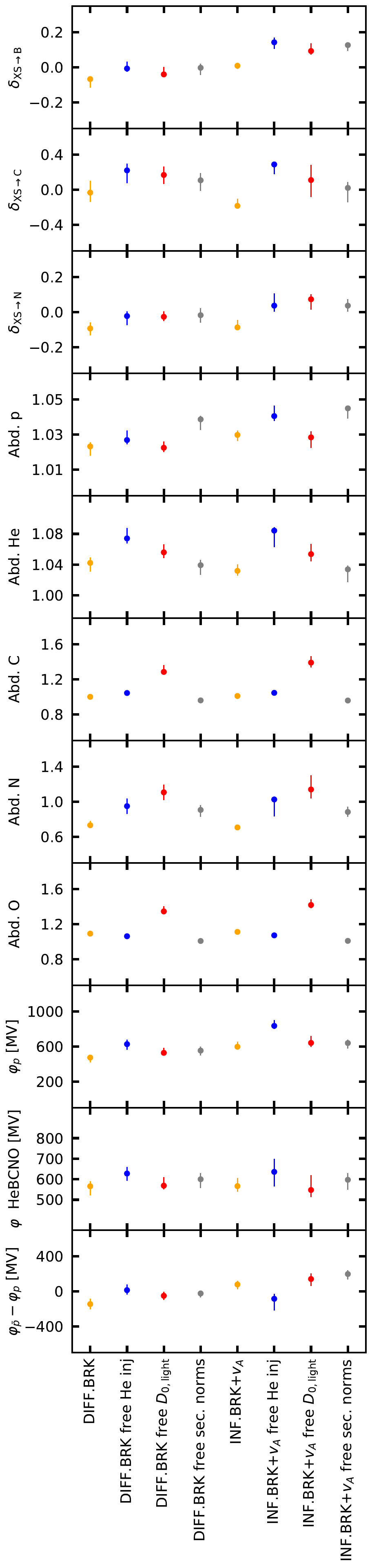}}
\end{picture}
        \caption{ 
          Results of the parameter fit in graphical form. In the left column plot we show the total $\chi^2$ and the separate $\chi^2$s for each species. 
          The remaining three column plots contain the best-fit value and the 2$\sigma$ uncertainty for each parameter. The results are provided for both propagation frameworks,
          \DB\ and \DR, and for the four fit setups:
          \emph{default} (yellow points), \emph{free He inj}  (blue points),  \emph{free $D_{0,\rm light}$} (red points), and  
          \emph{free secondary norm} (grey points). 
          \label{fig:Summary_param}
        }
\label{fig:fitresults}
\end{figure*}
%                                      \         |
%                                        \       |
%                                          \     |
%=====================

%===================================================================
\section{Results} \label{sec::results}
%===================================================================

\begin{figure*}[t]
\centering
\setlength{\unitlength}{1.0\textwidth}
\begin{picture}(1,0.75)
 \put(0.00, 0.0){\includegraphics[width=0.48\unitlength ]{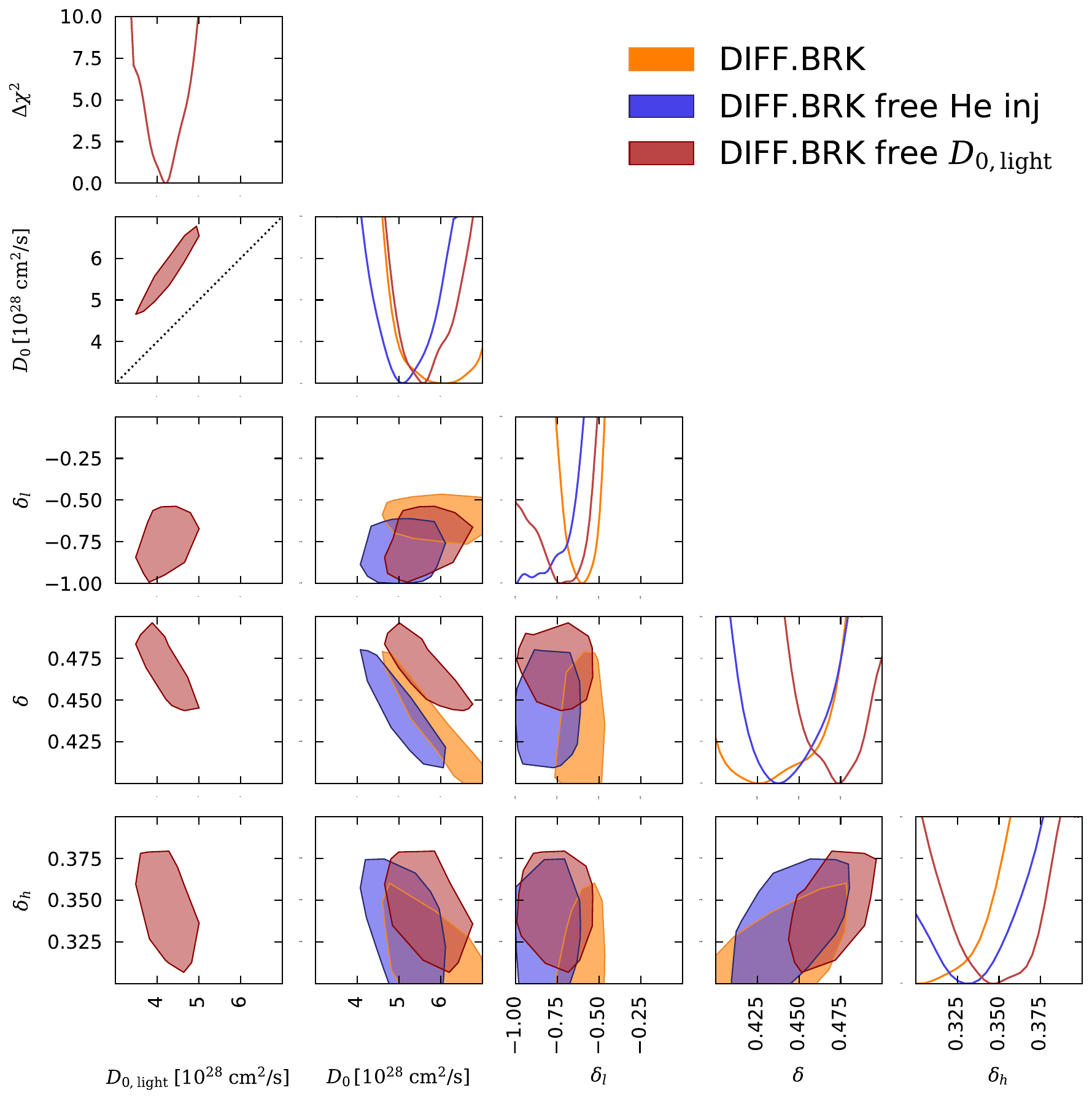}}
 \put(0.52, 0.0){\includegraphics[width=0.48\unitlength ]{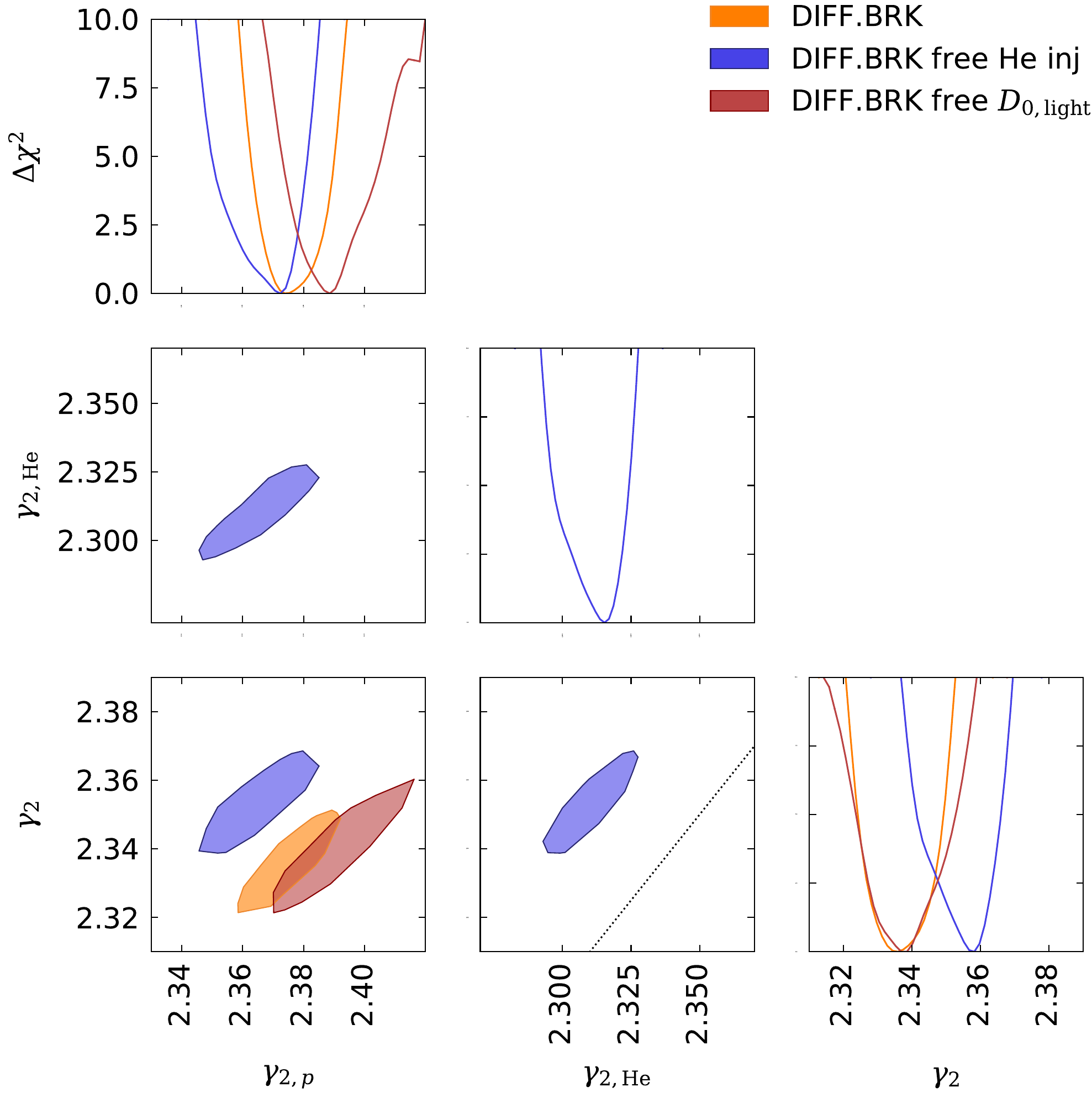}}
\end{picture}
\caption{ 
          Triangles of the best-fit region for the diffusion (left panel) and injection (right panel) parameters. The diagonal contains the 
          $\chi^2$-profile for each individual parameter, while the contours in the lower half show the 2$\sigma$ contours derived from the
          2-dimensional $\chi^2$-profiles. All setups refer to the \DB\ propagation framework. The different colors indicate the fit setup:
          \emph{default} (yellow), \emph{free He inj}  (blue),  and \emph{free $D_{0,\rm light}$} (red). 
          The full triangle with all propagation parameters is provided in the appendix.
          The dotted black lines mark the cases where $D_0 = D_{0,{\rm light}}$ in the left panel 
          and where $\gamma_2 = \gamma_{2,{\rm He}}$ in the right panel. Different values are clearly preferred.
        }
 \label{fig:DIFFBRK_smallTriangles}
\end{figure*}

\begin{figure*}[t]
\centering
\setlength{\unitlength}{1.0\textwidth}
\begin{picture}(1,0.75)
 \put(0.00, 0.0){\includegraphics[width=0.48\unitlength ]{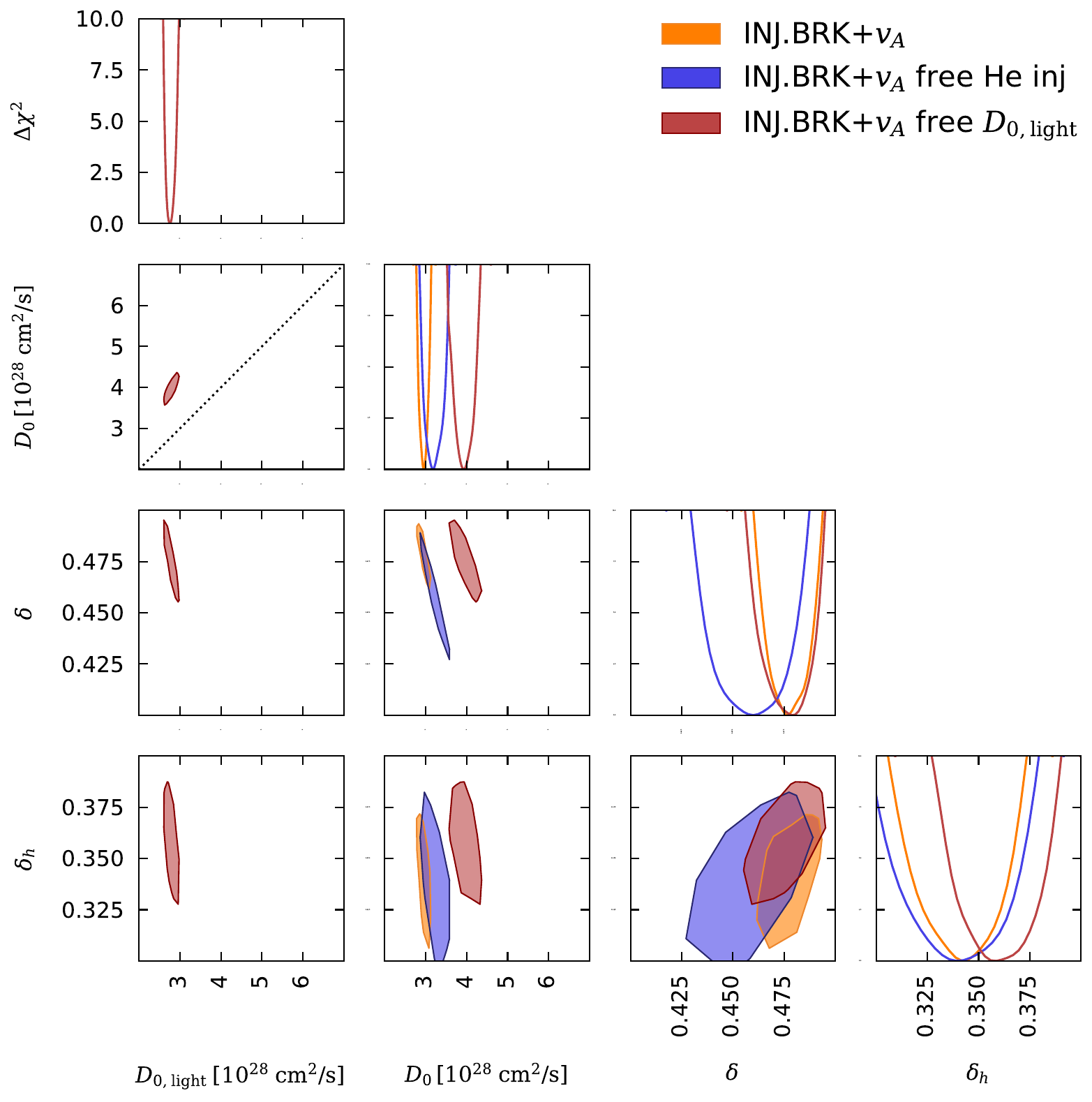}}
 \put(0.52, 0.0){\includegraphics[width=0.48\unitlength ]{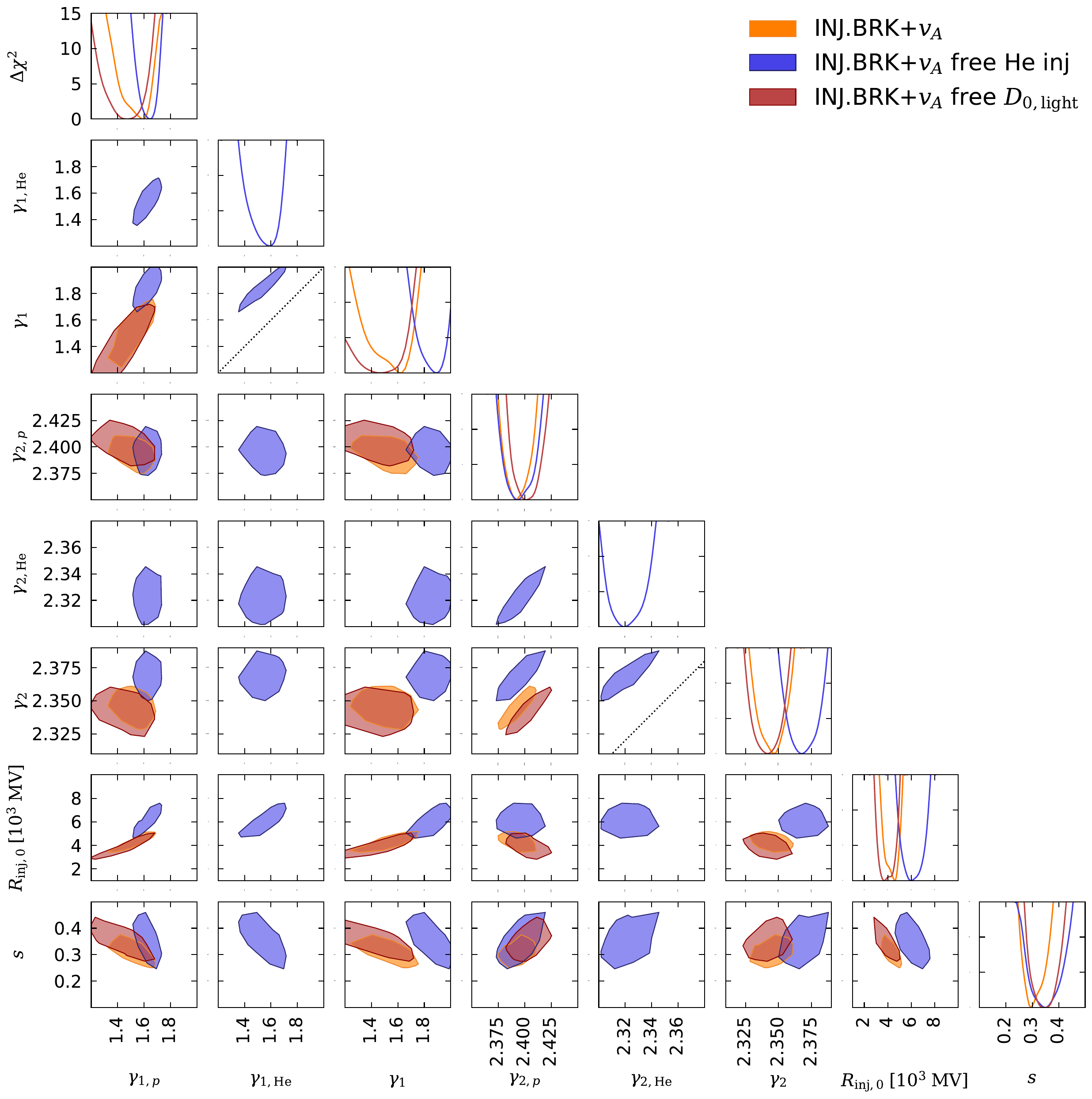}}
\end{picture}
\caption{ 
          Same as Fig.~\ref{fig:DIFFBRK_smallTriangles}, but for the propagation framework \DR.
        }
 \label{fig:INJBRKvA_smallTriangle}
\end{figure*}

The results of the fits are given in graphical form in Fig.~\ref{fig:fitresults}, while they are reported in tabular form in the appendix. Figures~\ref{fig:DIFFBRK_spectra} and \ref{fig:INJBRKvA_spectra}, instead, show the best fit spectra and residual for all the CR species considered in the fit ($p$, He, $\bar p/p$, $\mathrm{^3He/^4He}$, B/C, C, N and O ). Finally, Figs.~\ref{fig:DIFFBRK_smallTriangles} and \ref{fig:INJBRKvA_smallTriangle} show triangle plots for a subset of parameters, namely diffusion parameters and injection parameters separately. Bigger triangle plots with all the parameters including the cross-section-related ones are reported in the appendix. In the triangle plots and the spectral plots, we show only the three fits \emph{deault}, \emph{free He inj} and \emph{free $D_{0,{\rm light}}$} in order to not overcrowd the plots and keep good readability. All the triangle plots show $2\sigma$ contours.

As a preliminary consideration, we confirm as in \KCh\ that taking into account the cross-section uncertainties is crucial to correctly make inferences about CR propagation. For example, the B and $\mathrm{^3He}$ production cross sections require in various scenarios a $\sim 20\%$ renormalization, which is nonetheless, safely within the known uncertainties. More importantly, while the slope of the $\mathrm{^3He}$ production cross section is compatible with the nominal one in the \DB\ scenario, in the \DR\ scenario a systematic shift of about 0.2 is necessary, which, however, is still within the uncertainties. A similar shift, as can be seen in Fig.~\ref{fig:Summary_param}, is observed also for the slope of the B production cross section. Without the freedom in the $\mathrm{^3He}$ and B cross sections the $\mathrm{^3He}$ and B spectra would give a poor fit bringing to the incorrect conclusion that the \DR\ scenario is disfavoured, or excluded.

Regarding the individual setups, various conclusions can be drawn:

%=====================
%    \                                           |
%      \                                         |
%        \                                       |
\begin{figure*}[t]
\centering
\setlength{\unitlength}{0.75\textwidth}
\begin{picture}(1,1.66)
 \put(0.00, 1.15 ){\includegraphics[width=0.55\unitlength ]{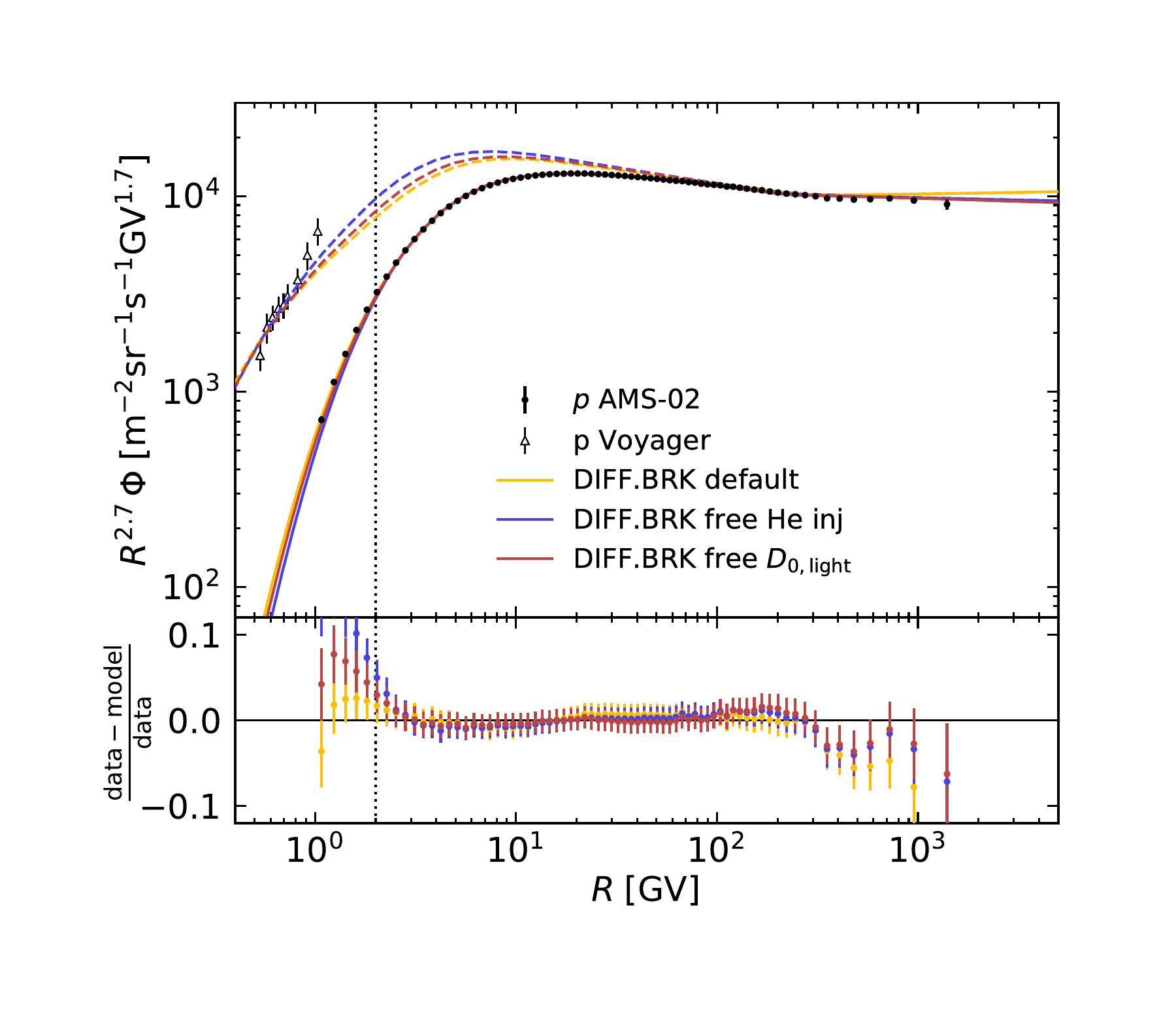}}
 \put(0.50, 1.15 ){\includegraphics[width=0.55\unitlength ]{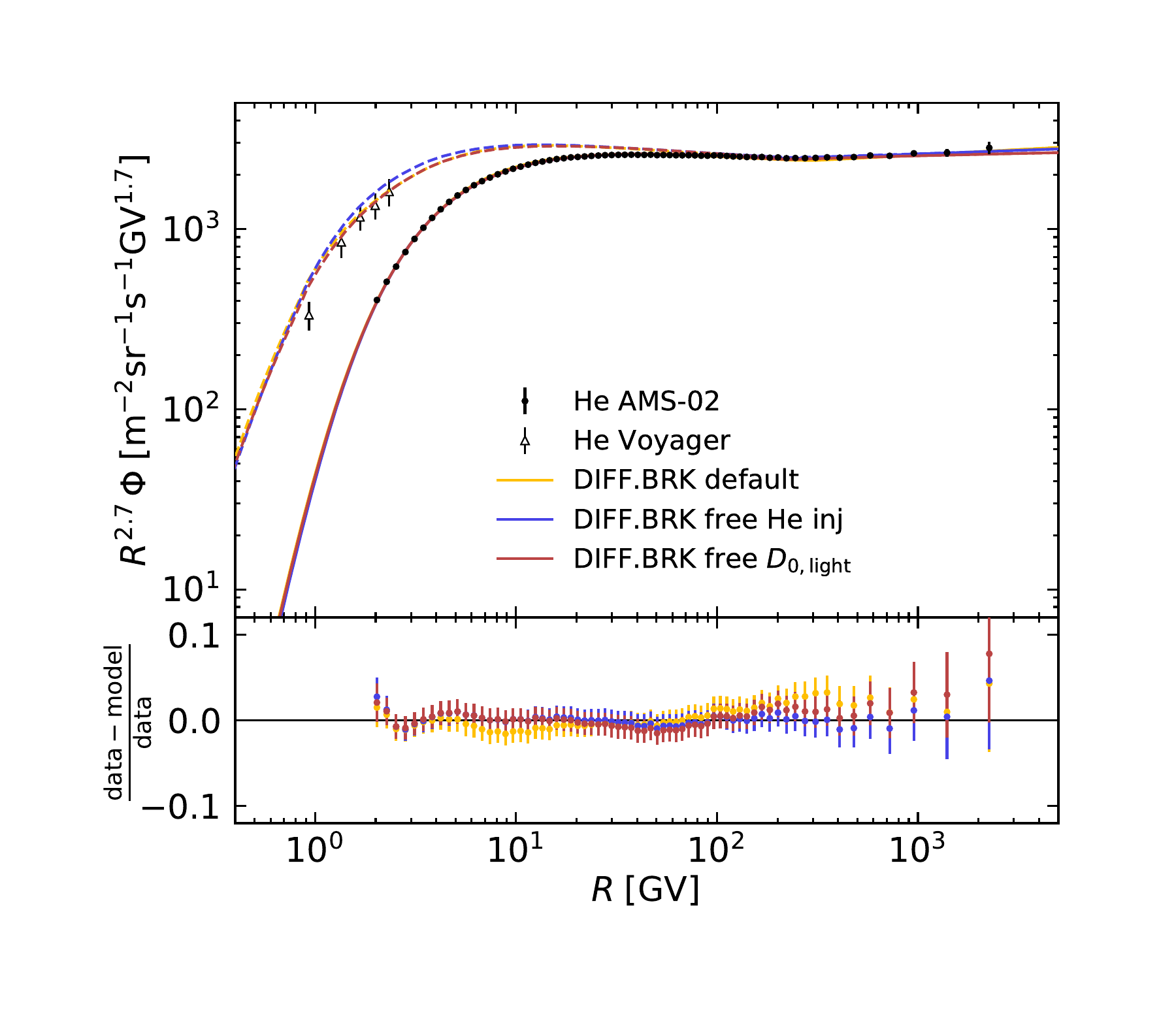}}
 \put(0.00, 0.75 ){\includegraphics[width=0.55\unitlength ]{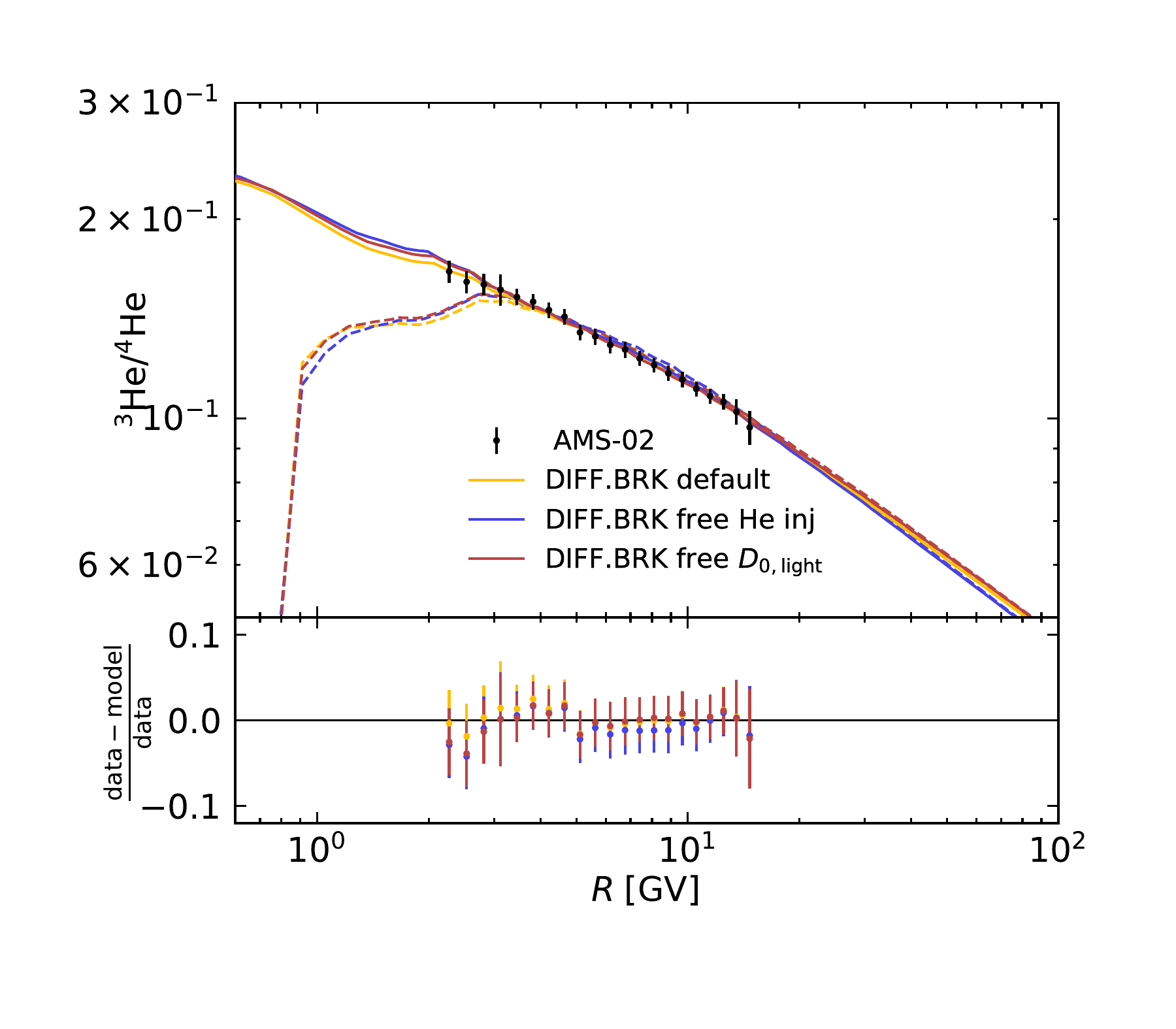}}
 \put(0.50, 0.75 ){\includegraphics[width=0.55\unitlength ]{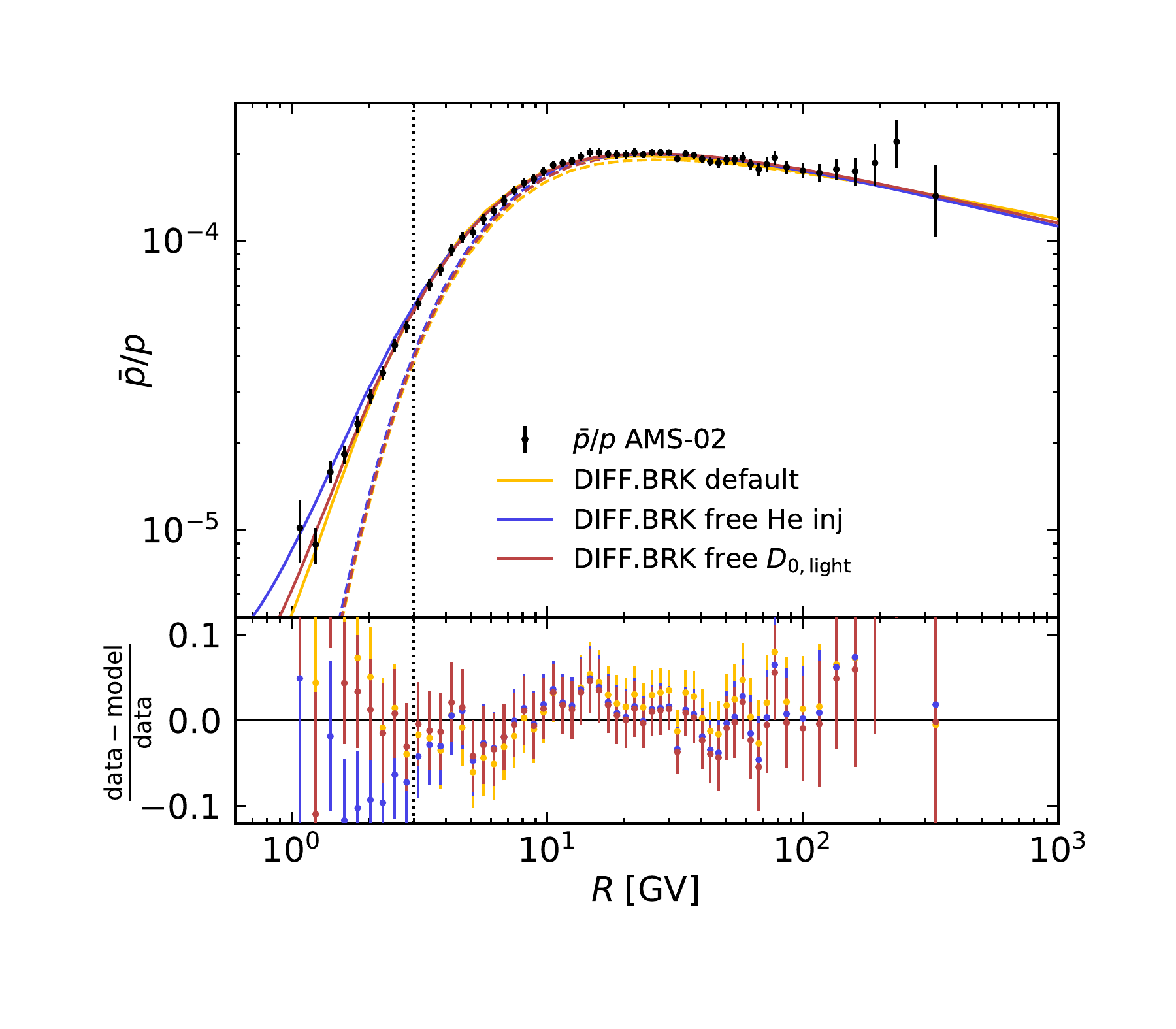}}
 \put(0.00, 0.35 ){\includegraphics[width=0.55\unitlength ]{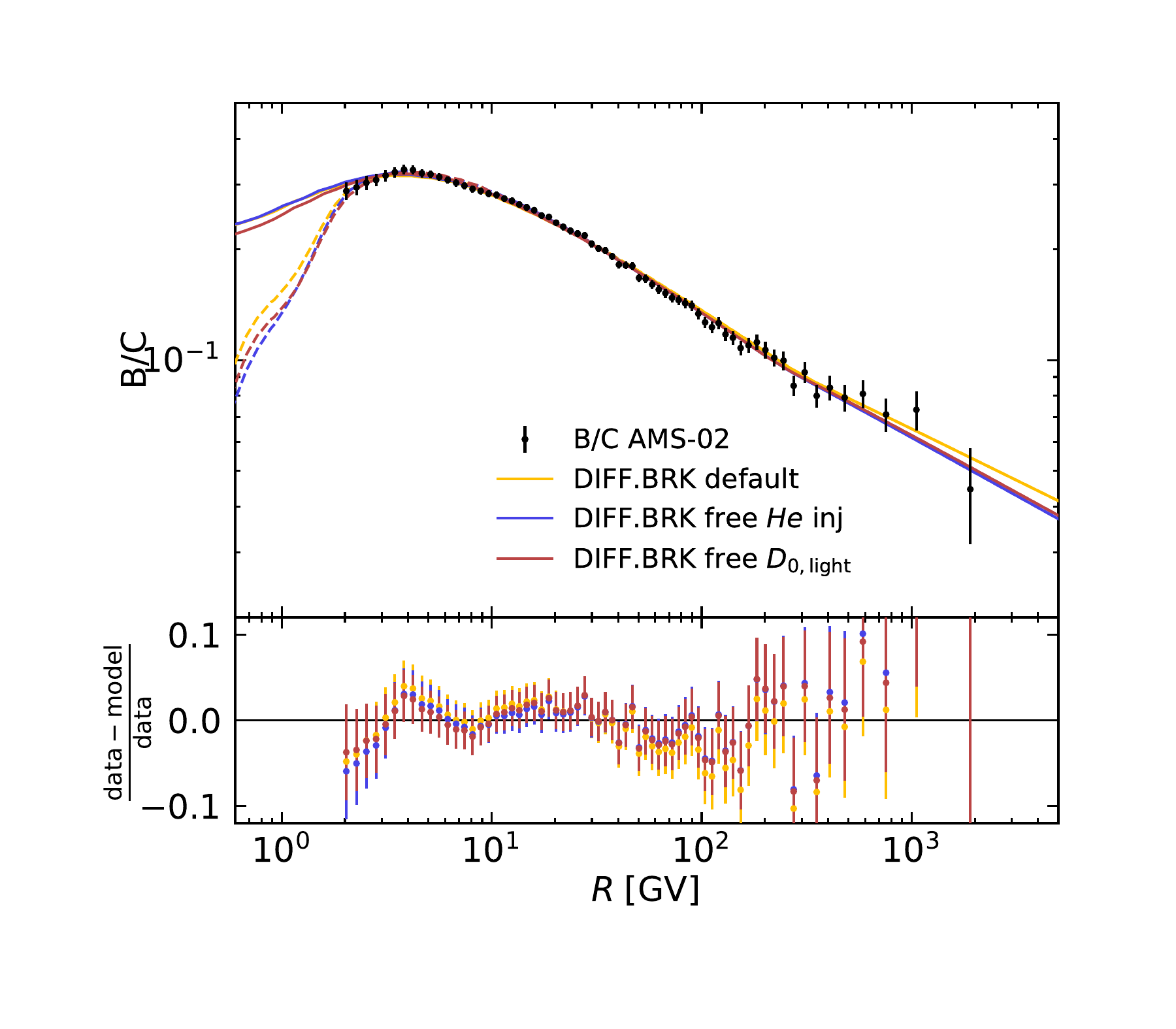}}
 \put(0.50, 0.35 ){\includegraphics[width=0.55\unitlength ]{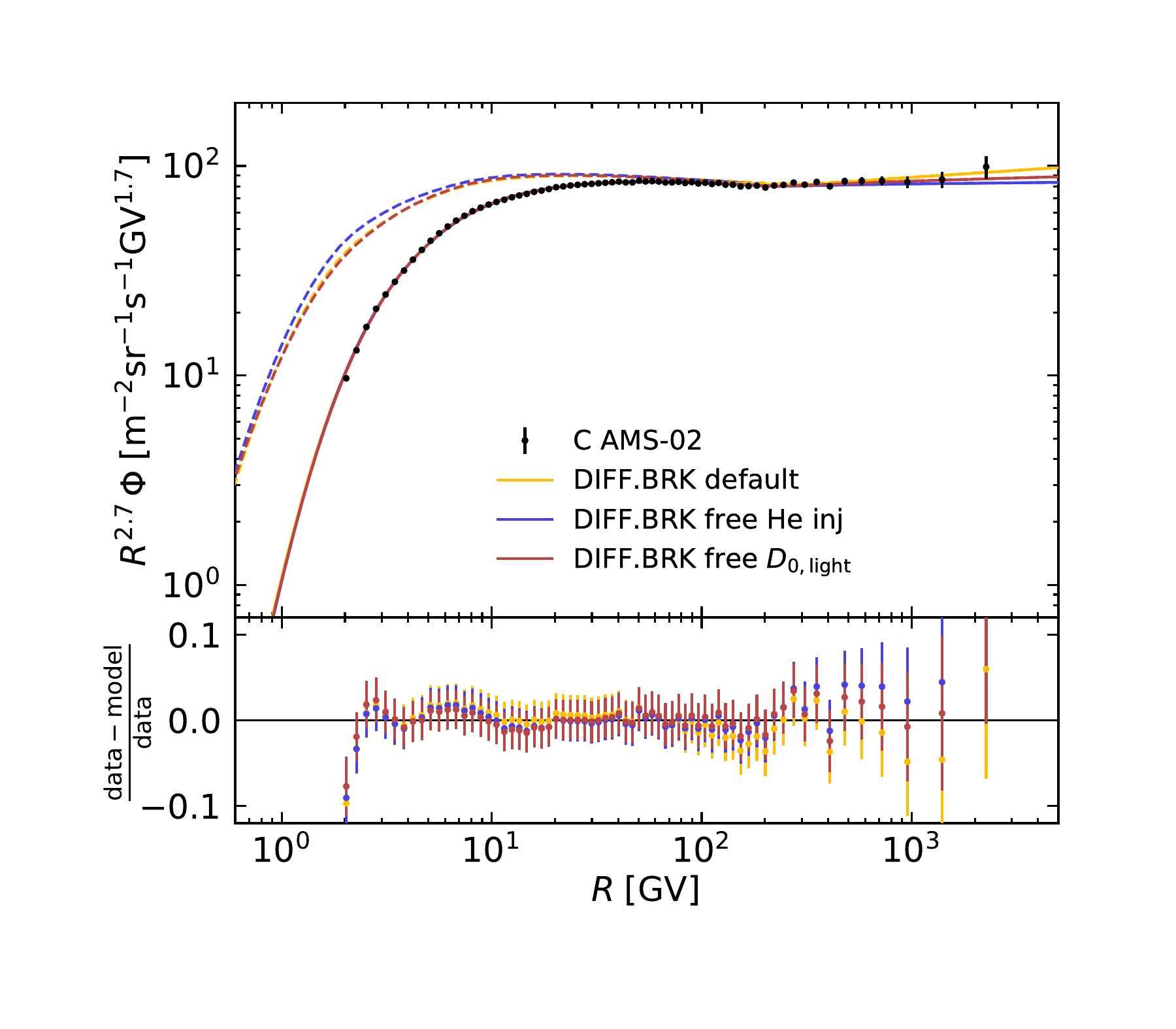}}
 \put(0.00, -0.05){\includegraphics[width=0.55\unitlength ]{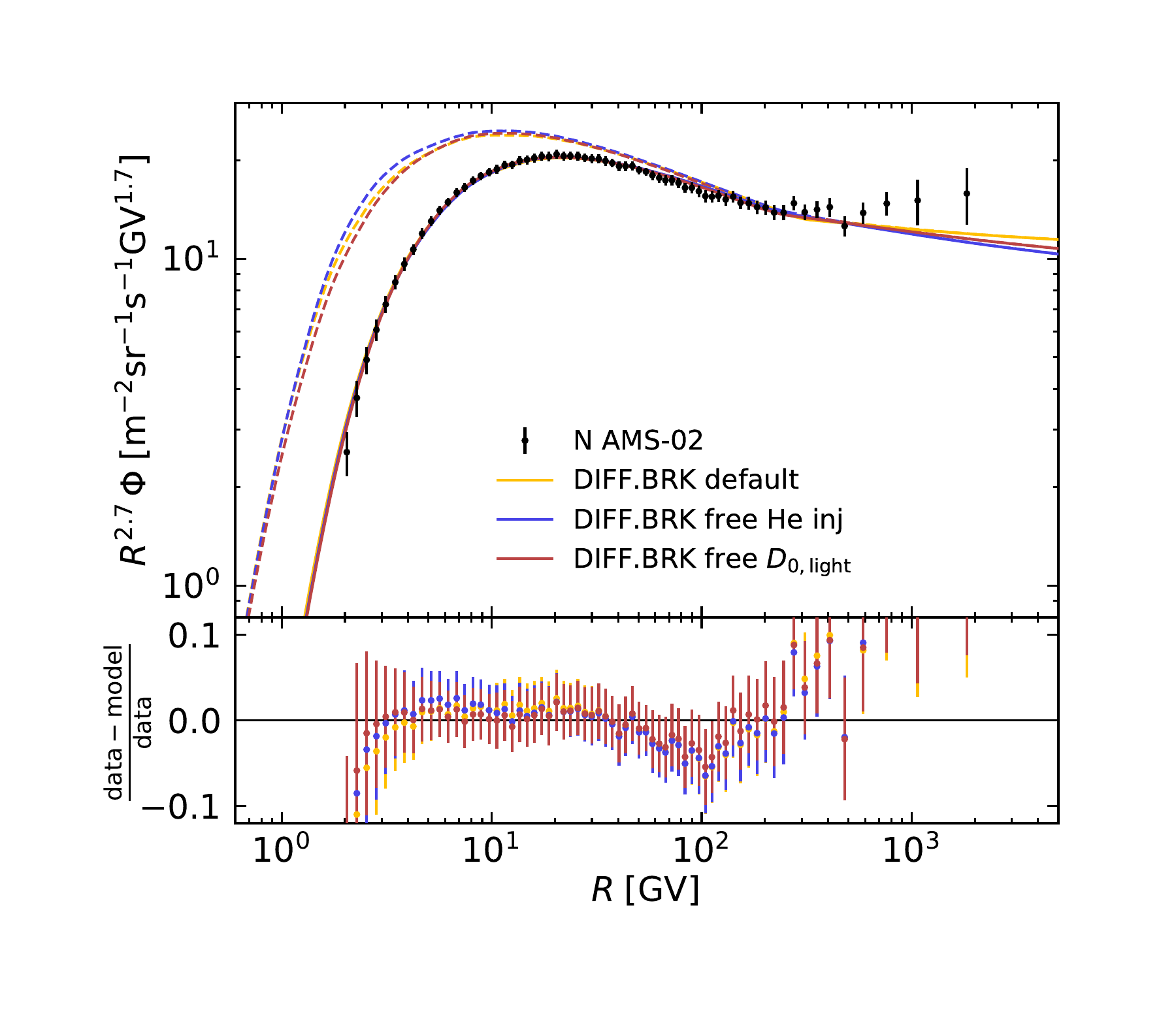}}
 \put(0.50, -0.05){\includegraphics[width=0.55\unitlength ]{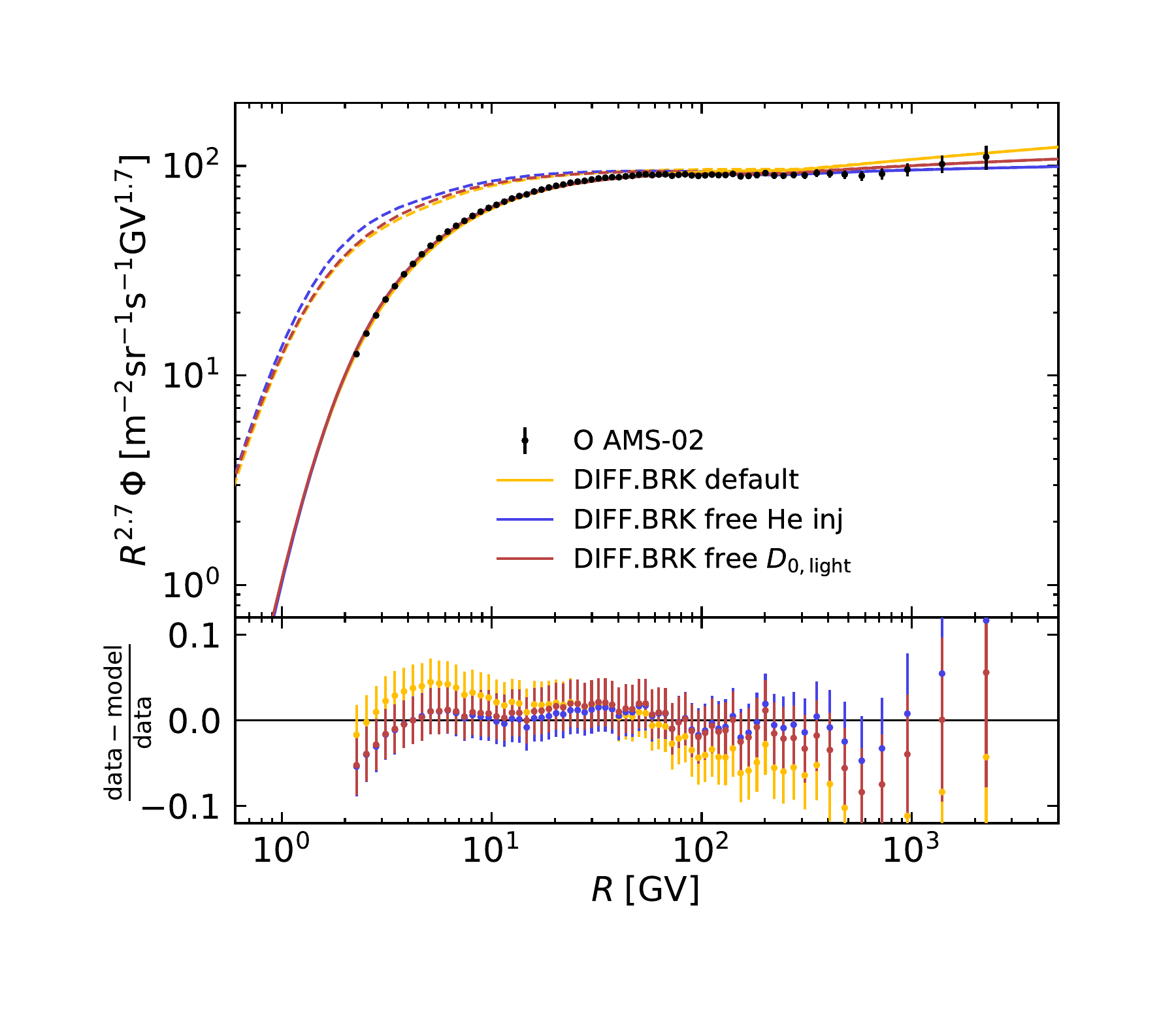}}
\end{picture}
\caption{ 
          Spectra of all CR species included in the fit compared with the data. Dashed lines show the interstellar spectra, while solid lines refer to the
          top-of-the-atmosphere flux. In the lower panels, we provide the residuals. The vertical dotted black lines indicate a 
          rigidity cut imposed on the AMS-02 data sets for protons and the antiproton-to-proton ratio.
          The three colors refer to the different fit setups:
          \emph{default} (yellow), \emph{free He inj} (blue), and \emph{free $D_{0,\rm light}$} (red).
          It can be seen  that the \emph{free He inj} and \emph{free $D_{0,\rm light}$} setups provide a good fit to all species, with flat residuals, 
          while the \emph{default} fit has difficulties in correctly reproducing the oxygen spectrum. See the main text for more details.
        }
 \label{fig:DIFFBRK_spectra}
\end{figure*}
%                                      \         |
%                                        \       |
%                                          \     |
%=====================

%====================
%    \                                           |
%      \                                         |
%        \                                       |
\begin{figure*}[t]
\centering
\setlength{\unitlength}{0.75\textwidth}
\begin{picture}(1,1.66)
 \put(0.00, 1.15 ){\includegraphics[width=0.55\unitlength ]{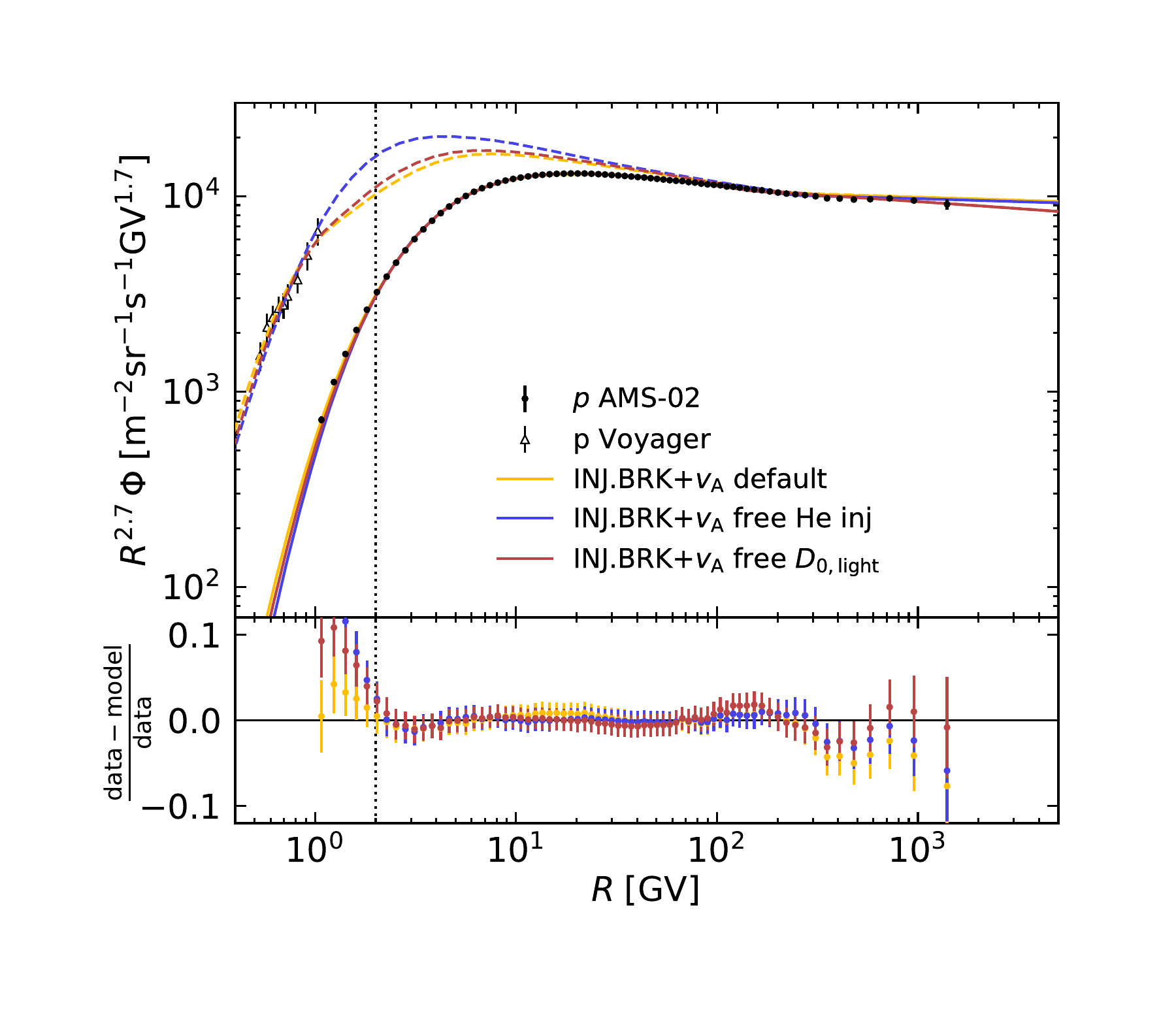}}
 \put(0.50, 1.15 ){\includegraphics[width=0.55\unitlength ]{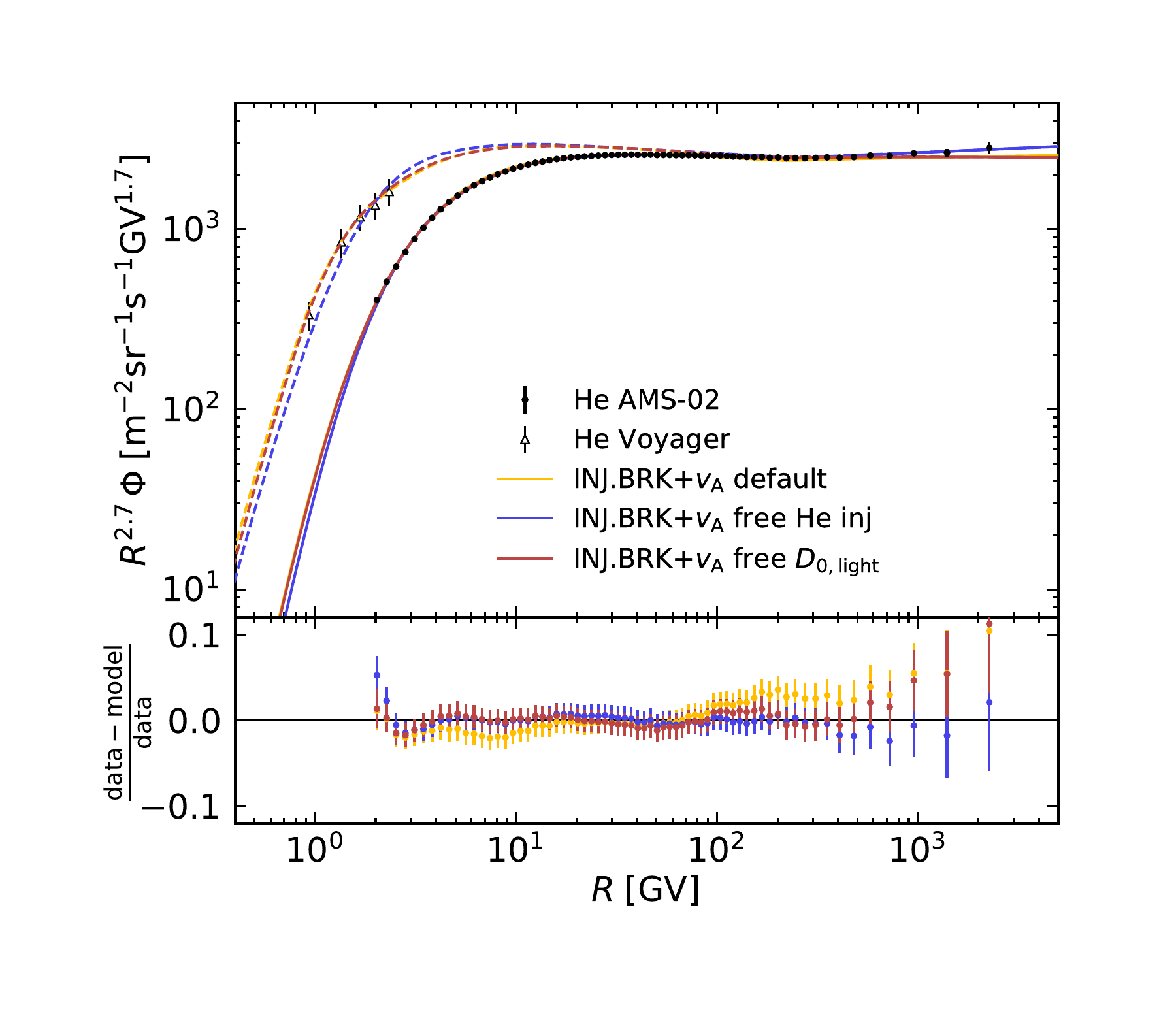}}
 \put(0.00, 0.75 ){\includegraphics[width=0.55\unitlength ]{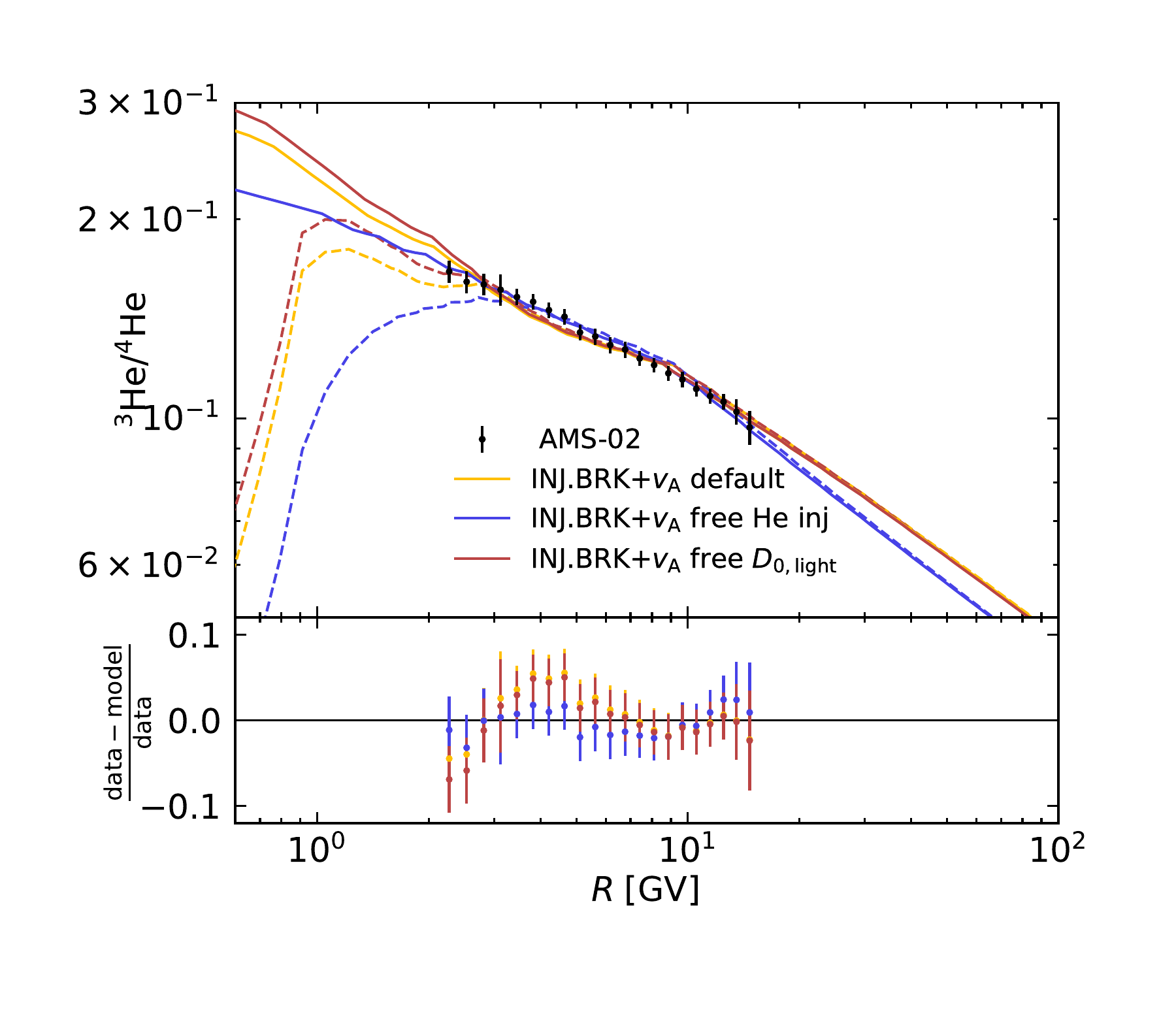}}
 \put(0.50, 0.75 ){\includegraphics[width=0.55\unitlength ]{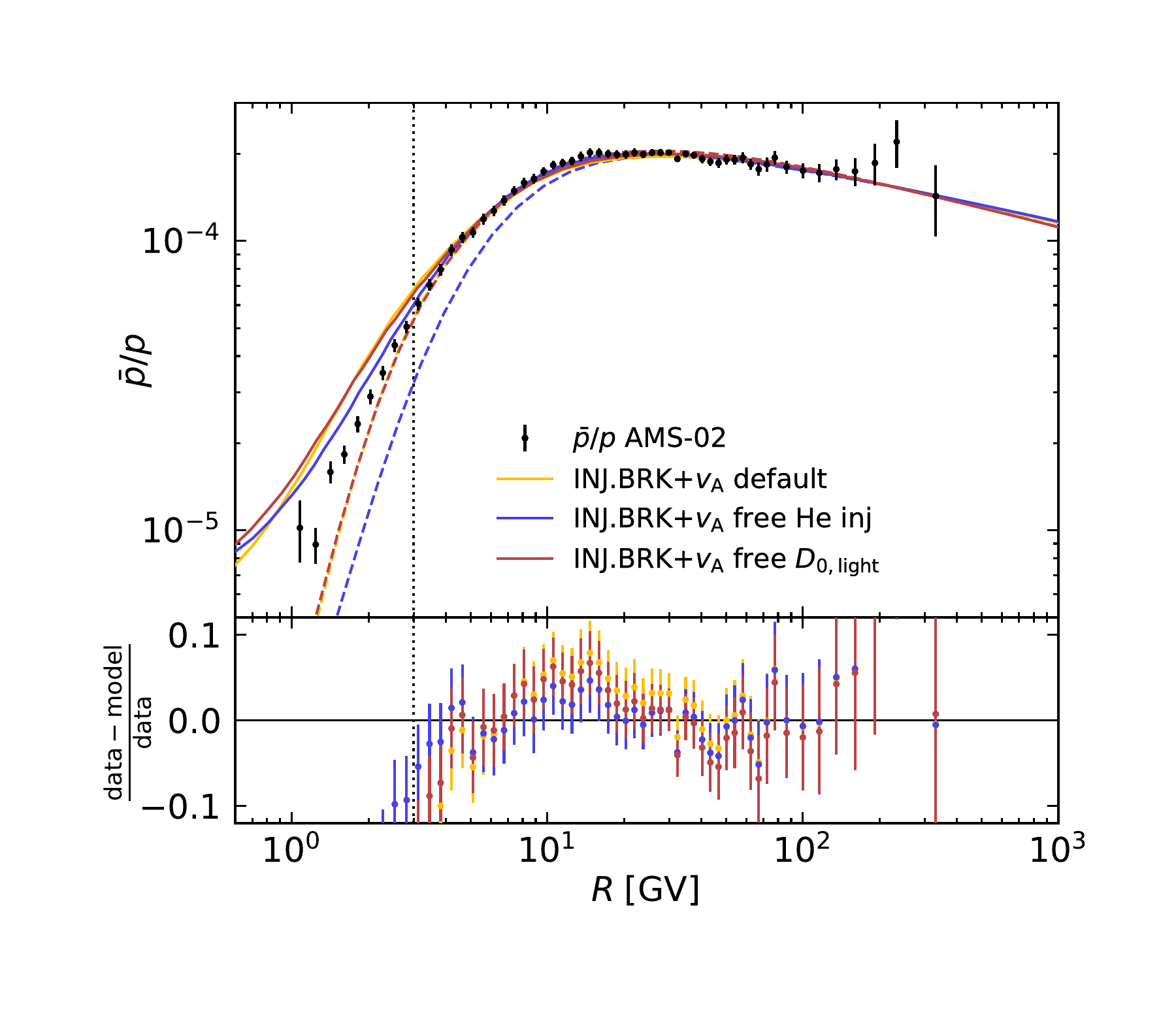}}
 \put(0.00, 0.35 ){\includegraphics[width=0.55\unitlength ]{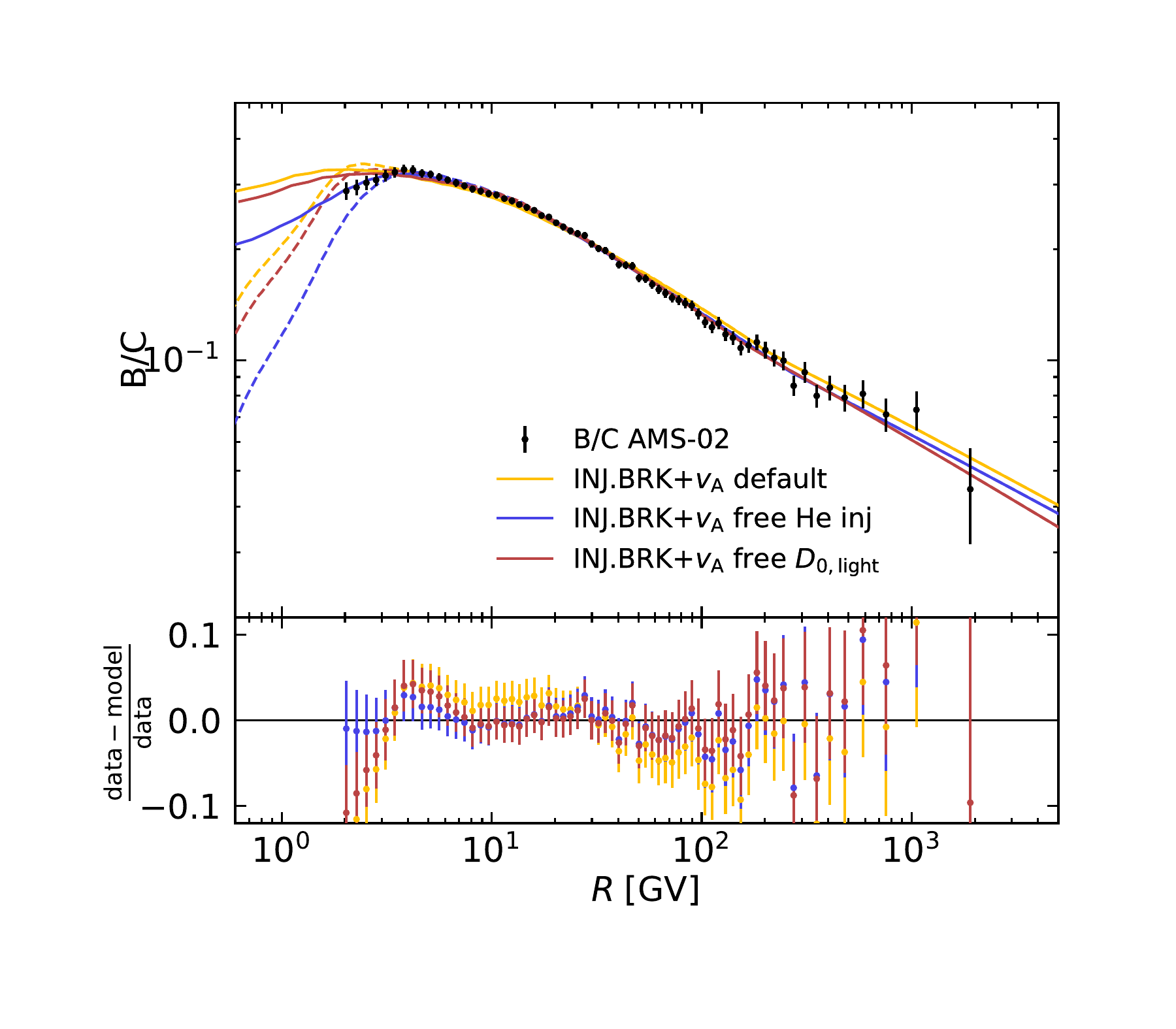}}
 \put(0.50, 0.35 ){\includegraphics[width=0.55\unitlength ]{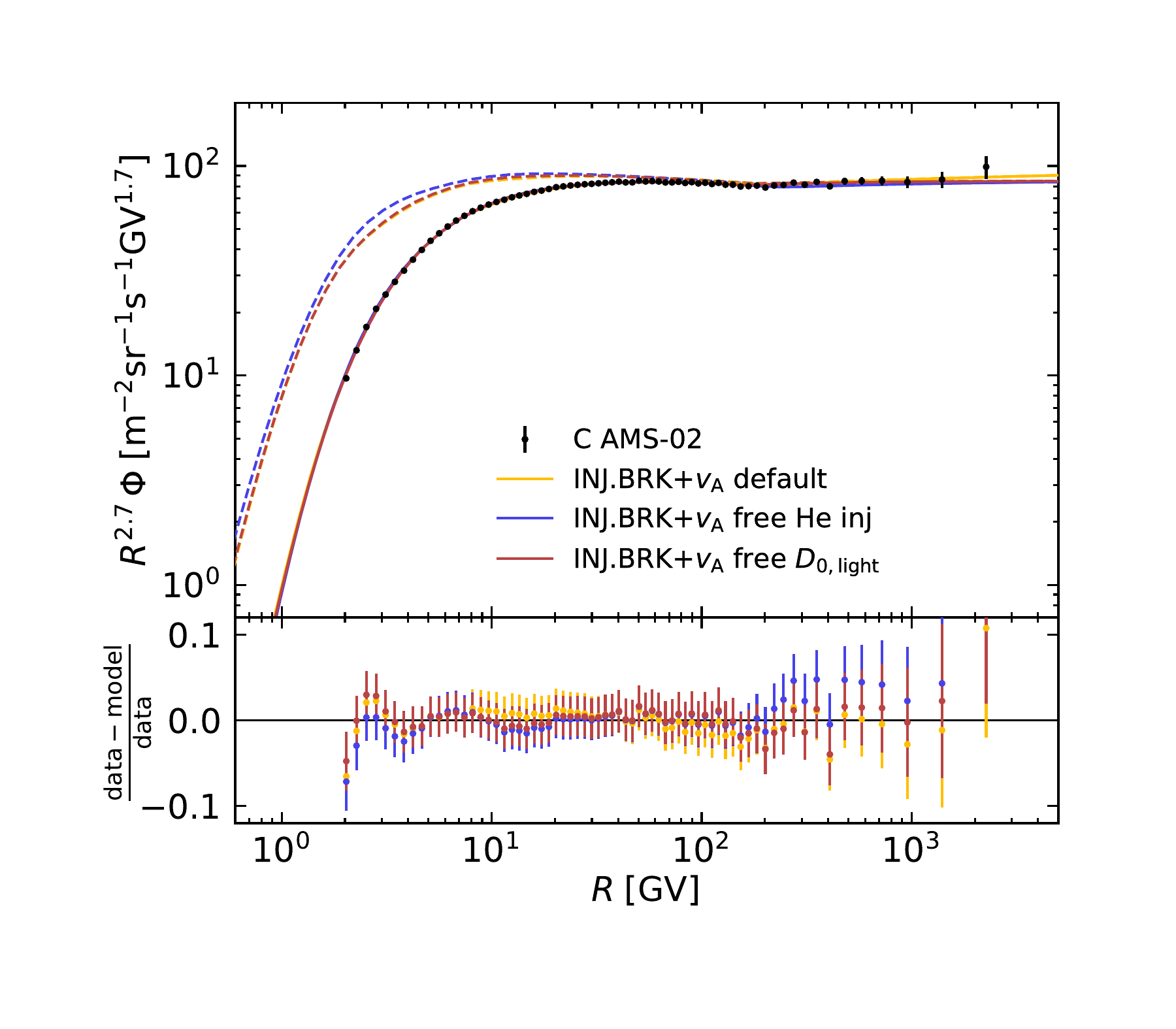}}
 \put(0.00, -0.05){\includegraphics[width=0.55\unitlength ]{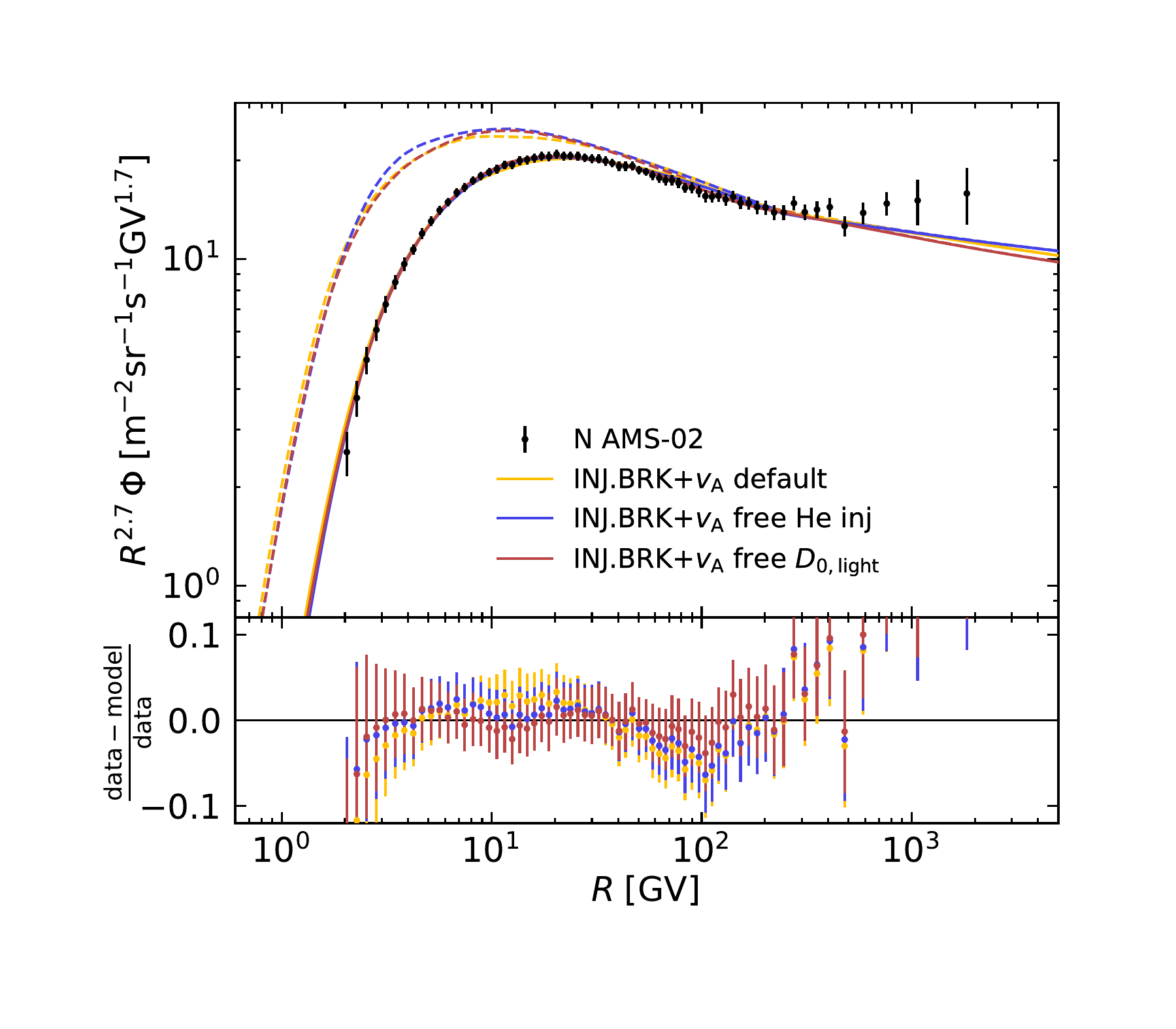}}
 \put(0.50, -0.05){\includegraphics[width=0.55\unitlength ]{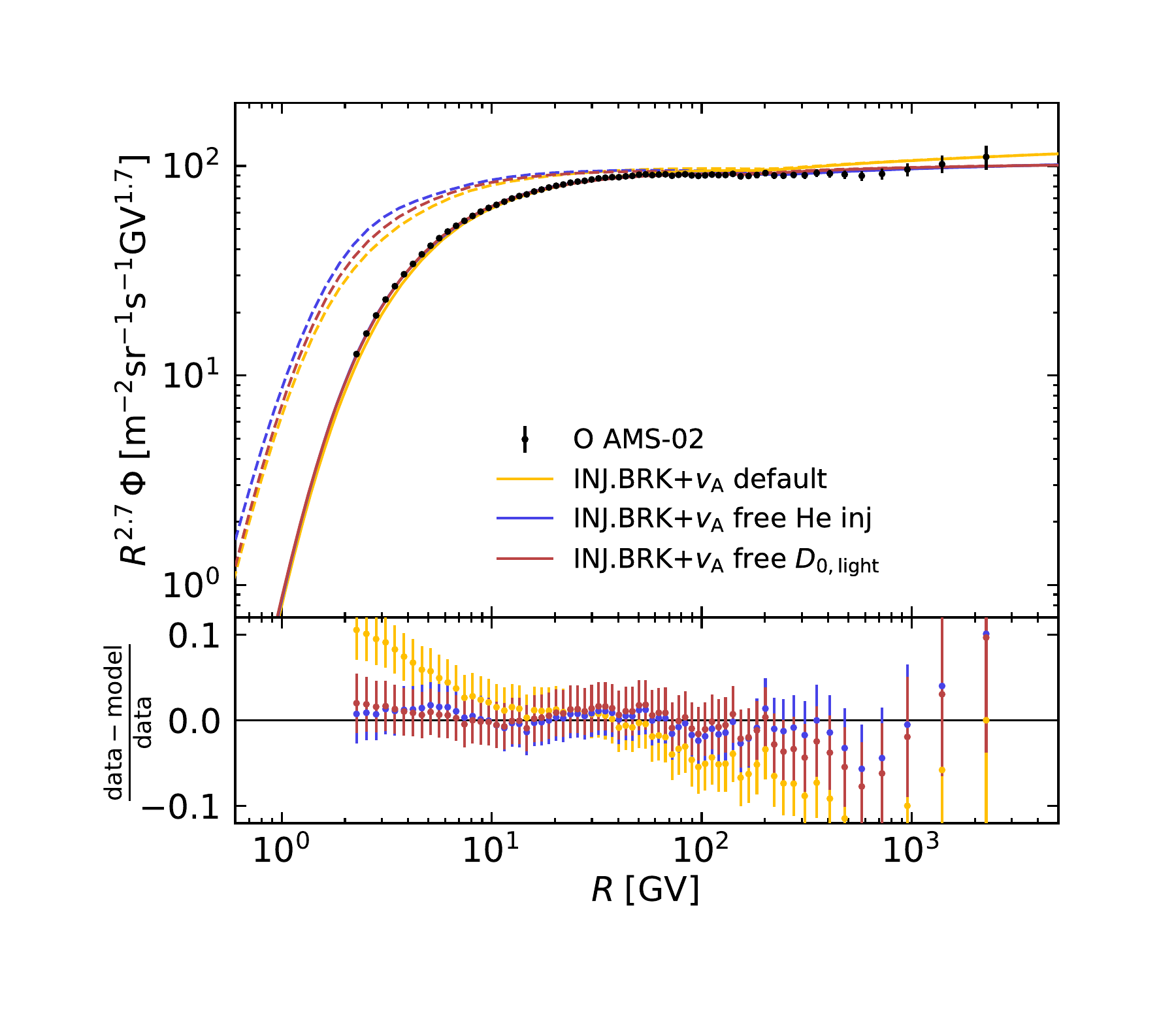}}
\end{picture}
\caption{ 
          Same as Fig.~\ref{fig:DIFFBRK_spectra}, but for the propagation framework \DR.
        }
 \label{fig:INJBRKvA_spectra}
\end{figure*}
%                                      \         |
%                                        \       |
%                                          \     |
%=====================

%====================
%    \                                           |
%      \                                         |
%        \                                       |
%                                      \         |
%                                        \       |
%                                          \     |
%=====================

%====================
%    \                                           |
%      \                                         |
%        \                                       |
%                                      \         |
%                                        \       |
%                                          \     |
%=====================

%=====================
%    \                                           |
%      \                                         |
%        \                                       |
\begin{figure*}[t]
\centering
\setlength{\unitlength}{0.75\textwidth}
\begin{picture}(1,0.37)
 \put(0.00, -0.05){\includegraphics[width=0.55\unitlength ]{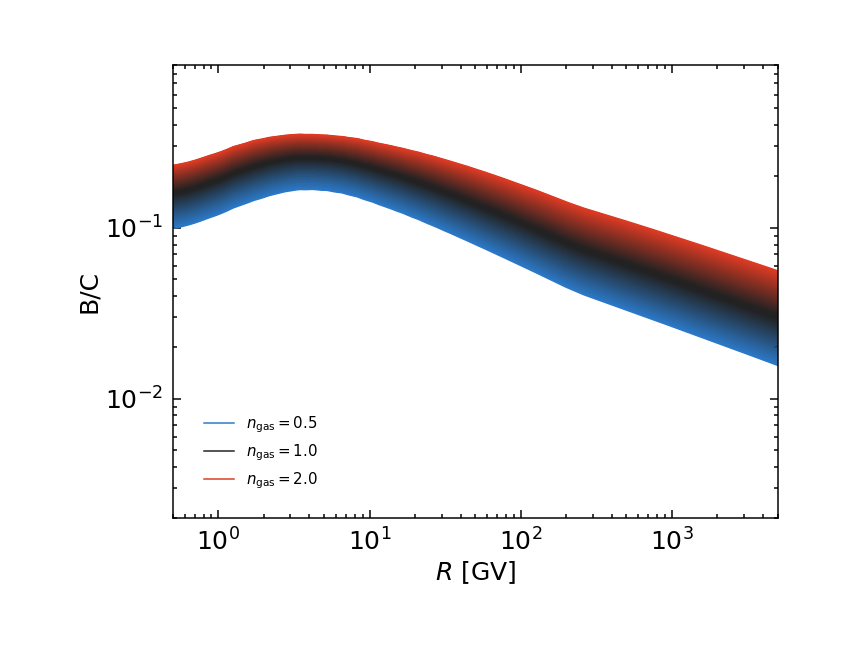}}
 \put(0.50, -0.05){\includegraphics[width=0.55\unitlength ]{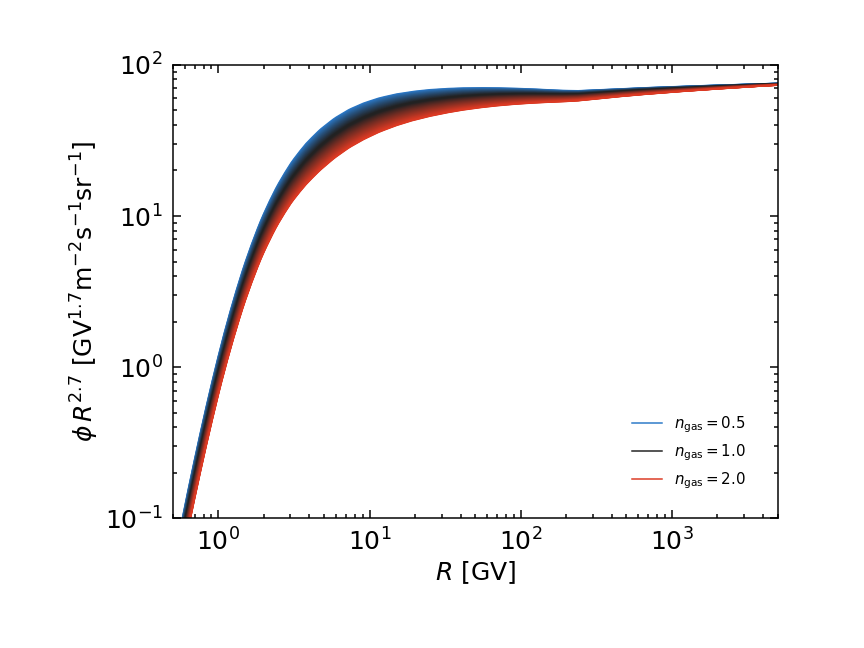}}
\end{picture}
\caption{ 
          Impact of the gas density on the secondary-to-primary CR flux ratio of B/C (left panel) and the primary CR flux of oxygen.
          The normalization of the default gas density is varied by a factor ranging from 0.5 to 2.
        }
 \label{fig:gas_density}
\end{figure*}
%                                      \         |
%                                        \       |
%                                          \     |
%=====================

\bigskip
\noindent
\textbf{\underline{Default fit}}\hspace{1.5mm}$\boldsymbol{\cdot}$ 
We have already noticed in Sec.~\ref{sec::motivation} that light and heavy nuclei are not fully compatible, in particular, regarding the slope of the primary injection spectrum ($\gamma_2$). Therefore, it is not very surprising that the naive approach of a combined fit (labeled \emph{default}) results in a relatively bad fit. The best-fit $\chi^2$s are 341 and 529 for the CR propagation setups \DB\ and \DR, respectively, for about 465 degrees of freedom in total. While overall this does not indicate a bad global fit, some species are obviously not well described. Figures~\ref{fig:DIFFBRK_spectra} and \ref{fig:INJBRKvA_spectra} show that in particular the O fluxes have a poor $\chi^2$ of 86 and 160 for the \DB\ and \DR\ setups respectively for 67 data points. The slope in particular is not reproduced correctly, which is completely expected given the fact that the light and heavy nuclei separately prefer different ones, as already discussed. A bit more surprising is, instead, the fact that C and N spectra are well reproduced. This is likely because they have a sizeable secondary component, especially N, and this can be partially adjusted by the fit to compensate for the different slope, also because of the freedom in the cross-section which we introduce. On the contrary, O is purely primary (or with only a sub-percent level of secondary), thus this adjustment is not possible. Finally, in the \DR\ setup also the antiproton flux at low energies has some difficulty to be reproduced, despite the cut at 3 GV which we use.

\bigskip
\noindent
\textbf{\underline{free He inj fit}}\hspace{1.5mm}$\boldsymbol{\cdot}$ 
Since the different injection slopes preferred by light and heavy nuclei seems to be the main problem of the \emph{default fit}, introducing as freedom different injections not surprisingly solves the issue and provides a much better fit, with a $\chi^2$ of 194 and 171 for the \DB\ and \DR\ setups. From figures~\ref{fig:DIFFBRK_spectra} and \ref{fig:INJBRKvA_spectra} it can be seen that all the species are well reproduced with flat residuals. We note that protons are nonetheless always allowed to have a steeper injection spectrum. With the additional freedom, He picks a slightly harder spectrum of roughly 0.05 in both the \DB\ and \DR\ setups. This is also visible in the right panels of Figs.~\ref{fig:DIFFBRK_smallTriangles} and \ref{fig:INJBRKvA_smallTriangle}. The drawback of this solution is of course the fact that CR universality for nuclei is violated. But, although this has little appeal and it is difficult to justify from a theoretical point of view, it remains in principle a viable possibility. The spectrum of C and N is also well reproduced, although the normalization of the secondary C cross-section production saturates our prior range and goes to a value $A_{\rm XS \rightarrow C}=0.5$ which is beyond the allowed experimental uncertainties. In principle, this can be solved by introducing further freedom also for the C injection spectrum. This feature would be reasonable in this scenario, since, once CR universality is violated, it becomes likely that each nucleus has its own injection spectrum, rather than having only a light vs heavy difference. Finally, the slight tension observed in the light vs heavy fit for the parameter $D_0$ does not translate into a serious problem in the joint fit, and intermediate values can provide a good fit for all species.

\bigskip
\noindent
\textbf{\underline{free $\boldsymbol{D_{0,{\rm light}}}$ fit}}\hspace{1.5mm}$\boldsymbol{\cdot}$ 
Since the final observed CR spectrum results from the effects of diffusion and energy losses on the injected spectrum, it is conceivable that, besides changing injection as in the above scenario, the tension can be relieved also allowing for a different diffusion. Hence, the free $D_{0,{\rm light}}$ case. We see indeed that this also works and a good fit is achieved with a $\chi^2$ of 206 and 230 for the \DB\ and \DR\ cases, with flat residuals for all the species, apart, perhaps, some tension in the $\bar{p}/p$ spectrum at low energy for the \DR\ case. Looking at the left panels of Figs.~\ref{fig:DIFFBRK_smallTriangles} and \ref{fig:INJBRKvA_smallTriangle}, we see that the diffusion coefficient of the light nuclei prefers smaller values than the one of heavier nuclei by roughly $1\times10^{28}\,\mathrm{cm^3/s}$, again consistently for both of the investigated propagation setups. A possible problem in this fit seems to be given by the large value of the secondary N production cross-section $A_{\rm XS \rightarrow N}\simeq 1.7-1.8$, which is significantly beyond the allowed (20-30\%) uncertainties. Also, the normalizations of C, N, and O are significantly different from 1 (around 1.2-1.3) which indicates that the fit is not fully self-consistent. 
To investigate this issue more thoroughly, we have performed fits using the physical abundances of C, N, and O as parameters rather than the normalizations thus removing the consistency problems although at the price of significantly computationally heavier fits. As expected, we observe that the fits prefer larger physical abundances for C, N, and O with respect to the values fixed in our standard setup. Furthermore, the normalization of the N production cross-section converges to a more reasonable value of $A_{\rm XS \rightarrow N}\simeq 1.4$ with the quality of the fit only worsening very slightly.

\bigskip
\noindent
\textbf{\underline{free sec. norms fit}}\hspace{1.5mm}$\boldsymbol{\cdot}$ 
The final scenario we consider is the one in which we leave free all the normalizations of the secondaries including $\bar{p}$. We do not show the results in the plots to not overcrowd them, but from Table~\ref{tab::fit_results}, and Fig.~\ref{fig:fitresults} it can be seen that also this scenario works, especially for the \DB\ case with a $\chi^2$ of 184, while in the \DR\ case it gives an improvement, with a $\chi^2$ of 276, although some tension remains both for $\bar{p}/p$ and mainly for O. This, however, comes at the price of very large production cross-sections for all the secondary species, in the range 1.5-1.9, which are, often, too much beyond the allowed uncertainties. For this reason, we deem this solution, ultimately, not realistic. It is interesting, however, to understand why this scenario works. The reason lies in the above-mentioned degeneracy between the secondary CR production cross-section and the diffusion coefficient $D_0$. The large cross-section is thus accompanied by large $D_0$ values of about $8.5\times10^{28}\,\mathrm{cm^3/s}$ in the \DB\ case. With a large $D_0$, i.e. with fast diffusion, the relative weight of inelastic energy losses is significantly decreased and the final observed spectrum results almost entirely from the effect of diffusion alone, with energy losses playing only a minor role. In this way, since the observed spectra of He, C and O have the same slope, they can be explained with the same injection and same diffusion, despite the energy losses being significantly different. This explains also why this solution does not work very well for \DR\ fit. In this case, the $D_0$ secondary cross section degeneracy is not perfect due to the presence of reacceleration and thus the above adjustments work less well.

%===================================================================
\subsection*{Further scenarios} \label{sec::morescenarios}
%===================================================================

We have also investigated further scenarios, besides the ones described above, although in a less systematic manner. The results are nonetheless worth mentioning at least briefly. All these scenarios, described below, are to be intended as extensions of the \emph{default} setup, and they all assume CR universality.

\begin{itemize}
\item \emph{free inelastic cross sections}\hspace{0.3cm}  
This is the third ingredient, after diffusion and injection which is natural to change to resolve the light vs heavy tension since it directly affects the observed spectrum of primaries. We have thus performed a fit including as free parameters the normalization and slopes of the inelastic fragmentation cross sections of He, C, N, and O using the same parametrization employed for the secondary CR production cross sections. This scenario can provide a good $\chi^2$ and flat residuals for all the species. However, this comes at the price of very low normalizations for the C, N, and O fragmentation cross sections, less than a factor of two with respect to the nominal values. Again, this is easy to understand, and the mechanism is similar to the \emph{free sec norm} case. Decreasing the energy losses, the observed spectrum is given basically only by the effect of diffusion, and thus it is possible to have the same injection for light and heavy nuclei. Nonetheless, as in \emph{free sec norm} case, the required down-scaling of the cross sections, of the order of 100\% or more, is significantly beyond the allowed experimental uncertainties, and thus, ultimately, this is not a realistic solution.
\item \emph{free gas density}\hspace{0.3cm} A further natural parameter to check is the normalization of the gas density. Indeed, the energy losses actually always depend on the product of the cross section and the gas density, so the overall effect can be controlled in both ways. Furthermore, the gas density in the Galaxy has indeed some sizeable uncertainty \cite{Johannesson:2018bit}. In practice, we retain the \textsc{Galprop} spatial model of the gas distribution in the Galaxy and we introduce the overall normalization as a parameter. The effect of varying the gas density with respect to the reference value is shown in Fig.~\ref{fig:gas_density}, and it can be seen, as expected, that it changes the slope of the primary CRs, in the same, degenerate, way as the inelastic cross section, producing a hardening of the spectrum when increasing the gas density, and thus increasing the energy losses. However, as shown in the figure, it affects also the normalization of the secondaries, which increases with increasing gas density. The effect is easily understood since also the production of secondaries is proportional to the product of the production cross section times the gas density. This attempt, however, has been unsuccessful and no good fit could be achieved. Ultimately the reason is given by the fact that both the primaries and secondaries are affected at the same time, while it is only the primaries which need \emph{fixing} in order to solve the light vs heavy tension.
\item \emph{free diffusion halo height}\hspace{0.3cm} In all the above fits we kept the half-halo size $z_h$ fixed to 4 kpc, because of the $z_h$-$D_0$ degeneracy. Nonetheless, this degeneracy is weakly broken by secondary effects and there is the possibility that leaving $z_h$ would help in resolving the light vs heavy nuclei tension. We thus tested a scenario in which we leave $z_h$ free to vary. This, however, did bring only a minor improvement in the fit without providing a solution to the above tension. 
\item  \emph{free $D_{0,{\rm light}}$ + free $D_{0,{\rm light},p}$.}\hspace{0.3cm} 
With the \emph{free $D_{0,{\rm light}}$} scenario, we saw that it is possible to reconcile the inconsistency between He and CNO with an effectively different value of the diffusion coefficient for light and heavier nuclei. In this spirit, it is tempting to see if the universality between $p$ and He can be restored by introducing a different diffusion coefficient for $p$ itself. We tested this possibility in a dedicated fit which is an extension of the \emph{free $D_{0,{\rm light}}$} scenario with the extra parameter $D_{0,{\rm light},p}$, and forcing $\gamma_{2,p} = \gamma_2$. Ultimately, however, the result is unsatisfactory and a good fit to $p$ and He spectra cannot be achieved in this scenario. Nonetheless, it might be worth testing this possibility more in the future by exploring a more self-consistent non-homogenous diffusion scenario instead of the effective approach used here.
\end{itemize}

%===================================================================
\section{Discussion and comparison with other works} \label{sec::comparison}
%===================================================================

To our knowledge, this is the first work carefully exploring the universality of CR nuclei by trying to consistently model both secondary and primary CRs and both light and heavy nuclei using the AMS-02 data. There are, nonetheless, numerous works in the literature that perform similar analyses, although with various limitations. 

The authors of Refs.~\cite{Boschini:2018baj,Boschini:2019gow,Boschini:2020jty} determine the local interstellar spectra of CR nuclei from proton to iron using a propagation setup similar to our \DR\ framework and the \texttt{Galprop} code. CR source injection and propagation parameters are determined using an iterative procedure. In more detail, at each iteration step first the CR propagation parameters are fitted and then the source parameters of the primary CRs are adjusted, individually for each species, together with the parameters for solar modulation which is treated using the numerical \textsc{HelMod} code. The injection spectra are modeled as a triple broken power law with smoothing at each break, and with a further fourth break in the case of $p$ and He. As a consequence, this model has more than 200 free parameters for the injection spectra of primary CRs providing in principle a maximum violation of CR universality. The result is, however, not further discussed in detail. 

Refs.~\cite{Luque:2021nxb,DeLaTorreLuque:2021yfq,Luque:2021ddh} focus on a similar subset of CRs as our work, although $^3$He data are not included. CR propagation is treated numerically with the \texttt{Dragon} code, and in the work a particular focus is given on the nuclear cross-section uncertainties. Also, this analysis relies on an iterative adjustment of propagation and source parameters, which leads indeed to different injection slopes for $p$, He, C, and O, although the implications for CR universality are not discussed. 

Ref.~\cite{Liu:2018ujp} performs an analysis similar to ours, using the \texttt{Dragon} code and a propagation model which has non-homogeneous diffusion in a thin region close to the Galactic plane. They use as CR data $\bar{p}$, $p$, He, Li, Be, B, C, and O, not including $\mathrm{^3He}$ and N. It is difficult, however, to comment on the results in detail, since a discussion on the quality of the fit and the uncertainties is missing. Also, the impact of cross-section uncertainties is not considered in the analysis.

The analysis in Refs.~\cite{Boudaud:2019efq,Weinrich:2020cmw,Weinrich:2020ftb} exploits a semi-analytical model of propagation and takes nuclear cross-section uncertainties into account. They fit the secondary CRs $\mathrm{^3He}$, Li, Be, B ($\bar{p}$ is not included) to determine the propagation parameters. The injection spectra of primaries are again adjusted in an iterative process. They use different propagation frameworks exploring setups very similar to our frameworks \DB\ and \DR and they find that both light and heavy secondaries are compatible with the same propagation. Regarding CR universality, they find a harder injection slope for He compared to C and O, in qualitative agreement with our findings, although numerically they find a smaller difference of $0.02\pm0.007$ \cite{Boudaud:2019efq}, compared to our $\sim 0.05$.

In the work in Refs.~\cite{Evoli:2019wwu,Evoli:2019iih,Schroer:2021ojh} both primary and secondary CRs are fitted using a semi-analytical approach with a propagation model similar to our \DR\ framework. Some dataset is not included in the analysis, namely $\mathrm{^3He}$ and $\bar{p}$. They use a lower limit of 10~GV in the fit to exclude the rigidity region most strongly affected by solar modulation, which, however, limits the possibility to test the presence of a break at few GVs in the diffusion coefficient or in the injection spectra. The issue of CR universality is discussed and they find that to explain the data a harder spectrum for He compared to C and O is required, by an amount $\sim 0.05$ \cite{Evoli:2019wwu}, in agreement with our findings, while using the same injection for He and C and O worsen significantly the fit. For all CR spectra heavier than He, they find that a single injection slope gives a reasonable fit to the AMS-02 CR data except iron.

The recent work by Ref.~\cite{Zhao:2021yzf} discusses a non-homogenous diffusion model with an inhibited diffusion coefficient around the galactic plane. The Be, B, and C data are fitted well by this model. Proton and helium data are then fitted by separately adjusting their injection parameters. However, the universality of injection parameters in not discussed.

%===================================================================
\section{Summary and Conclusions} \label{sec::conclusions}
%===================================================================

In this work, we have investigated the universality of CR nuclei regarding the two aspects of acceleration in the sources and subsequent propagation in the Galaxy. To this end, we have performed fits of CR data on antiprotons, and from protons to oxygen recently provided by the AMS-02 experiment. As already indicated by pre-AMS-02 CR measurements, universality is violated for protons and helium.
While, from a theoretical perspective, the basic theory of shock-acceleration predicts a universal CR spectrum produced in the sources, various ideas have been proposed to explain the proton and Helium spectra.  
For example, their difference could be explained by different populations of sources with different helium and hydrogen compositions \cite{Tomassetti:2015xem,Tomassetti:2016bcf,Lagutin:2019nfn,Ptuskin:2012qu},  
or by an $A/Z$-dependent  and shock Mach number-dependent efficiency of diffusive shock acceleration as indicated by recent numerical hybrid simulations~\cite{Hanusch:2018bsk,Caprioli:2017oun}.
The issue, however, is still not fully clear for heavier primaries, in particular helium, carbon, and oxygen, which we thus study in detail here. To explore the dependence of the results on the uncertainties in the modeling of CR propagation we employ two well-distinct propagation scenarios. In the first setup (labeled \DB) the diffusion coefficient has a break in rigidity between 5 and 10 GeV and a single power law is used for the injection spectrum of primaries. No reacceleration is present. In the second setup (labeled \DR) a break is present in the injection spectrum of primaries but not in the diffusion coefficient, while reacceleration is included. Both setups allow for convective winds driving CRs away from the Galactic plane. At the same time, we also include a treatment of uncertainties in the production cross sections of secondary CRs. These uncertainties are often larger than the uncertainties of the CR flux measurements and thus cannot be neglected. To this purpose, we model these uncertainties through the use of cross-section nuisance parameters which during the fit are treated on equal footing as the propagation and source parameters. In order to handle the large ($> 20$) dimensionality of the global parameter space Monte Carlo scanning techniques are employed.

The main result of the analysis is that a different source/injection spectra for He on one hand and C, N, and O, on the other hand, are required, i.e., universality is violated between He and C, N, and O, with the conclusion holding for both propagation setups. In particular, using the same source spectrum for all the nuclei makes it very difficult to describe the measured O spectrum, which is poorly fitted. The significance of the result is quantified using a fit in which we leave the injection spectral index of He free with respect to the one of C, N, and O. In this case, the quality of the fits improve significantly by a $\Delta \chi^2$ of 148 and 357 for the \DB\ and \DR, respectively. The fits prefer a source injection slope which is harder for He by about $\sim 0.05$ with respect to C, N, and O. In this respect, it is interesting to note that the measured spectra of He, C, and O have the same slope ($\sim 2.7$) within uncertainties. This means that propagation effects, in particular spallation, energy losses, and contributions from secondary components, which are different for different species, alter the source spectra to produce at the end similar propagated spectra. Thus, the propagation effects and spectral differences at the source seem to compensate each other in such a way that they produce the same observed slopes, which is certainly a curious outcome. Whether this is just a coincidence or it hides something more fundamental in nature is unclear.

In principle the same ideas used to explain the spectral difference of $p$ and He could be explored to explain the differences in He, C and O, except for the class of explanations which rely on the $A/Z$ dependence of the shock acceleration efficiency. Those are clearly ruled out since He, C, and O all have the same $A/Z$.
The explanation based on the stochasticity of CR sources and their chemical composition, instead, can be extended to include sources with different compositions in $p$, He, C, N, and O. Although, the increasing number of free parameters makes these scenarios rapidly unappealing.
We thus explore further alternatives that could save universality. In particular, we investigate a scenario in which diffusion is different for light and heavy nuclei, which is technically implemented by using different normalizations of the diffusion coefficient $D_0$ for light nuclei ($p$, He, $\bar p$) and heavier nuclei in the fit. A physical motivation for this scenario might be provided by inhomogeneous diffusion in the Galaxy. The propagation volume, in fact, depends on the mass of the CR nuclei. Thus, if the medium is non-homogenous, different species could effectively sample a different diffusion coefficient. We find indeed that this scenario is able to provide a good description of all the CR species, retaining at the same time the same source spectra for He, C, N, and O, i.e., maintaining CR universality for nuclei. In this case, we find that the light nuclei prefer a diffusion coefficient that is 25\% smaller than the one of the heavier nuclei. 
For future studies, It would be interesting to further investigate this result using physical models of inhomogeneous diffusion. A further alternative that we have explored is radical modifications of the CR production and/or spallation cross sections. We find, indeed, that large modifications of these cross-sections also provide another scenario that explains the data and preserves CR universality. However, the required modifications would be of the order of 100\% or more and are certainly beyond the (yet large) known uncertainties. We thus ultimately deem this solution unlikely.

In conclusion, both the viable scenarios which we have tested require a violation of CR universality, either of the universality of CR acceleration with consequently different source spectra for He and C, O, or of the universality of CR propagation with different diffusion for light and heavier nuclei. From a theoretical point of view, however, the first scenario has little theoretical appeal since it is at odds with the expectations from CR acceleration in SNR shocks. On the other hand, the second scenario is less problematic and can be reconciled in a framework of inhomogeneous diffusion in the Galaxy. Indeed, inhomogeneity and anisotropy are certainly present at some level, although mostly for simplicity of analysis, diffusion is typically taken homogeneous and isotropic. Whether a physically motivated scenario can provide the required amount of inhomogeneity indicated in the above results is an issue that requires dedicated theoretical studies. From the experimental and data analysis point of view, instead, further help in clarifying the picture can come from further studies of the newly released data from AMS-02 of nuclei up to iron, as well as from the expected measurements of isotopes as deuterium or the isotopes of beryllium.

\section*{Acknowledgments}

We thank Tim Linden for helpful comments and discussions as well as carefully reading the manuscript. 
The work of A.C. is supported by: ``Departments of Excellence 2018-2022'' grant awarded by the Italian Ministry of Education, University and Research (MIUR) L. 232/2016; Research grant ``The Dark Universe: A Synergic Multimessenger Approach'' No. 2017X7X85K, PRIN 2017, funded by MIUR; Research grant TAsP (Theoretical Astroparticle Physics) funded by INFN.
M.K. is partially supported by the Swedish National Space Agency under contract 117/19 and the European Research Council under grant 742104.
The majority of computations in this work were performed with computing resources granted by RWTH Aachen University under the project No. rwth0085. Furthermore, computations were enabled by resources provided by the Swedish National Infrastructure for Computing (SNIC) under the project No. 2020/5-463 partially funded by the Swedish Research Council through grant agreement no. 2018-05973.

\bibliography{bibliography}{}
\bibliographystyle{apsrev4-1.bst}

\clearpage
\newpage

\appendix*

%=====================
%    \                                           |
%      \                                         |
%        \                                       |
\begin{figure*}[b]
\centering
\includegraphics[width=0.99\textwidth]{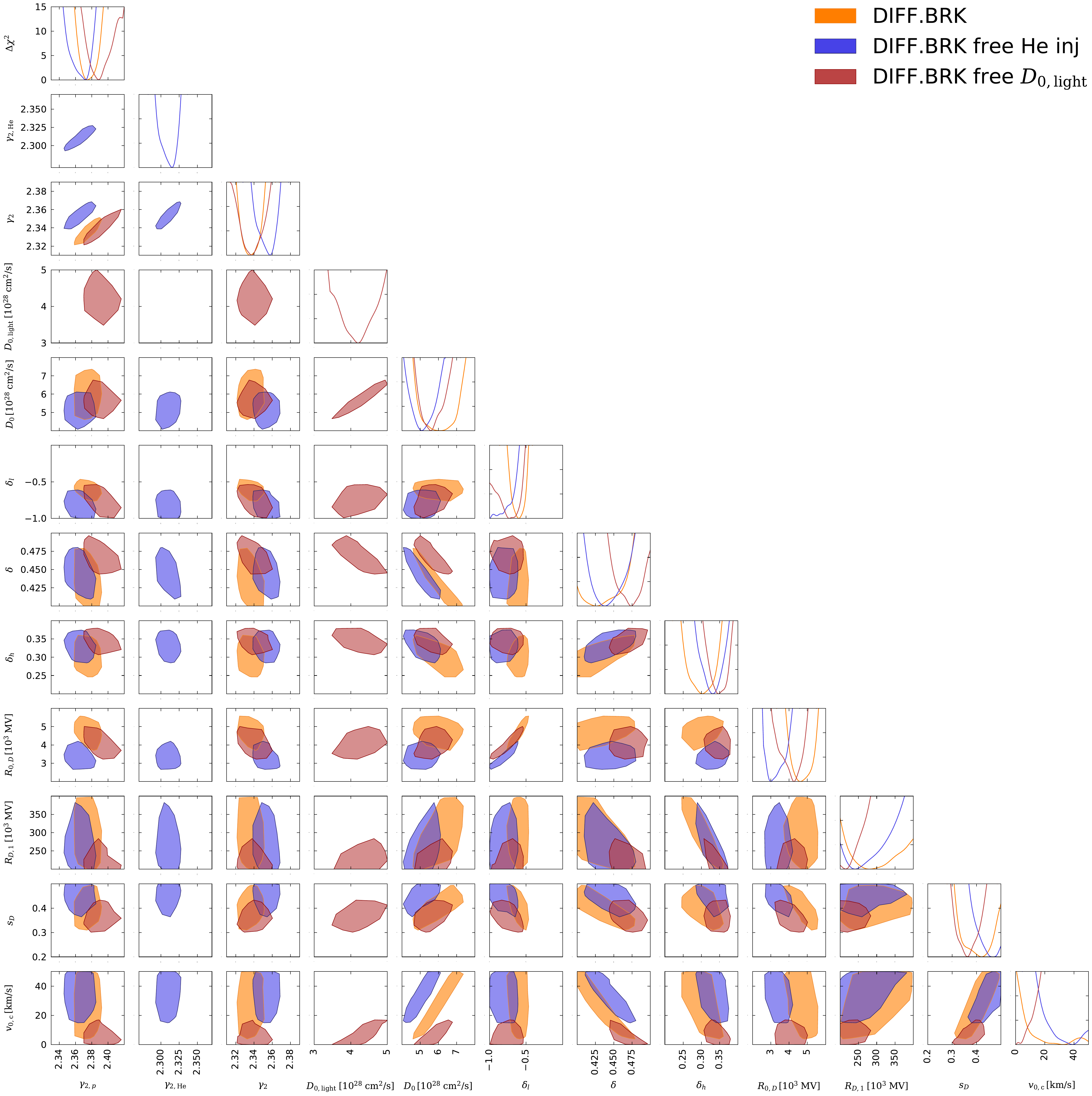}
\caption{ 
          Triangle plot of the best-fit region for the full set of CR propagation parameters. The diagonal contains the 
          $\chi^2$-profile for each individual parameter, while the contours in the lower half show the 2$\sigma$ contours derived from the
          2-dimensional $\chi^2$-profiles. All setups refer to the \DB\ propagation framework. The different colors indicate the fit setup:
          \emph{default} (yellow), \emph{free He inj}  (blue),  and \emph{free $D_{0,\rm light}$} (red). 
          \label{fig:BASE_PROP_triangele}
        }
\end{figure*}
%                                      \         |
%                                        \       |
%                                          \     |
%=====================

%=====================
%    \                                           |
%      \                                         |
%        \                                       |
\begin{figure*}[b]
\centering
\includegraphics[width=0.99\textwidth]{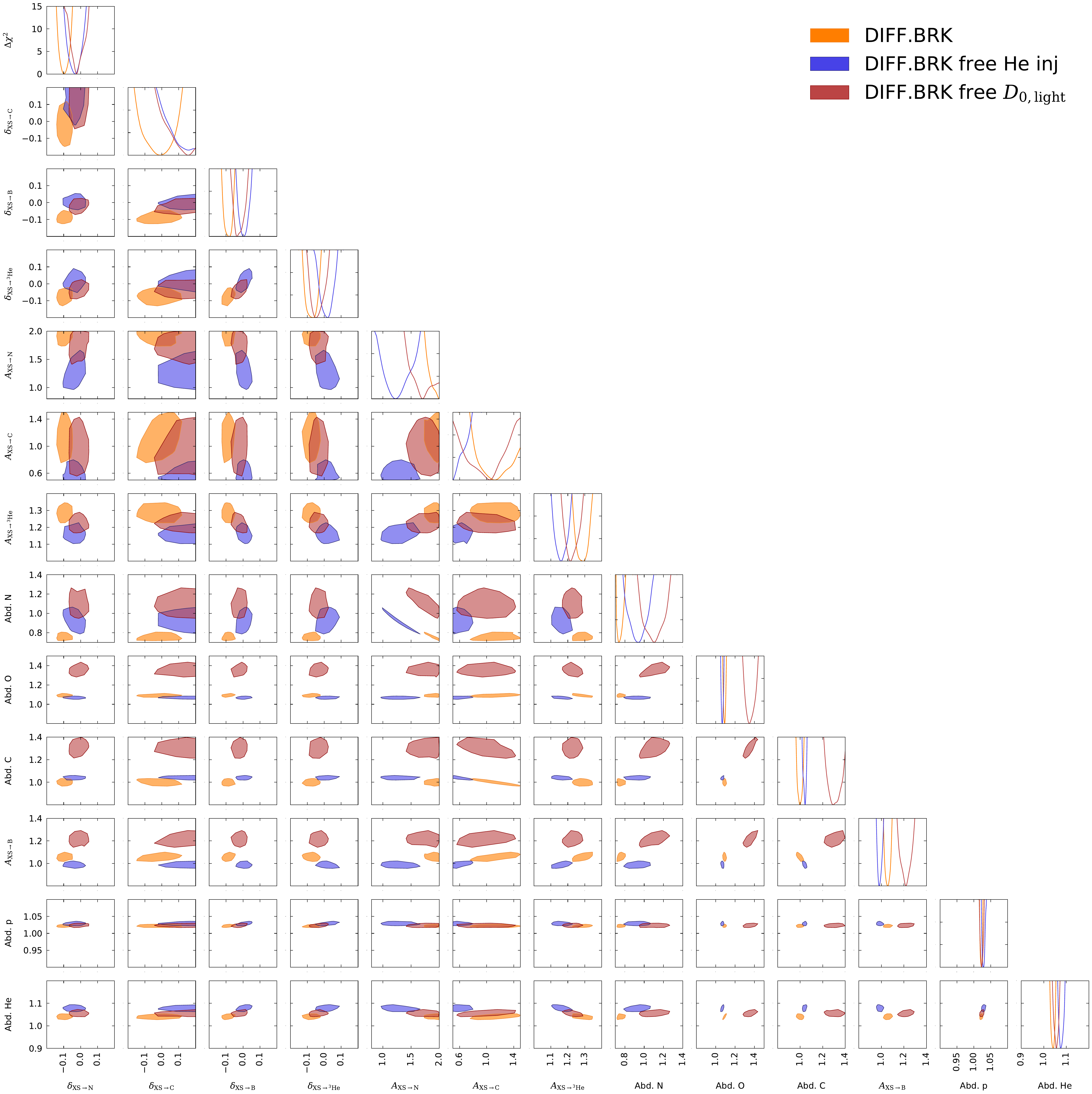}
\caption{ 
          Same as Fig.~\ref{fig:BASE_PROP_triangele}, but for the full set of cross section nuisance and CR abundance parameters.
          \label{fig:BASE_XS_triangele}
        }
\end{figure*}
%                                      \         |
%                                        \       |
%                                          \     |
%=====================

\section{Supplemental material}
\label{sec::app}

In this appendix, we collect supplemental material providing further details about the results described in the main text. In Tab.~\ref{tab::fit_results} we collect the values of all $\chi^2$s and the 
best-fit values including the $1\sigma$ uncertainties for all parameters and for all the eight main fits discussed in Sec.~\ref{sec::results}. The values from this 
table are presented in graphical form in Fig.~\ref{fig:Summary_param}. Then Figs.~\ref{fig:BASE_PROP_triangele} and~\ref{fig:BASE_XS_triangele}
show the triangle plots for the \DB\ propagation setup with the full set of propagation and cross-section nuisance parameters, respectively. We note that
Fig.~\ref{fig:DIFFBRK_smallTriangles}, contains a subset of the full triangles shown here.
Finally, in Tab.~\ref{tab::fit_results_light_heavy} we provide the $\chi^2$s and best-fit parameter values of fits performed with only the light nuclei. For comparison 
we also report the parameter values derived from the fit of the heavy nuclei  (taken from \KCh). 
These results are discussed in Sec.~\ref{sec::motivation}.

%=====================================================================
%    \                                                                                        |
%      \                                                                                      |
%        \                                                                                    |
\begin{table}    
   \caption{
        For all 8 fits discussed in Sec.~\ref{sec::strategy} we report the total $\chi^2$, the degree-of-freedom (dof), 
        the contribution to the $\chi^2$ form each single species, 
        and the best-fit value and 1 $\sigma$ error for each parameter. 
        The number of data point for each species are reported again in brackets next to the $\chi^2$s in the first column\@.  
        \label{tab::fit_results}
    }
  \centering
\scalebox{0.85}{
\renewcommand{\arraystretch}{1.5}
\begin{tabular}{cccccccccccccccccccccccc}
\hline \hline
Parameter                                                         &&&  \multicolumn{7}{c}{ $\mathrm{DIFF.BRK}$ }                                                                                                                                                       &$\;$& \multicolumn{7}{c}{ $\mathrm{INJ.BRK}+v_A$      }                                                                                                                                                \\ \cline{4-10} \cline{12-18}
$\,$                                                              &&& default                                    &    &  free He inj                               &    & free $D_{0,{\rm light}}$                   &    &  free sec. norms                           &    & default                                    &    &  free He inj                               &    & free $D_{0,{\rm light}}$                   &    &  free sec. norms                           \\ \cline{1-1} \cline{4-4} \cline{6-6} \cline{8-8} \cline{10-10} \cline{12-12} \cline{14-14} \cline{16-16} \cline{18-18}
$\chi^2$                                                          &&& $                                   341.0$ &    & $                                   193.8$ &    & $                                   205.9$ &    & $                                   184.3$ &    & $                                   528.9$ &    & $                                   171.5$ &    & $                                   254.4$ &    & $                                   276.2$ \\
dof                                                               &&& $                                     466$ &    & $                                     465$ &    & $                                     465$ &    & $                                     465$ &    & $                                     464$ &    & $                                     462$ &    & $                                     463$ &    & $                                     456$ \\
$\chi^2$  $p$ AMS-02    (67)                                      &&& $                                 {36.81}$ &    & $                                 {29.82}$ &    & $                                 {19.67}$ &    & $                                  {5.88}$ &    & $                                 {27.24}$ &    & $                                 {12.44}$ &    & $                                 {19.82}$ &    & $                                 {16.32}$ \\
$\chi^2$  $p$ Voyager   (9)                                       &&& $                                 {10.82}$ &    & $                                  {5.60}$ &    & $                                  {9.32}$ &    & $                                  {8.07}$ &    & $                                  {3.22}$ &    & $                                  {5.31}$ &    & $                                  {2.94}$ &    & $                                  {2.38}$ \\
$\chi^2$  He AMS-02     (68)                                      &&& $                                 {39.64}$ &    & $                                  {9.27}$ &    & $                                 {21.30}$ &    & $                                 {12.02}$ &    & $                                 {74.05}$ &    & $                                 {16.06}$ &    & $                                 {19.88}$ &    & $                                 {20.49}$ \\
$\chi^2$  He Voyager    (5)                                       &&& $                                  {6.88}$ &    & $                                 {10.04}$ &    & $                                  {3.68}$ &    & $                                  {7.98}$ &    & $                                  {0.33}$ &    & $                                  {5.07}$ &    & $                                  {0.78}$ &    & $                                  {1.10}$ \\
$\chi^2$  3He/4He       (26)                                      &&& $                                  {2.85}$ &    & $                                  {4.78}$ &    & $                                  {3.26}$ &    & $                                  {1.87}$ &    & $                                 {17.71}$ &    & $                                  {5.84}$ &    & $                                 {17.83}$ &    & $                                 {16.79}$ \\
$\chi^2$  $\bar p/p$    (48)                                      &&& $                                 {31.44}$ &    & $                                 {23.57}$ &    & $                                 {20.75}$ &    & $                                 {20.17}$ &    & $                                 {72.17}$ &    & $                                 {22.67}$ &    & $                                 {52.25}$ &    & $                                 {50.26}$ \\
$\chi^2$  C             (68)                                      &&& $                                 {25.95}$ &    & $                                 {23.42}$ &    & $                                 {16.88}$ &    & $                                 {17.97}$ &    & $                                 {18.40}$ &    & $                                 {23.53}$ &    & $                                 {14.16}$ &    & $                                 {13.40}$ \\
$\chi^2$  N             (67)                                      &&& $                                 {38.86}$ &    & $                                 {38.13}$ &    & $                                 {28.52}$ &    & $                                 {27.55}$ &    & $                                 {47.10}$ &    & $                                 {33.85}$ &    & $                                 {25.76}$ &    & $                                 {25.95}$ \\
$\chi^2$  O             (67)                                      &&& $                                 {86.21}$ &    & $                                 {13.80}$ &    & $                                 {23.61}$ &    & $                                 {31.26}$ &    & $                                {160.17}$ &    & $                                 {10.95}$ &    & $                                 {17.13}$ &    & $                                 {54.46}$ \\
$\chi^2$  B/C           (67)                                      &&& $                                 {49.63}$ &    & $                                 {34.37}$ &    & $                                 {34.87}$ &    & $                                 {30.89}$ &    & $                                 {97.94}$ &    & $                                 {26.46}$ &    & $                                 {41.09}$ &    & $                                 {44.69}$ \\
$\gamma_{1,p}$                                                    &&& -                                          &    & -                                          &    & -                                          &    & -                                          &    & $               {1.612}^{+0.010}_{-0.048}$ &    & $                  {1.64}^{+0.04}_{-0.05}$ &    & $                  {1.43}^{+0.16}_{-0.08}$ &    & $                  {1.21}^{+0.17}_{-0.04}$ \\
$\gamma_{2,p}$                                                    &&& $               {2.372}^{+0.010}_{-0.002}$ &    & $               {2.373}^{+0.002}_{-0.006}$ &    & $               {2.390}^{+0.002}_{-0.004}$ &    & $               {2.395}^{+0.005}_{-0.005}$ &    & $               {2.391}^{+0.004}_{-0.003}$ &    & $               {2.393}^{+0.008}_{-0.006}$ &    & $               {2.404}^{+0.007}_{-0.010}$ &    & $               {2.408}^{+0.006}_{-0.007}$ \\
$\gamma_{1,\mathrm{He}}$                                          &&& -                                          &    & -                                          &    & -                                          &    & -                                          &    & -                                          &    & $                  {1.58}^{+0.06}_{-0.06}$ &    & -                                          &    & -                                          \\
$\gamma_{2,\mathrm{He}}$                                          &&& -                                          &    & $               {2.315}^{+0.003}_{-0.004}$ &    & -                                          &    & -                                          &    & -                                          &    & $               {2.318}^{+0.008}_{-0.004}$ &    & -                                          &    & -                                          \\
$\gamma_1$                                                        &&& -                                          &    & -                                          &    & -                                          &    & -                                          &    & $                  {1.64}^{+0.02}_{-0.06}$ &    & $                  {1.90}^{+0.05}_{-0.06}$ &    & $                  {1.48}^{+0.16}_{-0.20}$ &    & $                  {1.19}^{+0.17}_{-0.14}$ \\
$\gamma_2$                                                        &&& $               {2.334}^{+0.009}_{-0.004}$ &    & $               {2.360}^{+0.001}_{-0.004}$ &    & $               {2.339}^{+0.002}_{-0.003}$ &    & $               {2.322}^{+0.006}_{-0.004}$ &    & $               {2.347}^{+0.002}_{-0.003}$ &    & $               {2.367}^{+0.008}_{-0.003}$ &    & $               {2.342}^{+0.007}_{-0.008}$ &    & $               {2.325}^{+0.006}_{-0.004}$ \\
$R_{inj,0} \;\mathrm{[ 10^{3}\;MV]}$                              &&& -                                          &    & -                                          &    & -                                          &    & -                                          &    & $                  {4.70}^{+0.06}_{-0.23}$ &    & $                  {5.91}^{+0.81}_{-0.36}$ &    & $                  {3.66}^{+0.87}_{-0.39}$ &    & $                  {3.19}^{+0.64}_{-0.21}$ \\
$s$                                                               &&& -                                          &    & -                                          &    & -                                          &    & -                                          &    & $               {0.296}^{+0.011}_{-0.007}$ &    & $                  {0.35}^{+0.02}_{-0.03}$ &    & $                  {0.36}^{+0.02}_{-0.04}$ &    & $                  {0.42}^{+0.01}_{-0.03}$ \\
$D_{0,{\rm light}}\;\mathrm{[ 10^{28}\;cm^2/s]}$                  &&& -                                          &    & -                                          &    & $                  {4.21}^{+0.08}_{-0.10}$ &    & -                                          &    & -                                          &    & -                                          &    & $                  {2.69}^{+0.15}_{-0.01}$ &    & -                                          \\
$D_{0}\,\mathrm{[ 10^{28}\;cm^2/s]}$                              &&& $                  {6.06}^{+0.82}_{-0.78}$ &    & $                  {5.13}^{+0.19}_{-0.19}$ &    & $                  {5.57}^{+0.13}_{-0.12}$ &    & $                  {8.52}^{+0.24}_{-0.70}$ &    & $                  {2.96}^{+0.02}_{-0.03}$ &    & $                  {3.15}^{+0.11}_{-0.10}$ &    & $                  {3.85}^{+0.22}_{-0.05}$ &    & $                  {5.21}^{+0.06}_{-0.08}$ \\
$\delta_{l}$                                                      &&& $                 {-0.61}^{+0.06}_{-0.04}$ &    & $                 {-0.83}^{+0.07}_{-0.12}$ &    & $                 {-0.74}^{+0.04}_{-0.04}$ &    & $                 {-0.86}^{+0.06}_{-0.07}$ &    & -                                          &    & -                                          &    & -                                          &    & -                                          \\
$\delta_{}$                                                       &&& $                  {0.43}^{+0.02}_{-0.02}$ &    & $               {0.437}^{+0.007}_{-0.006}$ &    & $               {0.472}^{+0.005}_{-0.004}$ &    & $               {0.431}^{+0.017}_{-0.005}$ &    & $               {0.476}^{+0.003}_{-0.003}$ &    & $               {0.463}^{+0.007}_{-0.010}$ &    & $               {0.483}^{+0.003}_{-0.011}$ &    & $               {0.463}^{+0.003}_{-0.007}$ \\
$\delta_{h}$                                                      &&& $                  {0.30}^{+0.02}_{-0.02}$ &    & $               {0.339}^{+0.005}_{-0.018}$ &    & $               {0.343}^{+0.008}_{-0.003}$ &    & $               {0.332}^{+0.024}_{-0.009}$ &    & $               {0.343}^{+0.005}_{-0.005}$ &    & $                  {0.34}^{+0.01}_{-0.01}$ &    & $               {0.371}^{+0.004}_{-0.023}$ &    & $                  {0.36}^{+0.01}_{-0.01}$ \\
$R_{0,D}\,\mathrm{[ 10^{3}\;MV]}$                                 &&& $                  {4.58}^{+0.52}_{-0.21}$ &    & $                  {3.27}^{+0.26}_{-0.32}$ &    & $                  {4.19}^{+0.09}_{-0.17}$ &    & $                  {3.35}^{+0.31}_{-0.16}$ &    & -                                          &    & -                                          &    & -                                          &    & -                                          \\
$R_{1,D}\,\mathrm{[ 10^{3}\;MV]}$                                 &&& $              {280.91}^{+58.39}_{-23.87}$ &    & $               {228.49}^{+24.56}_{-6.68}$ &    & $               {219.14}^{+6.29}_{-11.36}$ &    & $              {267.68}^{+15.07}_{-54.64}$ &    & $                {206.94}^{+7.00}_{-6.51}$ &    & $              {240.73}^{+20.09}_{-17.34}$ &    & $               {158.83}^{+44.83}_{-3.13}$ &    & $               {212.50}^{+9.07}_{-17.24}$ \\
$s_{D}$                                                           &&& $                  {0.42}^{+0.03}_{-0.06}$ &    & $                  {0.46}^{+0.02}_{-0.01}$ &    & $               {0.370}^{+0.007}_{-0.013}$ &    & $               {0.490}^{+0.005}_{-0.048}$ &    & -                                          &    & -                                          &    & -                                          &    & -                                          \\
$v_{0,\mathrm{c}}\,\mathrm{[km/s]}$                               &&& $               {30.74}^{+16.91}_{-17.17}$ &    & $                 {33.84}^{+4.27}_{-3.03}$ &    & $                  {2.89}^{+1.64}_{-0.80}$ &    & $                {32.26}^{+4.79}_{-19.16}$ &    & $                  {0.06}^{+0.23}_{-0.03}$ &    & $               {0.014}^{+2.732}_{-0.004}$ &    & $                  {0.18}^{+0.58}_{-0.12}$ &    & $                  {0.60}^{+0.63}_{-0.59}$ \\
$v_{\mathrm{A}}\,\mathrm{[km/s]}$                                 &&& -                                          &    & -                                          &    & -                                          &    & -                                          &    & $                  {1.53}^{+0.56}_{-0.72}$ &    & $                 {17.33}^{+1.38}_{-1.09}$ &    & $                  {0.86}^{+3.45}_{-0.85}$ &    & $                  {1.80}^{+2.62}_{-1.30}$ \\
$\delta_{\mathrm{XS}\rightarrow \mathrm{^3He}}$                   &&& $                 {-0.06}^{+0.01}_{-0.04}$ &    & $               {0.006}^{+0.021}_{-0.009}$ &    & $              {-0.061}^{+0.011}_{-0.004}$ &    & $                 {-0.03}^{+0.01}_{-0.03}$ &    & $               {0.222}^{+0.005}_{-0.012}$ &    & $                  {0.29}^{+0.01}_{-0.02}$ &    & $                  {0.19}^{+0.02}_{-0.02}$ &    & $               {0.227}^{+0.020}_{-0.008}$ \\
$\delta_{\mathrm{XS}\rightarrow \mathrm{B}}$                      &&& $              {-0.067}^{+0.003}_{-0.030}$ &    & $              {-0.007}^{+0.017}_{-0.004}$ &    & $              {-0.040}^{+0.011}_{-0.003}$ &    & $                 {-0.00}^{+0.02}_{-0.02}$ &    & $               {0.009}^{+0.004}_{-0.005}$ &    & $               {0.147}^{+0.005}_{-0.013}$ &    & $               {0.111}^{+0.005}_{-0.024}$ &    & $               {0.127}^{+0.004}_{-0.021}$ \\
$\delta_{\mathrm{XS}\rightarrow \mathrm{C}}$                      &&& $                 {-0.03}^{+0.08}_{-0.03}$ &    & $                  {0.22}^{+0.04}_{-0.07}$ &    & $                  {0.17}^{+0.02}_{-0.03}$ &    & $                  {0.11}^{+0.02}_{-0.09}$ &    & $              {-0.184}^{+0.020}_{-0.004}$ &    & $               {0.298}^{+0.001}_{-0.041}$ &    & $                  {0.16}^{+0.05}_{-0.10}$ &    & $                  {0.02}^{+0.02}_{-0.08}$ \\
$\delta_{\mathrm{XS}\rightarrow \mathrm{N}}$                      &&& $              {-0.093}^{+0.009}_{-0.020}$ &    & $              {-0.022}^{+0.007}_{-0.017}$ &    & $              {-0.027}^{+0.008}_{-0.006}$ &    & $                 {-0.02}^{+0.02}_{-0.02}$ &    & $              {-0.087}^{+0.012}_{-0.004}$ &    & $               {0.075}^{+0.008}_{-0.027}$ &    & $                  {0.06}^{+0.01}_{-0.02}$ &    & $               {0.037}^{+0.024}_{-0.010}$ \\
$A_{\mathrm{XS}\rightarrow \bar p}$                               &&& $               {1.000}^{+0.000}_{-0.000}$ &    & $               {1.000}^{+0.000}_{-0.000}$ &    & $               {1.000}^{+0.000}_{-0.000}$ &    & $                  {1.61}^{+0.02}_{-0.02}$ &    & $               {1.000}^{+0.000}_{-0.000}$ &    & $               {1.000}^{+0.000}_{-0.000}$ &    & $               {1.000}^{+0.000}_{-0.000}$ &    & $               {1.748}^{+0.005}_{-0.043}$ \\
$A_{\mathrm{XS}\rightarrow \mathrm{^3He}}$                        &&& $                  {1.29}^{+0.02}_{-0.03}$ &    & $                  {1.17}^{+0.01}_{-0.01}$ &    & $               {1.219}^{+0.009}_{-0.008}$ &    & $               {1.900}^{+0.000}_{-0.022}$ &    & $               {1.213}^{+0.008}_{-0.009}$ &    & $               {1.063}^{+0.009}_{-0.020}$ &    & $               {1.148}^{+0.033}_{-0.007}$ &    & $               {1.898}^{+0.001}_{-0.010}$ \\
$A_{\mathrm{XS}\rightarrow \mathrm{B}}$                           &&& $               {1.061}^{+0.009}_{-0.018}$ &    & $               {0.979}^{+0.006}_{-0.004}$ &    & $               {1.215}^{+0.008}_{-0.008}$ &    & $               {1.474}^{+0.009}_{-0.023}$ &    & $               {0.996}^{+0.006}_{-0.006}$ &    & $               {0.961}^{+0.004}_{-0.008}$ &    & $                  {1.19}^{+0.03}_{-0.01}$ &    & $               {1.493}^{+0.004}_{-0.033}$ \\
$A_{\mathrm{XS}\rightarrow \mathrm{C}}$                           &&& $                  {1.10}^{+0.16}_{-0.08}$ &    & $               {0.500}^{+0.039}_{-0.000}$ &    & $                  {1.07}^{+0.03}_{-0.07}$ &    & $                  {1.44}^{+0.04}_{-0.23}$ &    & $                  {0.96}^{+0.05}_{-0.08}$ &    & $                  {0.56}^{+0.02}_{-0.05}$ &    & $                  {0.70}^{+0.31}_{-0.08}$ &    & $                  {1.38}^{+0.03}_{-0.24}$ \\
$A_{\mathrm{XS}\rightarrow \mathrm{N}}$                           &&& $               {1.996}^{+0.002}_{-0.076}$ &    & $                  {1.19}^{+0.07}_{-0.04}$ &    & $                  {1.70}^{+0.05}_{-0.04}$ &    & $                  {1.76}^{+0.13}_{-0.07}$ &    & $               {2.000}^{+0.000}_{-0.028}$ &    & $                  {1.17}^{+0.08}_{-0.10}$ &    & $                  {1.72}^{+0.15}_{-0.10}$ &    & $                  {1.85}^{+0.13}_{-0.07}$ \\
Abd. $p$                                                          &&& $               {1.023}^{+0.001}_{-0.002}$ &    & $               {1.027}^{+0.002}_{-0.001}$ &    & $               {1.022}^{+0.001}_{-0.001}$ &    & $               {1.039}^{+0.000}_{-0.003}$ &    & $               {1.030}^{+0.001}_{-0.001}$ &    & $               {1.042}^{+0.002}_{-0.002}$ &    & $               {1.024}^{+0.006}_{-0.001}$ &    & $               {1.045}^{+0.000}_{-0.004}$ \\
Abd. He                                                           &&& $               {1.042}^{+0.003}_{-0.005}$ &    & $               {1.074}^{+0.003}_{-0.002}$ &    & $               {1.056}^{+0.003}_{-0.002}$ &    & $               {1.039}^{+0.002}_{-0.008}$ &    & $               {1.032}^{+0.003}_{-0.002}$ &    & $               {1.079}^{+0.003}_{-0.008}$ &    & $               {1.054}^{+0.008}_{-0.005}$ &    & $               {1.034}^{+0.001}_{-0.008}$ \\
Abd. C                                                            &&& $               {1.001}^{+0.010}_{-0.012}$ &    & $               {1.044}^{+0.002}_{-0.004}$ &    & $               {1.285}^{+0.020}_{-0.007}$ &    & $               {0.960}^{+0.009}_{-0.006}$ &    & $               {1.011}^{+0.010}_{-0.004}$ &    & $               {1.033}^{+0.006}_{-0.004}$ &    & $                  {1.41}^{+0.02}_{-0.05}$ &    & $               {0.959}^{+0.013}_{-0.004}$ \\
Abd. N                                                            &&& $               {0.733}^{+0.018}_{-0.005}$ &    & $                  {0.95}^{+0.02}_{-0.03}$ &    & $                  {1.11}^{+0.02}_{-0.02}$ &    & $                  {0.91}^{+0.02}_{-0.05}$ &    & $               {0.708}^{+0.008}_{-0.001}$ &    & $                  {0.95}^{+0.05}_{-0.04}$ &    & $                  {1.17}^{+0.05}_{-0.06}$ &    & $                  {0.88}^{+0.02}_{-0.04}$ \\
Abd. O                                                            &&& $               {1.092}^{+0.006}_{-0.006}$ &    & $               {1.062}^{+0.008}_{-0.001}$ &    & $               {1.345}^{+0.014}_{-0.007}$ &    & $               {1.008}^{+0.003}_{-0.009}$ &    & $               {1.112}^{+0.003}_{-0.002}$ &    & $               {1.070}^{+0.004}_{-0.009}$ &    & $                  {1.44}^{+0.02}_{-0.03}$ &    & $               {1.009}^{+0.002}_{-0.008}$ \\
$\varphi_p\,\mathrm{[GV]}$                                        &&& $              {474.40}^{+12.06}_{-25.70}$ &    & $              {627.32}^{+30.14}_{-19.69}$ &    & $               {528.37}^{+14.86}_{-6.64}$ &    & $              {554.03}^{+17.65}_{-18.70}$ &    & $               {597.58}^{+14.42}_{-4.95}$ &    & $              {857.43}^{+11.88}_{-15.74}$ &    & $              {664.03}^{+22.70}_{-53.84}$ &    & $              {639.57}^{+14.17}_{-35.62}$ \\
$\varphi_{\rm HeBCNO}\;\mathrm{[GV]}$                             &&& $               {565.54}^{+7.08}_{-21.99}$ &    & $              {627.71}^{+19.09}_{-14.32}$ &    & $               {568.55}^{+10.21}_{-4.70}$ &    & $              {599.84}^{+11.00}_{-12.75}$ &    & $               {566.27}^{+11.62}_{-7.02}$ &    & $              {647.58}^{+29.37}_{-20.84}$ &    & $               {582.80}^{+9.34}_{-48.79}$ &    & $               {597.13}^{+9.55}_{-30.30}$ \\
$\varphi_{\bar p}-\varphi_p\,\mathrm{[GV]}$                       &&& $             {-144.87}^{+39.49}_{-25.97}$ &    & $               {13.94}^{+16.85}_{-12.30}$ &    & $              {-50.34}^{+12.89}_{-12.23}$ &    & $              {-23.60}^{+14.13}_{-26.56}$ &    & $               {78.24}^{+10.83}_{-14.21}$ &    & $             {-123.25}^{+44.19}_{-31.95}$ &    & $              {141.75}^{+21.38}_{-69.97}$ &    & $              {197.22}^{+20.63}_{-23.57}$ \\
\hline
\hline
\end{tabular}
}
\renewcommand{\arraystretch}{1.0}
\end{table}
%                                                                                  \         |
%                                                                                    \       |
%                                                                                      \     |
%=====================================================================

%=====================================================================
%    \                                                                                        |
%      \                                                                                      |
%        \                                                                                    |
\begin{table}
   \caption{
        Comparison of best-fit CR propagation parameters and uncertainties at the 1$\sigma$ C.L., as well as the $\chi^2$s for fits, performed only on the light nuclei ($p$, He, $\bar p/p$, and $^3$He/$^4$He) 
        or the heavier nuclei (B/C, C, N, O). The results on the heavier nuclei are taken from \KCh.
        Results are provided for two CR propagation frameworks, \DB\ and \DR.
        \label{tab::fit_results_light_heavy}
    }
  \centering
\scalebox{0.85}{
\renewcommand{\arraystretch}{1.5}
\begin{tabular}{ccccccccc}
\hline \hline
Parameter && \multicolumn{3}{c}{\DB}  &$\;$& \multicolumn{3}{c}{\DR} \\ \cline{3-5} \cline{7-9} 
                                                              &$\;$& BCNO                                   &$\;$& $p\,$He$\,\bar p\,$$^3$He                  && BCNO                                   &$\;$& $p\,$He$\,\bar p\,$$^3$He                  \\ \cline{1-1} \cline{3-3} \cline{5-5} \cline{7-7} \cline{9-9} 
$\chi^2$                                                          && $                                    72.4$ && $                                    67.3$ && $                                    74.2$ && $                                    88.8$ \\
dof                                                               && $                                     252$ && $                                     207$ && $                                     251$ && $                                    205 $ \\
$\chi^2$  $p$ AMS-02 (67)                                         && -                                          && $                                  {14.3}$ && -                                          && $                                  {19.1}$ \\
$\chi^2$  $p$ Voyager (9)                                         && -                                          && $                                   {6.0}$ && -                                          && $                                   {7.4}$ \\
$\chi^2$  He AMS-02 (68)                                          && -                                          && $                                  {11.6}$ && -                                          && $                                  {14.7}$ \\
$\chi^2$  He Voyager (5)                                          && -                                          && $                                   {5.4}$ && -                                          && $                                   {9.2}$ \\
$\chi^2$  $\bar p/p$ AMS-02 (48)                                  && -                                          && $                                  {23.9}$ && -                                          && $                                  {34.7}$ \\
$\chi^2$  $^3$He/$^4$He (26)                                      && -                                          && $                                   {4.9}$ && -                                          && $                                   {2.6}$ \\
$\chi^2$  B/C (67)                                                && $                                  {27.8}$ && -                                          && $                                  {25.9}$ && -                                          \\
$\chi^2$  C (68)                                                  && $                                  {13.1}$ && -                                          && $                                  {13.7}$ && -                                          \\
$\chi^2$  N (67)                                                  && $                                  {15.9}$ && -                                          && $                                  {17.2}$ && -                                          \\
$\chi^2$  O (67)                                                  && $                                  {14.0}$ && -                                          && $                                  {14.8}$ && -                                          \\
$\gamma_{1,p}$                                                    && -                                          && -                                          && -                                          && $                  {1.68}^{+0.01}_{-0.08}$ \\
$\gamma_{2,p}$                                                    && -                                          && $                  {2.36}^{+0.02}_{-0.01}$ && -                                          && $               {2.420}^{+0.007}_{-0.018}$ \\
$\gamma_1$                                                        && -                                          && -                                          && $                  {1.20}^{+0.42}_{-0.16}$ && $                  {1.76}^{+0.02}_{-0.06}$ \\
$\gamma_2$                                                        && $               {2.357}^{+0.003}_{-0.005}$ && $                  {2.31}^{+0.02}_{-0.01}$ && $               {2.362}^{+0.016}_{-0.004}$ && $               {2.377}^{+0.006}_{-0.015}$ \\
$R_{{\rm inj},0} \;\mathrm{[ 10^{3}\;MV]}$                        && -                                          && -                                          && $                  {3.28}^{+1.82}_{-0.59}$ && $                  {6.76}^{+0.28}_{-0.87}$ \\
$s$                                                               && -                                          && -                                          && $               {0.490}^{+0.009}_{-0.052}$ && $                  {0.35}^{+0.03}_{-0.04}$ \\
$D_{0}\,\mathrm{[ 10^{28}\;cm^2/s]}$                              && $                  {5.05}^{+0.99}_{-1.34}$ && $                  {3.54}^{+0.53}_{-0.15}$ && $                  {4.16}^{+0.33}_{-0.88}$ && $                  {3.31}^{+0.07}_{-0.25}$ \\
$\delta_{l}$                                                      && $                 {-0.98}^{+0.22}_{-0.01}$ && $                 {-0.71}^{+0.05}_{-0.17}$ && -                                          && -                                          \\
$\delta_{}$                                                       && $                  {0.49}^{+0.03}_{-0.04}$ && $                  {0.50}^{+0.01}_{-0.04}$ && $                  {0.45}^{+0.02}_{-0.02}$ && $               {0.423}^{+0.021}_{-0.007}$ \\
$\delta_{h}$                                                      && $               {0.315}^{+0.045}_{-0.008}$ && $                  {0.38}^{+0.02}_{-0.02}$ && $                  {0.30}^{+0.04}_{-0.02}$ && $                  {0.34}^{+0.02}_{-0.02}$ \\
$R_{D,0}  \;\mathrm{[ 10^{3}\;MV]}$                               && $                  {3.94}^{+0.52}_{-0.35}$ && $                  {3.82}^{+0.21}_{-0.52}$ && -                                          && -                                          \\
$s_{D}$                                                           && $                  {0.38}^{+0.06}_{-0.11}$ && $                  {0.39}^{+0.06}_{-0.02}$ && -                                          && -                                          \\
$R_{D,1}  \;\mathrm{[ 10^{3}\;MV]}$                               && $              {180.24}^{+13.05}_{-29.71}$ && $              {226.89}^{+52.22}_{-38.44}$ && $              {214.34}^{+16.02}_{-39.90}$ && $              {234.07}^{+41.40}_{-14.34}$ \\
$v_{0,\mathrm{c}}\,\mathrm{[km/s]}$                               && $                 {3.34}^{+21.76}_{-2.49}$ && $                 {4.29}^{+10.85}_{-2.84}$ && $                  {0.34}^{+3.88}_{-0.23}$ && $                  {0.18}^{+0.77}_{-0.13}$ \\
$v_{\mathrm{A}}\,\mathrm{[km/s]}$                                 && -                                          && -                                          && $                 {19.23}^{+3.65}_{-3.77}$ && $                 {18.17}^{+0.40}_{-2.25}$ \\
$\delta_{\mathrm{XS}\rightarrow \mathrm{^3He}}$                   && -                                          && $                  {0.01}^{+0.01}_{-0.03}$ && -                                          && $                  {0.08}^{+0.01}_{-0.04}$ \\
$\delta_{\mathrm{XS}\rightarrow \mathrm{B}}$                      && $              {-0.065}^{+0.084}_{-0.008}$ && -                                          && $                  {0.16}^{+0.03}_{-0.04}$ && -                                          \\
$\delta_{\mathrm{XS}\rightarrow \mathrm{C}}$                      && $                 {-0.08}^{+0.23}_{-0.08}$ && -                                          && $                  {0.28}^{+0.02}_{-0.09}$ && -                                          \\
$\delta_{\mathrm{XS}\rightarrow \mathrm{N}}$                      && $                 {-0.08}^{+0.07}_{-0.03}$ && -                                          && $                  {0.10}^{+0.02}_{-0.04}$ && -                                          \\
$A_{\mathrm{XS}\rightarrow \bar p/p}$                             && -                                          && $               {1.000}^{+0.000}_{-0.000}$ && -                                          && $               {1.000}^{+0.000}_{-0.000}$ \\
$A_{\mathrm{XS}\rightarrow \mathrm{^3He}}$                        && -                                          && $                  {1.13}^{+0.03}_{-0.02}$ && -                                          && $                  {0.83}^{+0.01}_{-0.02}$ \\
$A_{\mathrm{XS}\rightarrow \mathrm{B}}$                           && $                  {1.11}^{+0.04}_{-0.13}$ && -                                          && $                  {1.16}^{+0.01}_{-0.17}$ && -                                          \\
$A_{\mathrm{XS}\rightarrow \mathrm{N}}$                           && $                  {1.18}^{+0.04}_{-0.16}$ && -                                          && $                  {1.19}^{+0.03}_{-0.17}$ && -                                          \\
$A_{\mathrm{XS}\rightarrow \mathrm{C}}$                           && $                  {0.55}^{+0.04}_{-0.04}$ && -                                          && $                  {0.54}^{+0.04}_{-0.04}$ && -                                          \\
Abd. p                                                            && -                                          && $               {1.026}^{+0.004}_{-0.002}$ && -                                          && $               {1.036}^{+0.002}_{-0.003}$ \\
Abd. He                                                           && -                                          && $                  {1.07}^{+0.01}_{-0.01}$ && -                                          && $               {1.051}^{+0.013}_{-0.005}$ \\
Iso. Abd.  C                                                      && $                      {3592}^{+82}_{-21}$ && -                                          && $                      {3583}^{+65}_{-36}$ && -                                          \\
Iso. Abd.  N                                                      && $                        {325}^{+18}_{-6}$ && -                                          && $                       {337}^{+26}_{-39}$ && -                                          \\
Iso. Abd.  O                                                      && $                     {4345}^{+181}_{-21}$ && -                                          && $                      {4312}^{+181}_{-4}$ && -                                          \\
$\varphi$ AMS-02                                                  && $              {613.67}^{+44.11}_{-13.66}$ && $              {587.28}^{+25.73}_{-17.22}$ && $              {590.95}^{+56.33}_{-13.03}$ && $              {761.88}^{+13.11}_{-20.95}$ \\
$\varphi_{\bar p}-\varphi_p$ AMS-02                               && -                                          && $                 {23.85}^{+1.57}_{-1.90}$ && -                                          && $                 {34.71}^{+8.40}_{-1.09}$ \\
\hline
\hline
\end{tabular}
\renewcommand{\arraystretch}{1.0}
}
\end{table}
%%                                                                                  \         |
%%                                                                                    \       |
%%                                                                                      \     |
%%=====================================================================

\end{document}